\documentclass[
aps,prx,twocolumn,english,showpacs,superscriptaddress,amssymb,amsfonts]{revtex4-1}
\usepackage{graphicx}
\usepackage{epsfig}
\usepackage{color}
\usepackage{amsmath}
\usepackage{amsfonts}
\usepackage{mathrsfs}
\usepackage{txfonts}
\usepackage{color}
\usepackage{bm}

\usepackage[normalem]{ulem}

\usepackage{xcolor}

\usepackage[breaklinks]{hyperref} 
\hypersetup{colorlinks=true, linkcolor=blue, citecolor=blue, filecolor=blue, urlcolor=blue}

\renewcommand{\mathbf}{\bm}

\begin{document}

	\title{Floquet chiral hinge modes and their interplay with Weyl physics in a three-dimensional lattice}
	
	\author{Biao Huang}
	\email{phys.huang.biao@gmail.com} 
	\affiliation{Max Planck Institute for the Physics of Complex Systems, N\"{o}thnitzer Stra{\ss}e 38, 01069 Dresden, Germany}
	
	\author{Viktor Novi\v{c}enko}
		\email{viktor.novicenko@tfai.vu.lt}
	\affiliation{Institute of Theoretical Physics and Astronomy,
		Vilnius University, Saul\.{e}tekio 3, LT-10257 Vilnius, Lithuania}
	
	\author{Andr\'{e} Eckardt}
	\email{eckardt@tu-berlin.de}
	\affiliation{Max Planck Institute for the Physics of Complex Systems, N\"{o}thnitzer Stra{\ss}e 38, 01069 Dresden, Germany}
	\affiliation{Institut f\"{u}r Theoretische Physik, Technische Universit\"{a}t Berlin, Hardenbergstra{\ss}e 36, 10623 Berlin, Germany}
	
	\author{Gediminas Juzeli\={u}nas}
	\email{gediminas.juzeliunas@tfai.vu.lt} 
	\affiliation{Institute of Theoretical Physics and Astronomy,
		Vilnius University, Saul\.{e}tekio 3, LT-10257 Vilnius, Lithuania}

	\date{\today}

	\begin{abstract}
 	We demonstrate that a three dimensional time-periodically driven (Floquet) lattice  can exhibit   chiral hinge states and describe   their interplay with Weyl physics.   A peculiar type of   the hinge states   are enforced by the repeated boundary reflections with lateral   Goos-H\"{a}nchen like shifts   occurring at the second-order boundaries of our system. Such chiral hinge modes  coexist in a wide range of parameters regimes with Fermi arc surface states connecting a pair of Weyl points in a two-band model. 
	 We  find numerically that these modes still preserve their locality along the hinge and their chiral nature in the presence of local defects and   other parameter changes. We trace the robustness of such chiral hinge modes to special band structure unique in a Floquet system allowing  all the eigenstates to be localized in  quasi-one-dimensional regions parallel to each other   when open hinge boundaries are introduced.
 	The implementation of a model featuring both the second-order Floquet skin effect and the Weyl physics is straightforward with ultracold atoms in optical superlattices.
	\end{abstract}
	\maketitle

	\section{Introduction}

	In recent years, researches have demonstrated that time-periodically driven  (Floquet) systems can show intriguing and unique effects that find no counterparts in non-driven systems. Examples include anomalous Floquet topological insulators featuring robust chiral edge modes for vanishing Chern numbers~\cite{Kitagawa2010,Rudner2013,Roy2016,Yao2017,Mukherjee2017,Peng2016,Maczewsky2016,Keyserlingk2016a,Else2016,Potirniche2016,Wintersperger2020} and discrete time  crystals~\cite{Sacha2015,Khemani2016,Else2016a,Yao2017a,Zhang2017,Choi2017,Rovny2018,Ho2017,Huang2018,Sacha2017,Yao2018,Sacha2020Book}. The periodic driving shifts the fundamental theoretical framework from focusing on Hamiltonian eigenproblems to the unitary evolution operators genuinely depending on time, resulting in a plethora of new concepts and methods such as space-time winding numbers~\cite{Rudner2013} and spectral pairing~\cite{Keyserlingk2016}.  In this context, it appears natural and tantalizing to explore possible new classes of periodically driven systems that go beyond descriptions by traditional theories.
	
	In this paper, we show that time-periodic driving  of a three dimensional (3D) lattice can give rise to a new type of chiral hinge  (i.e. second-order boundary) states.  
	Such hinge modes are associated with a reorganization of the Floquet eigenstates in response to shifting from periodic to open boundary conditions. The phenomena may be intuitive understood starting from a fine-tuned point, where the quasi-energy bands  have linear dispersion along only one direction and remain flat along the other two.  When open hinge boundaries are introduced, this  leads to reorganization of eigenstates  of the system  into a set states of the quasi-one-dimensional nature orienting parallel to each other and located at various distances from the hinge. Among them, the modes closest to certain lattice hinge correspond to trajectories with uni-directional (chiral) motion within each driving period. 
	Importantly, the counterpart of such a chiral hinge mode with opposite chirality resides on the other side of the lattice.   Therefore the chiral transport carried by these hinge modes would be preserved under  a small perturbation  assuming  it does not mix eigenstates spatially separating far away from each other. Such a robustness is verified numerically by observing  the particle transport encountering perturbations in the form of parameter change and local defects.
	
	Surface modes exist widely under various circumstances, and it is helpful to mention what may be different in our system. First, unlike the case of (Floquet) higher-order topological phases~\cite{Benalcazar2017,Benalcazar2017a,Khalaf2018,Yan2018,Wang2018b,Huang2020,Peng2018,Bomantara2019,RodriguezVega2019,Plekhanov2019}, the 
	hinge states studied here  are formed inside the bulk spectrum near   zero quasi-energy, as can be seen in Fig.~\ref{fig:spectrum} column (2). Their existence and robustness, therefore, requires a new theoretical explanation without invoking a bulk gap.   Secondly, the localization of modes at the boundaries resembles the non-Hermitian skin effect   \cite{Yao18PRL,budich2020RMP,Ueda2020AdvPhys,Kawabata2020PRB,Wu2019,Xiao2020,Song2019}, with a notable difference being the reduced dimensionality which may be thought of as a higher-order skin effect. Thirdly, different from one-dimensional  (1D) periodically driven lattices~\cite{Budich2017}   exhibiting helical modes,   
in our system the  modes with opposite chirality   residing at opposite hinges   are well spatially separated,   so the scattering due to a local perturbation would not reverse the chiral direction of the particle transport.

	The aforementioned hinge states coexist with Weyl physics~\cite{Armitage2018, Anderson2012, Anderson2013, Dubcek2015, Sun2018PRL, Higashikawa2019, Lu2020, Wang2020, Zhu2020,Lang2017} when parameters are tuned away from the  limit of uni-directional dispersionless motion.  In that case a pair of Weyl points is created at quasienergy $\pi$ leading  to a Fermi-arc surface (i.e. first-order boundary) states. We  find numerically that the parameter regime  of the hinge states overlaps considerably with that for  the Weyl physics, signaling a simultaneous observation of both phenomena in a unified experimental setting.

 The model system proposed and studied here consists of a  simple, stepwise modulation of tunneling matrix elements  involving six steps. It generalizes   to three dimensions a two-dimensional lattice model introduced by Rudner {\it et. al.}~\cite{Rudner2013}  for studying anomalous Floquet topolgical insulators (see also Ref.~\cite{Kitagawa2010}).  The latter 2D tunnel modulation has been recently applied to ultracold atoms for the realization of anomalous topological band structures~\cite{Wintersperger2020}.   Our 3D model can equally be implemented with ultracold atoms using such a stepwise tunnel modulation, now in 3D optical superlattices.   Furthermore, besides giving rise to a new phenomenon, 
the  unconventional chiral second-order Floquet  skin effect,  the model proposed here also provides a simple recipe for the implementation of Weyl physics  by means of  time-periodic driving.

	  This paper is structured as follows. 
	In the next Section~\ref{sec:Model} a 3D periodically driven lattice is defined.  
	Subsequently the characteristic features of the bulk and hinge physics are considered in Secs.~\ref{sec:Periodic}  and \ref{sec:Hinge}. 
	The experimental implementation of our model,  using ultracold atoms in modulated superlattices, is discussed in Sec.~\ref{sec:Experimental},   followed by  the Concluding Section~\ref{sec:Conclusion}.  Additional technical details are presented in four Appendices \ref{Appendix:Evolution-operator-along-diagonal} -- \ref{Appendix:topo}.

	\begin{figure}[h]
		\parbox{8.5cm}{
			\parbox{6.7cm}{
			\parbox{3.2cm}{
				\includegraphics[width=3.0cm]{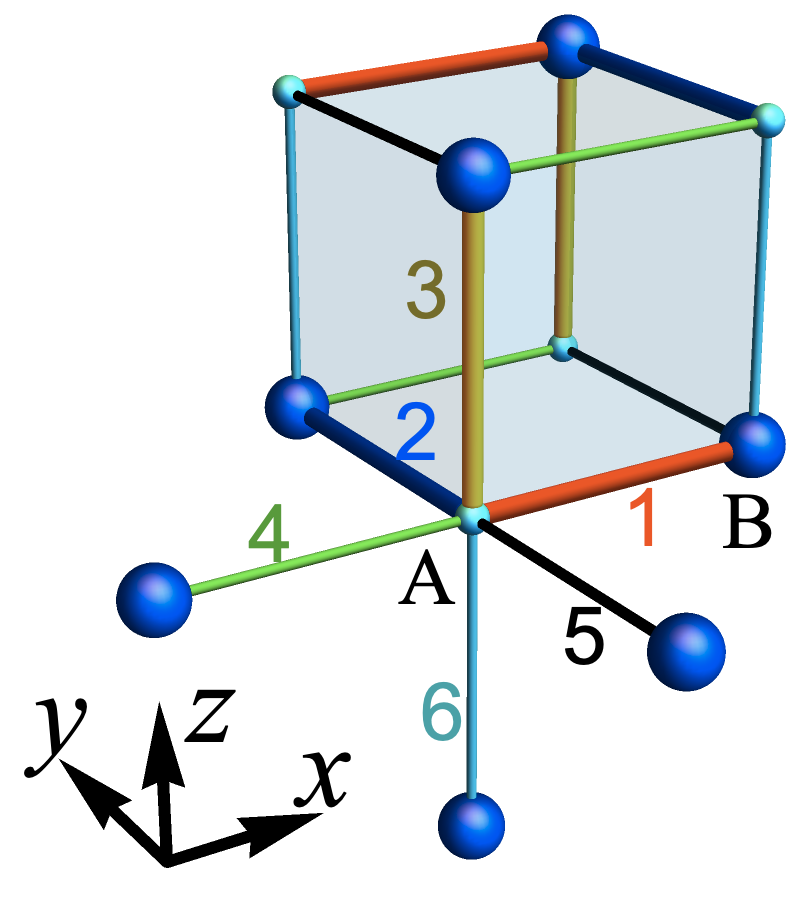}
				\\
				(a) Driving 
			}
			\parbox{3.4cm}{ 
				\includegraphics[width=3.5cm]{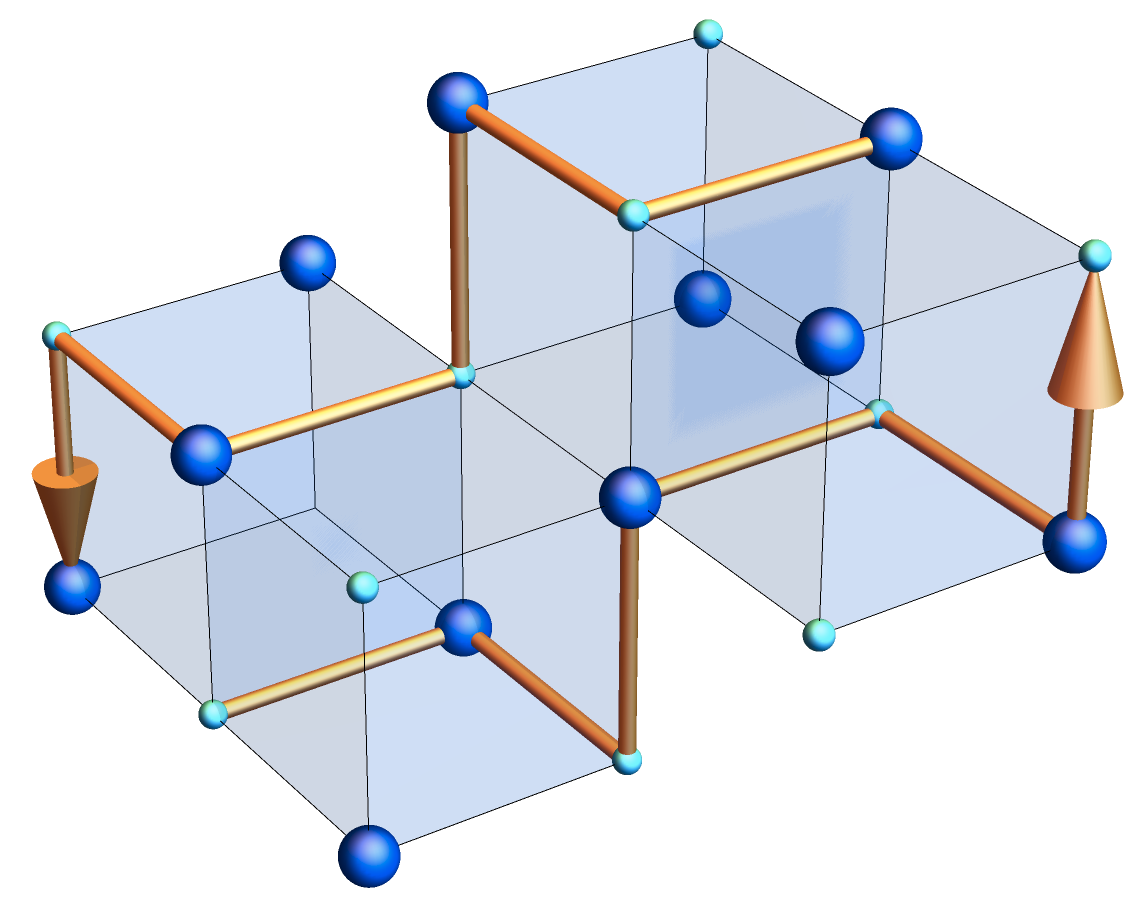}\\
				\quad \\ \quad \\
				(b) Bulk dynamics  
			}			
		\\
		\parbox{6.6cm}{
		\includegraphics[width=3.15cm]{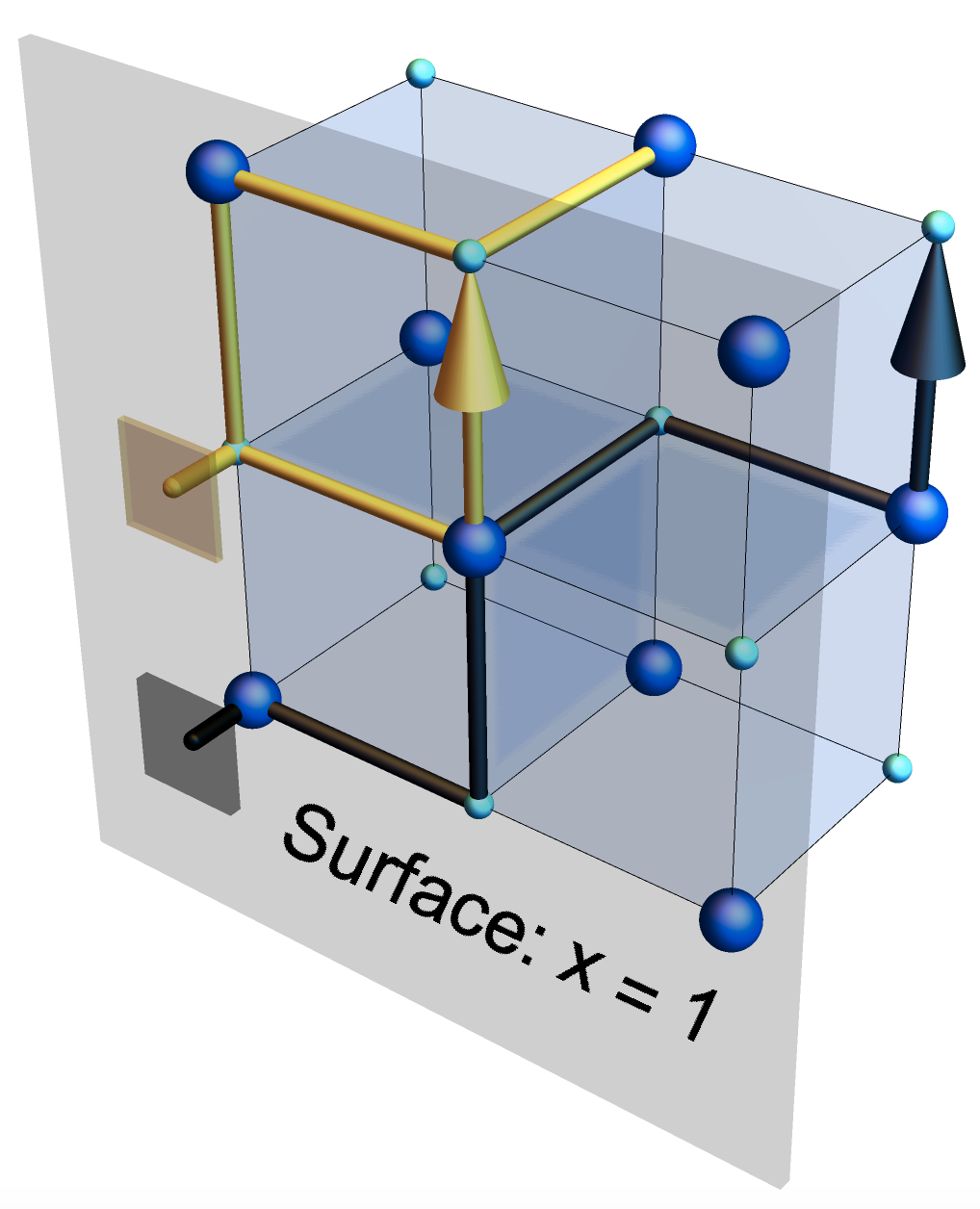}
		\includegraphics[width=3.25cm]{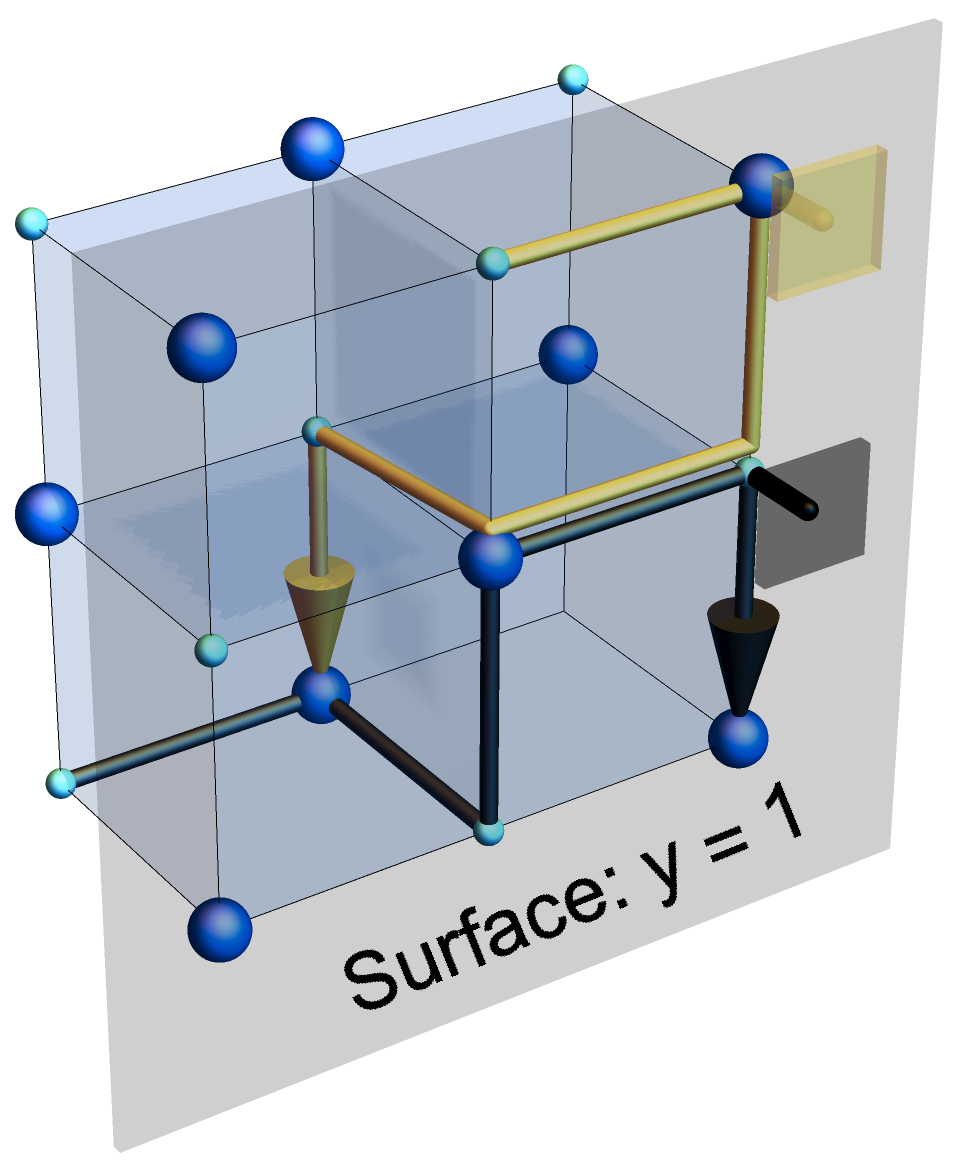}\\
		(c) Reflection by open boundary surfaces at  $ x=1 $  and  $ y=1 $  accompanied by  sublattice changes  $B \rightarrow A$ and $A \rightarrow B$, respectively.} 
		}
		\parbox{1.7cm}{
			\includegraphics[width=1.7cm]{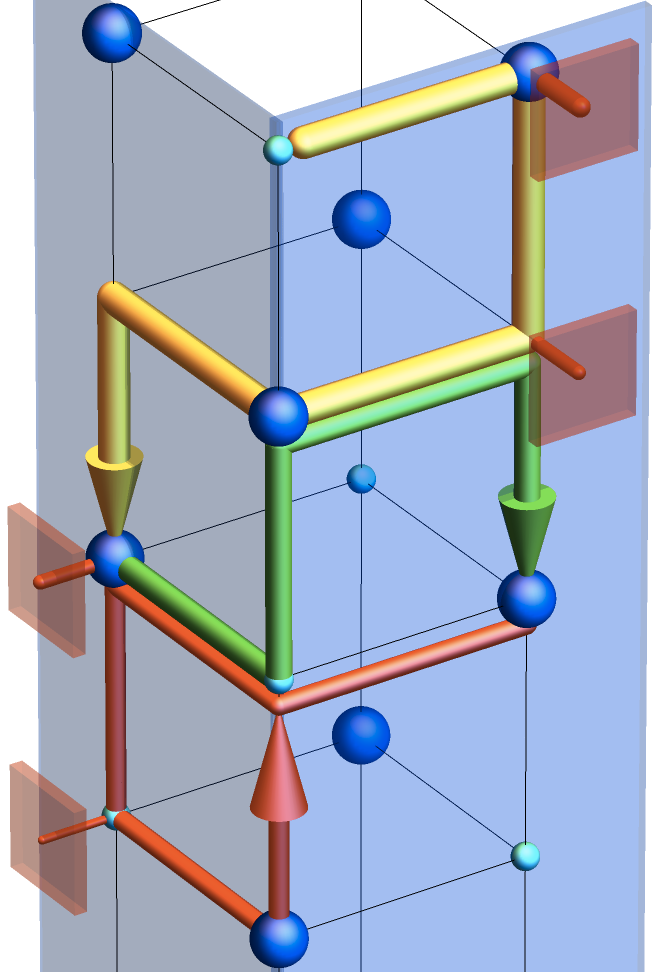}\\
			\quad \\ \quad \\
			\includegraphics[width=1.6cm]{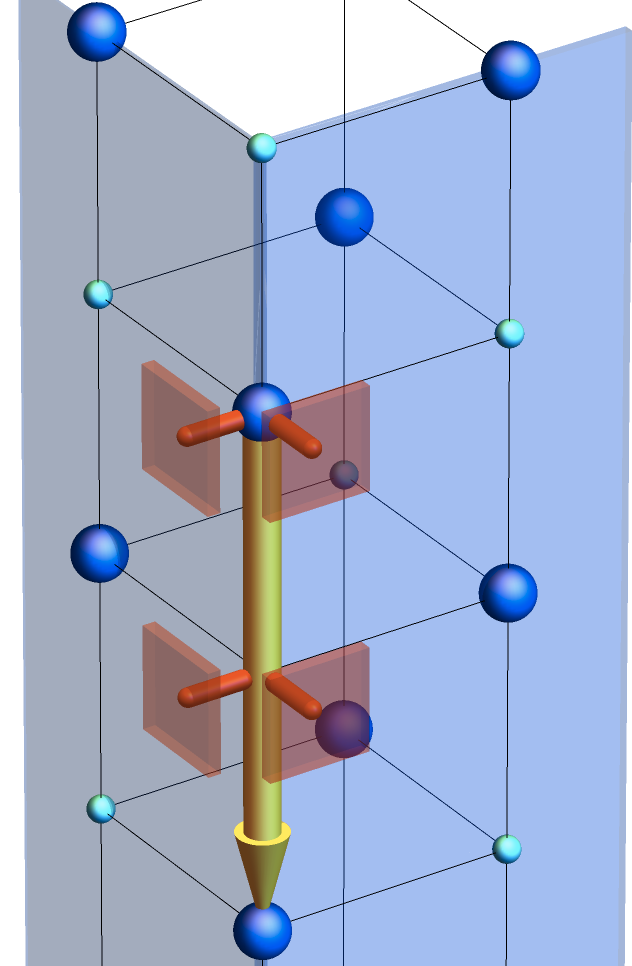}\\ \quad \\
			(d) Hinge dynamics 
		}
		}
		\caption{\label{fig:model} 
			{\bf (a)}  Bonds connected during the driving steps 1 to 6.		{\bf (b)} Bulk  trajectories within a Floquet cycle at the fine tuned point $ \phi=\pi/2 $. Depending on the starting sublattice, particles will travel in opposite directions along the cubic diagonal  $\mathbf{d} = (1,-1,1) $ within a period. 
			{\bf (c)} Trajectories at the same $ \phi=\pi/2 $ but with a surface termination (open boundary). Particles starting from sublattice $A$ (or $B$) near the  $ y=1 $ (or   $ x=1 $) surface would have their dynamics impeded by the open boundary at a certain driving step. That results in a switch of sublattice after a Floquet cycle, and therefore the direction of the trajectory is reversed after the reflection from the surface. 
			{\bf (d)}  The hinge formed by two intersecting terminating surfaces renders uni-directional modes   at $ \phi=\pi/2 $. The figure shows two of the  modes closest to the hinge starting  at a site of  $B$  (lower plot) or  $A$ (upper plot) sublattices  directly at the hinge. Each color for the arrow denotes one Floquet cycle.  The lower-panel of (d) shows the mode situated directly at the hinge   which undergoes a uni-directional motion along $ z $ within a period.
 }
	\end{figure}

	\section{Model \label{sec:Model}}
	 Consider a bipartite cubic lattice with alternating A-B sublattices in all  three Cartesian directions. 
	The lattice is described by a time-periodic Hamiltonian  $ H(t+T)=H(t) $, with the driving period $T$ divided into 6 steps. In each step  tunneling matrix elements $-J$ between sites $\mathbf{r}_A$ of sublattice $A$ and neighboring sites $\mathbf{r}_A\pm a\mathbf{e}_\mu$ of sublattice $B$ are switched on, with $\mu=x,y,z$. During the driving period $T$ the tunneling steps appear in a sequence $  \mu\pm = x+,\, y+,\, z+,\, x-,\, y-,\, z-$, as illustrated in Fig.~\ref{fig:model}(a) \footnote{Without including the tunneling along the $z$ direction described by third and the sixth driving steps, the dynamics reduces to a 2D motion in a periodically driven square lattice considered in refs. \cite{Rudner2013,Mukherjee2017,Wintersperger2020}.}. 	
	Within each step the evolution is determined by a coordinate-space Hamiltonian 
	$ H_{\pm\mu} = 
		-J\sum_{\boldsymbol{r}_A} 
		(|\boldsymbol{r}_A \rangle 
		\langle \boldsymbol{r}_A\pm a \boldsymbol{e}_\mu| 
		+ 
		|  \boldsymbol{r}_A\pm a \boldsymbol{e}_\mu\rangle 
		\langle \boldsymbol{r}_A|  ) $, 
	where   $-J $  is the tunneling matrix element, 
	$ \boldsymbol{r}_A $ specifies the location of sublattices $ A $, and $ a $ is the lattice  spacing such that $ \boldsymbol{r}_A \pm a\boldsymbol{e}_\mu $ denotes the locations of sites in sublattice $ B $ neighboring to the sublattice $ A $ site  $ \boldsymbol{r}_A $.  The tunneling processes occurring in each of the driving steps are characterized by a single dimensionless parameter, the  phase
	\begin{equation}
	\phi = -\frac{JT}{6\hbar} .
	\end{equation}

 The one-cycle evolution operator (or Floquet operator),
\begin{equation}
U_F = {\cal T} e^{-(i/\hbar) \int_0^T dt H(t) },
\end{equation}
whose repeated application describes the time-evolution in stroboscopic steps of the driving period  $T$, can be decomposed into terms corresponding to the six driving stages,  
	\begin{equation} \label{eq:uf}
	U_F = U_{z-}U_{y-}U_{x-} U_{z+} U_{y+} U_{x+}.
	\end{equation}

When dealing with the bulk dynamics we impose periodic boundary conditions in all three spatial directions. The evolution operators for the individual driving steps can then be represented as:
	\begin{equation} \label{eq:u_mu}
 U_{\mu\pm} (\boldsymbol{k})= e^{-\frac{iT}{6\hbar}H_{\mu\pm}(\boldsymbol{k}) } = e^{-i \phi ( \tau_1 \cos k_\mu  \mp \tau_2 \sin k_\mu ) }\,, 
	\end{equation}
where  $ \tau_{1,2,3} $ are Pauli matrices associated with the sublattice states $A$ (corresponds to spin-up state) and $B$ (corresponds to spin-down state) and where $k_{\mu}$  with $\mu = x,y,z$  denotes the  Cartesian components of the quasimomentum vector  $\mathbf{k}$. 
 Here and in the following, we will use a dimensionless description, where time, energy, length and quasimomentum are given in units of $ T, \hbar/T $, $a$, and $\hbar/a$, respectively.  The quasienergies $E_{n,\boldsymbol{k}}$ and the Floquet modes $ \left|n,\boldsymbol{k}\right\rangle $  are defined via  the eigenvalue  equation   for the Floquet operator 
\begin{equation} \label{eq:uf-eigen-equation}
U_F | u_{n,\boldsymbol{k}} \rangle  =\exp(-i E_{n,\boldsymbol{k}} ) | u_{n,\boldsymbol{k}} \rangle .
\end{equation}  

	We first note that the only global symmetry satisfied by the Floquet operator \eqref{eq:uf}-\eqref{eq:u_mu} 
	is  a  particle-hole flip $ \Gamma = CK $, where $ Ki=-iK $ is  complex conjugation  and  $ C=\tau_3 $ the third Pauli matrix. Thus, the system belongs to class D in Altland-Zirnbauer notation~\cite{Altland1997,Chiu2016}. The Floquet operator satisfies $ C U_F(\boldsymbol{k}) C^{-1} = U_F^*(-\boldsymbol{k})  $~\cite{Roy2017,Yao2017},  and therefore the  quasienergies must appear in pairs $ E_{1,\boldsymbol{k}} =  -E_{2,-\boldsymbol{k}}  $. Meanwhile, the system obeys  the inversion symmetry $ PU_F(\boldsymbol{k})P^{-1} = U_F(-\boldsymbol{k}) $, with $ P=\tau_1 $, enforcing that for each band  one has $ E_{n,\boldsymbol{k}} = E_{n,-\boldsymbol{k}} $.  
	Together, we see that  the Floquet spectrum has pairs of states with quasi-energies $ E_{1,\boldsymbol{k}} = - E_{2,\boldsymbol{k}} $. This means possible gaps or nodal points/lines can,  modulo $2\pi$, appear only at quasienergy $0$ or $\pi$. 
	
\begin{figure*}[t]
		\parbox{18cm}{
			\parbox{1cm}{\includegraphics[width=1.1cm]{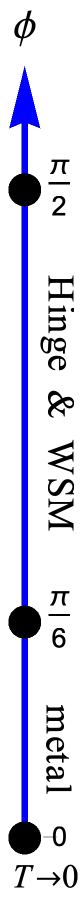}
			}
			\parbox{16cm}{ 
				\begin{tabular}{|c||c||c|c|}\hline
					{\large $ \phi $} & (1) PBC $ x,y,z $ &  (2) PBC $ z $, OBC $ x,y $ & (3) OBC $ x,y,z $ \\ \hline \hline
					{\Large $ \frac{\pi}{2} $}  & 
					\parbox{4.2cm}{\includegraphics[width=4.2cm]{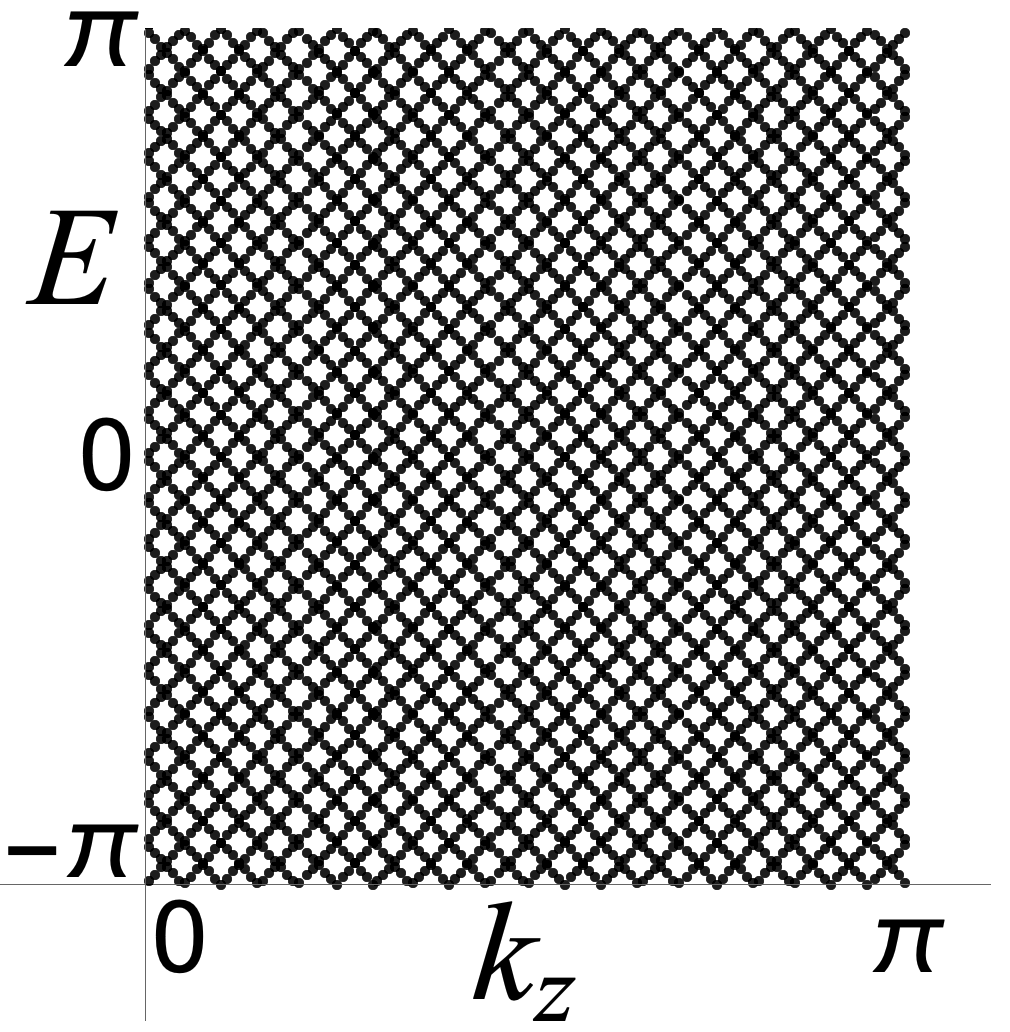} }
					&
					\parbox{3.9cm}{\includegraphics[width=4.2cm]{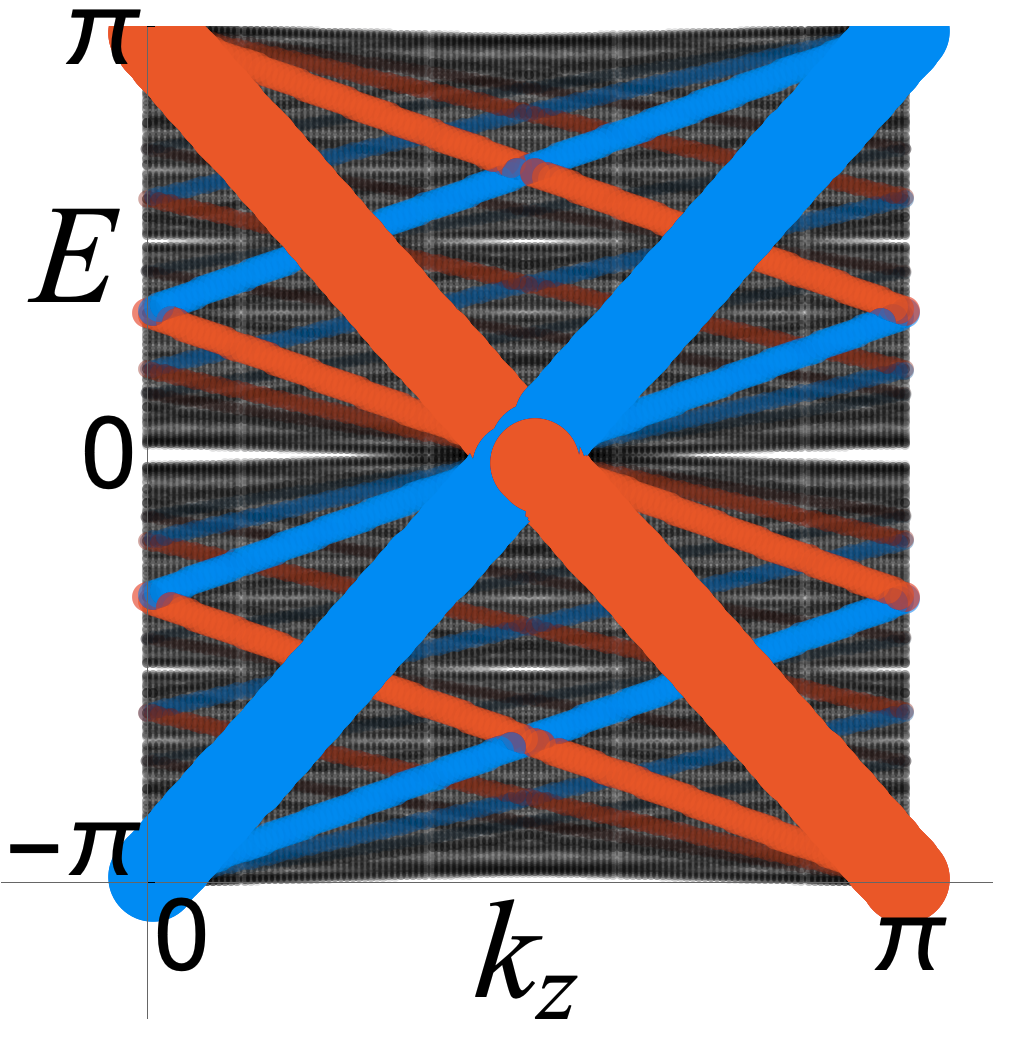} }		
					\parbox{3.8cm}{\includegraphics[width=3.8cm]{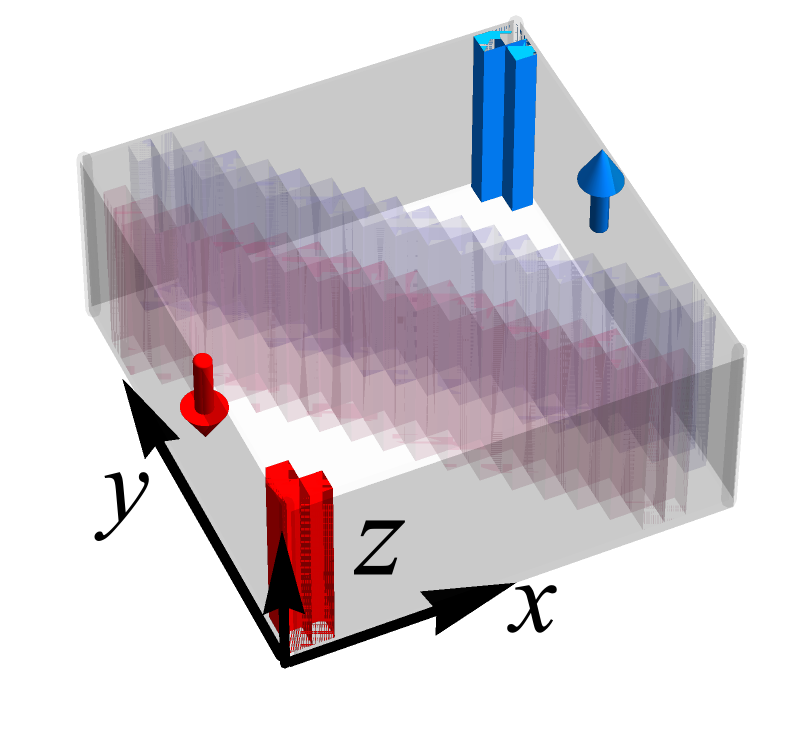} \\ 4 eigenstates }	
					&
					\parbox{3.2cm}{\includegraphics[width=3.2cm]{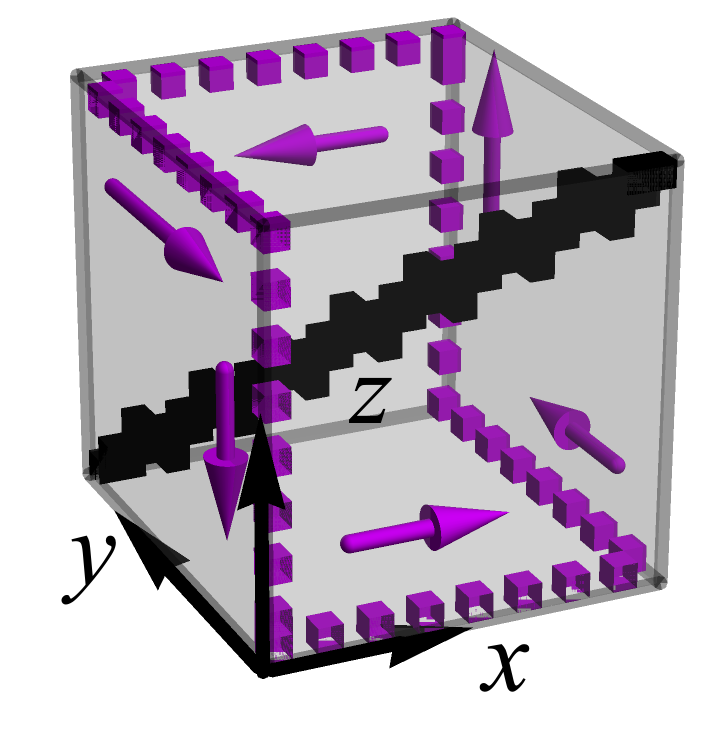} \\ 2 eigenstates }						
					\\ \hline
					{\Large $ \frac{\pi}{3} $} &			
					\parbox{4.2cm}{\includegraphics[width=4.2cm]{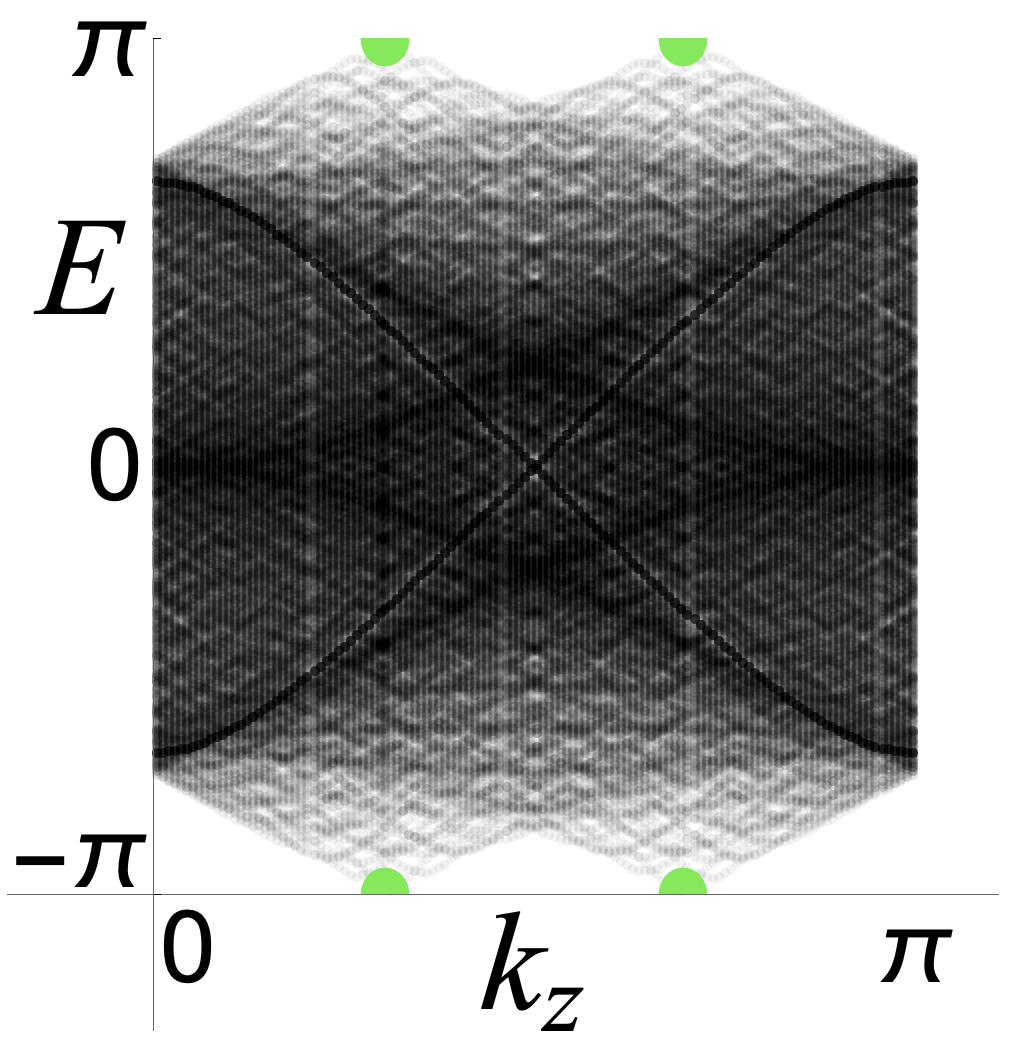} }
					&
					\parbox{3.9cm}{\includegraphics[width=4.2cm]{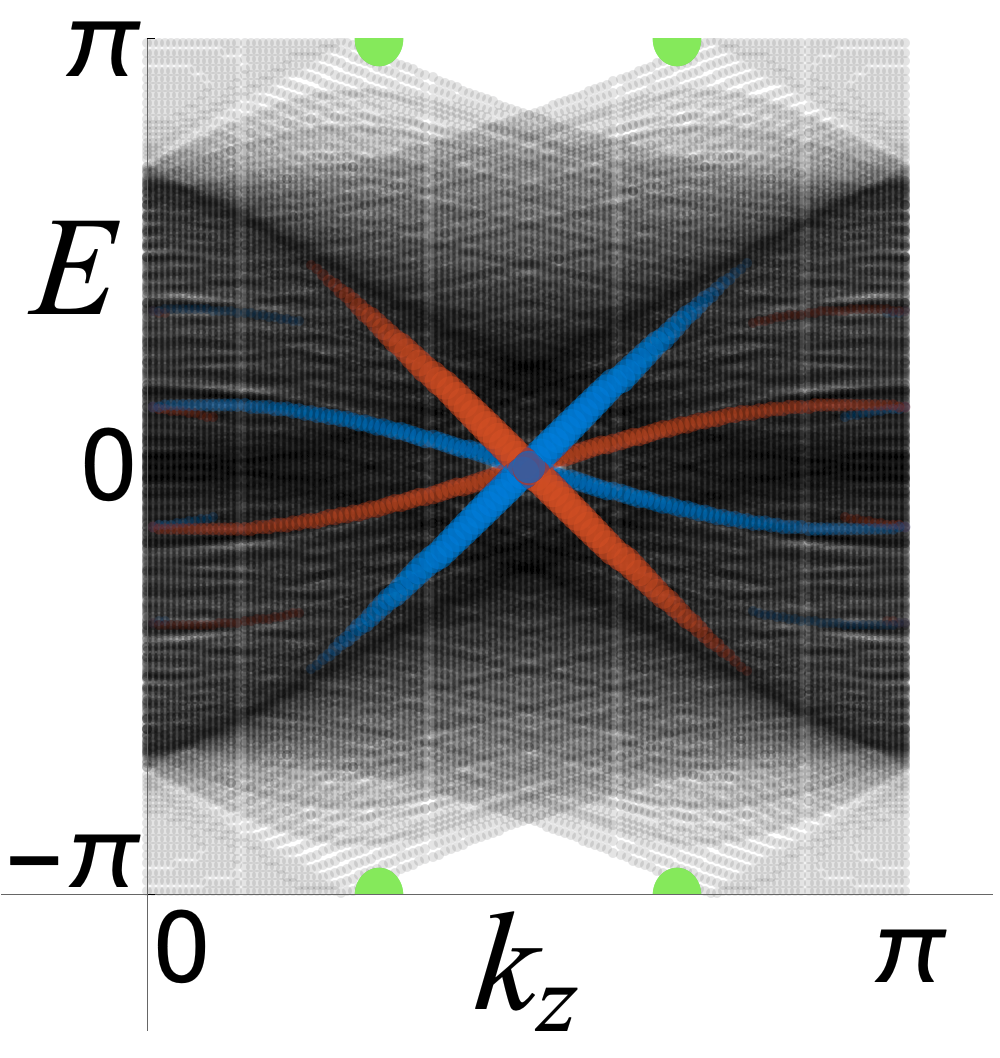} }					
					\parbox{3.8cm}{\includegraphics[width=3.8cm]{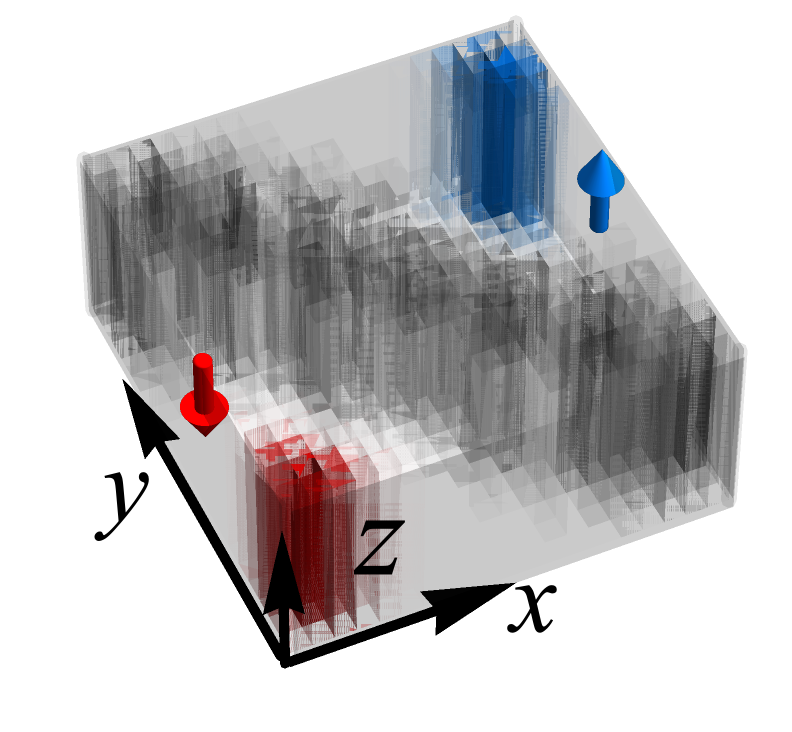} \\ 3 eigenstates
					}
					&
					\parbox{3.2cm}{\includegraphics[width=3.2cm]{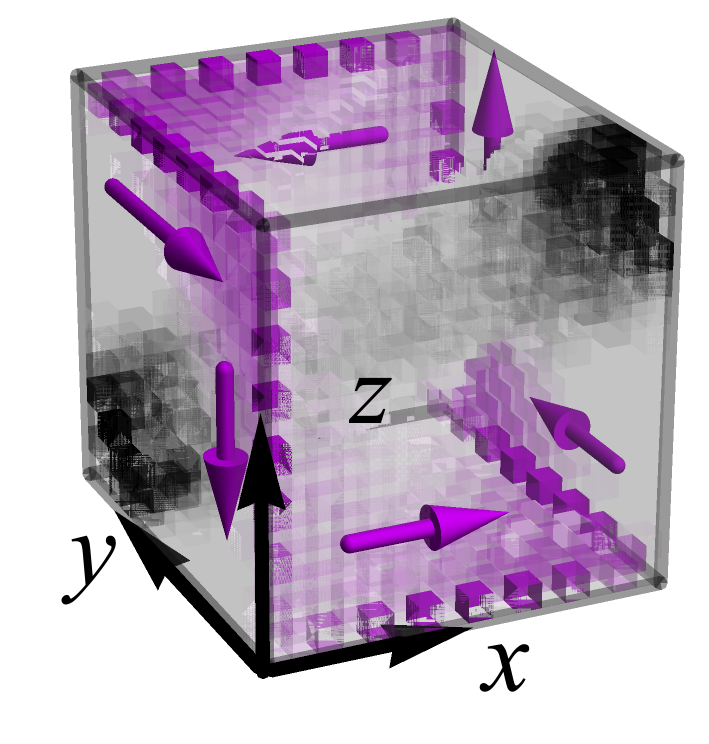} \\ 2 eigenstates }						
					\\ \hline
					{\Large $ \frac{\pi}{8} $} &	
					\parbox{4.2cm}{\includegraphics[width=4.2cm]{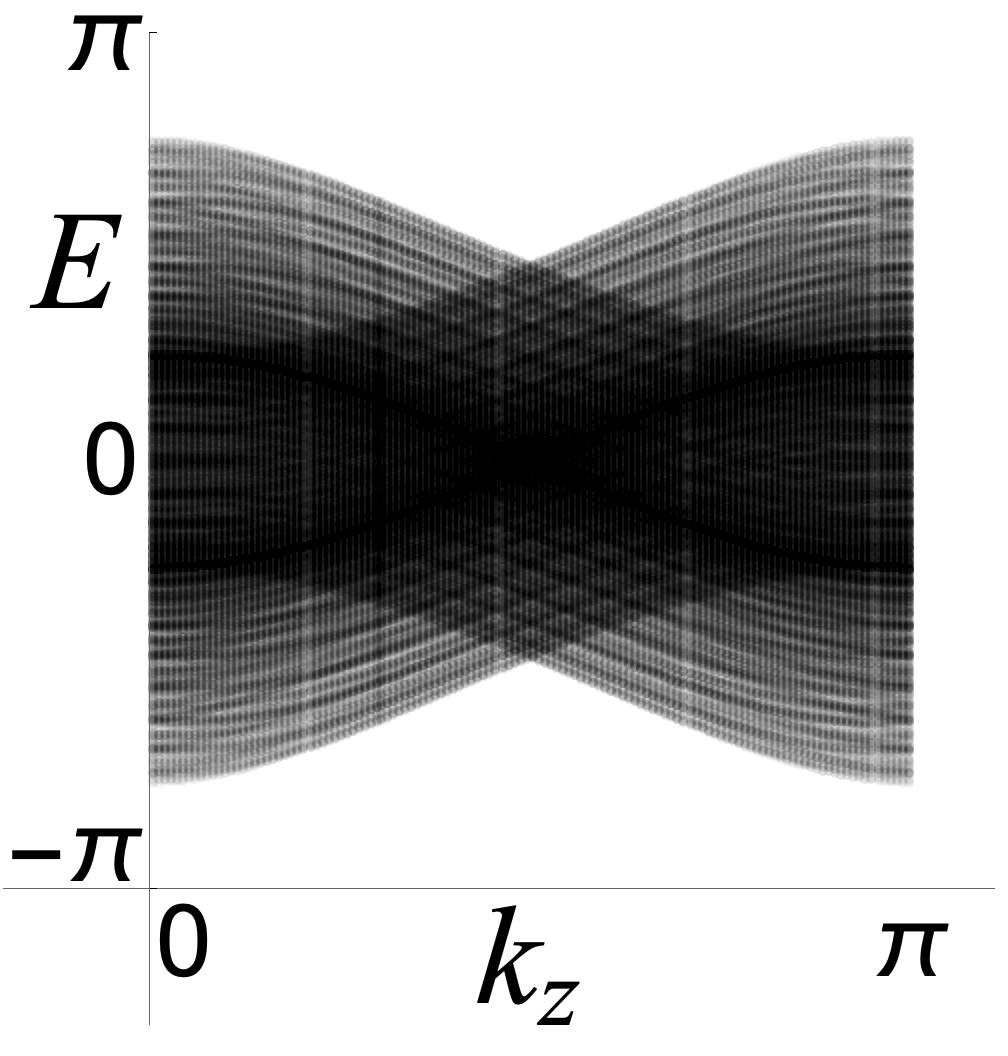} }
					&
					\parbox{3.9cm}{\includegraphics[width=4.2cm]{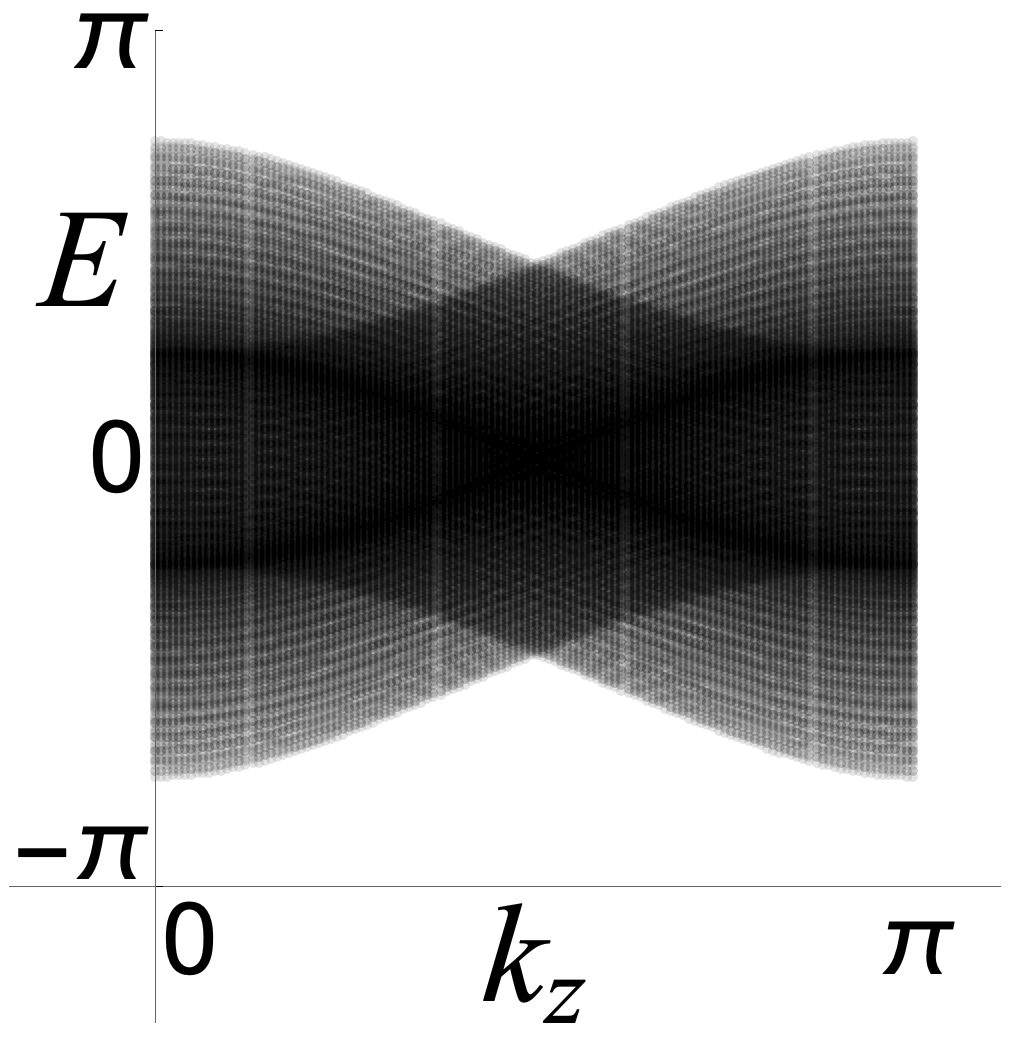} }		
					\parbox{3.8cm}{\includegraphics[width=3.8cm]{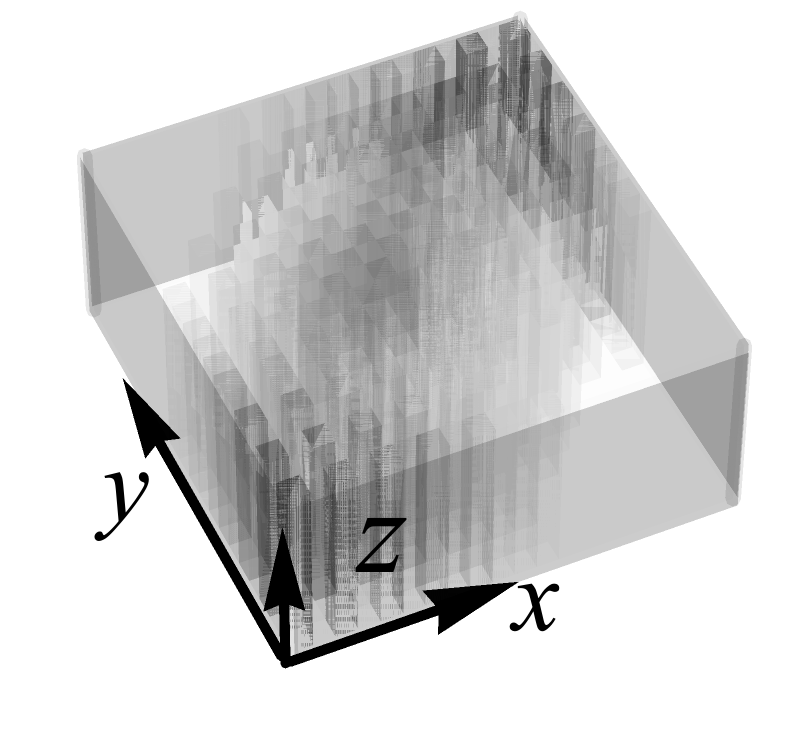} \\ 
						1 eigenstate }
					&
					\parbox{3.2cm}{\includegraphics[width=3.2cm]{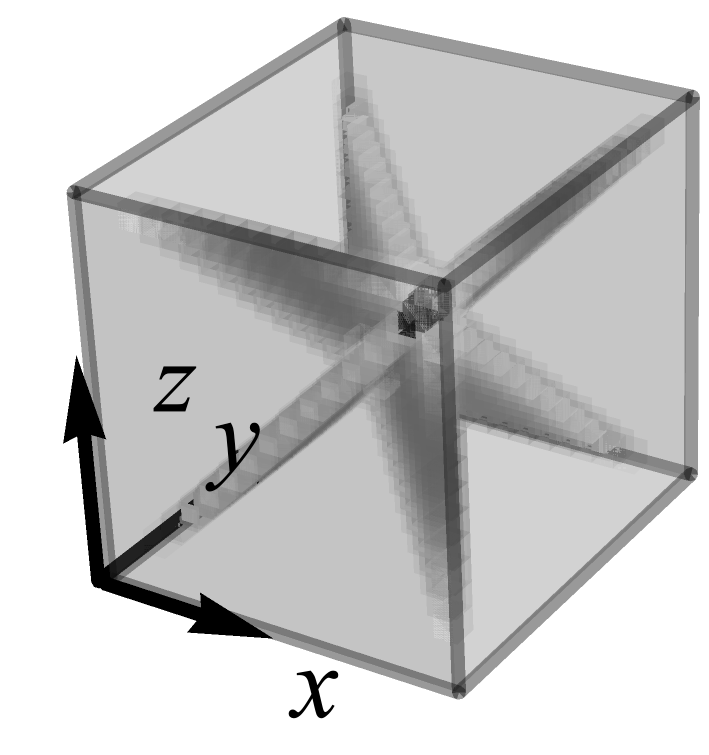} \\ 1 eigenstate }							
					\\ \hline
				\end{tabular}
			}
		}
	\caption{\label{fig:spectrum}     Table showing energy spectra and Floquet modes.  The phases of the system are indicated in the vertical axis on the left.
	The table's rows refer to the driving parameters $\phi=\pi/8, \pi/3, \pi/2$, corresponding to the metallic phase, the Weyl semimetal/hinge phase, and the fine-tuned point, respectively. The columns represent periodic/open boundary conditions (PBC/OBC) along the specified directions. 
The system sizes are $ L_x = L_y = 40 $ in column (1), 20 in (2), and 16 in (3).
Columns (1) and (2) contain energy spectra projected onto $k_z$.  Although the spectra for fully periodic and open boundary conditions in $x$ and $z$ direction are almost identical  in the metallic phase 
($\phi=\pi/8$), they are very different in the WSM/hinge phase ($\phi=\pi/3$) and at fine tuning ($\phi=\pi/2$).  The difference is a result of the formation of chiral hinge modes for open boundary conditions in $x$ and $z$ directions (column 2).   This can be inferred also from the dot size  reflecting the inverse participation ratio with respect to the site basis as a measure for localization,  as well as from the color code indicating the mean distance to two opposite hinges and interpolating from blue for one hinge over black near the center of the system to red for the other hinge. The green dots mark the Weyl points. 
We also plot the real-space densities of various Floquet modes for open boundary conditions along $x$ and $y$ (column 2) or all (column 3) directions.
	}
	\end{figure*}

 	At $ k_{x}=k_{y}=k_{z} = \pm\frac{\pi}{2} \, (\text{mod } \pi) $ Eqs.~\eqref{eq:uf} and \eqref{eq:u_mu}  yield $ U_F = 1 $, so that $ E_{n,\boldsymbol{k}} = 0 $. Therefore the quasienergy spectrum is always gapless at quasienergy  $0$ (modulo $2\pi $), regardless of the driving strength $ \phi $.   Then, the two-band spectrum could only possibly  open up a gap at quasienergy equal to  $ \pi$ (modulo $ 2\pi$). 
	To draw a complete phase diagram, we first note that flipping the sign of $\phi$ amounts to a particle-hole transformation $ U_F|_{-\phi} = C U_F|_\phi C^{-1} $, and from the previous analysis we see that such a flip does not change the spectrum. Furthermore, from  $ e^{-i\phi \hat{n}\cdot\boldsymbol{\tau}} = \tau_0 \cos\phi -i \hat{n}\cdot\boldsymbol{\tau} \sin\phi  $, where $\tau_0$ is the identity matrix, the periodicity of the  Floquet operator with respect to the parameter $\phi$ is clearly seen: $U_F|_\phi = U_F|_{\pi+\phi} $.  In this way, the irreducible parameter range is $ \phi\in[0,\pi/2] $ as illustrated in Fig.~\ref{fig:spectrum}.
		
	\section{Phase diagram for periodic boundary conditions
	\label{sec:Periodic}}

		\begin{figure}
		[t]
		\parbox{4.1cm}{
			\parbox{4cm}{
				\begin{tabular}{|c||c|}
					\hline
					{\Large $ \phi $} 
					&
					Weyl points
					\\ \hline 
					{\Large $ \frac{\pi}{2} $}
					&
					\parbox{3cm}{\includegraphics[width=3cm]{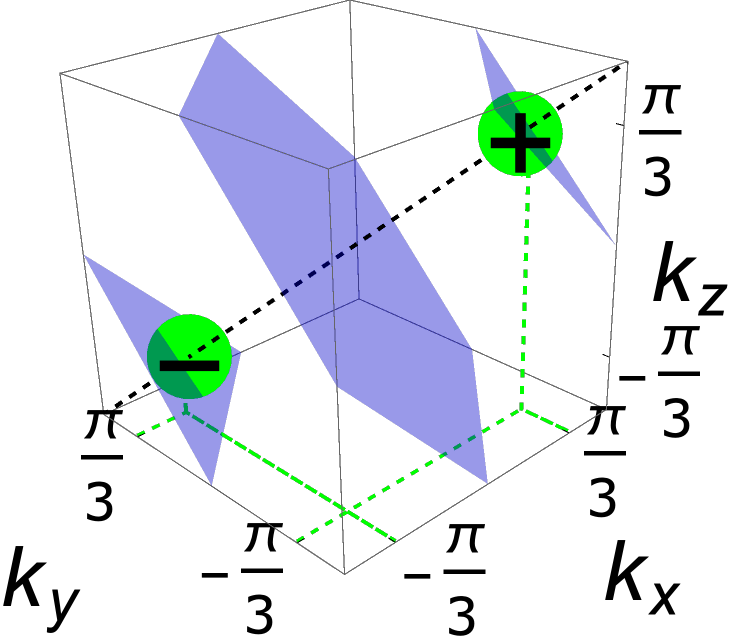}}
					\\ \hline
					{\Large $ \frac{\pi}{3} $} 
					&
					\parbox{3.2cm}{\includegraphics[width=3.2cm]{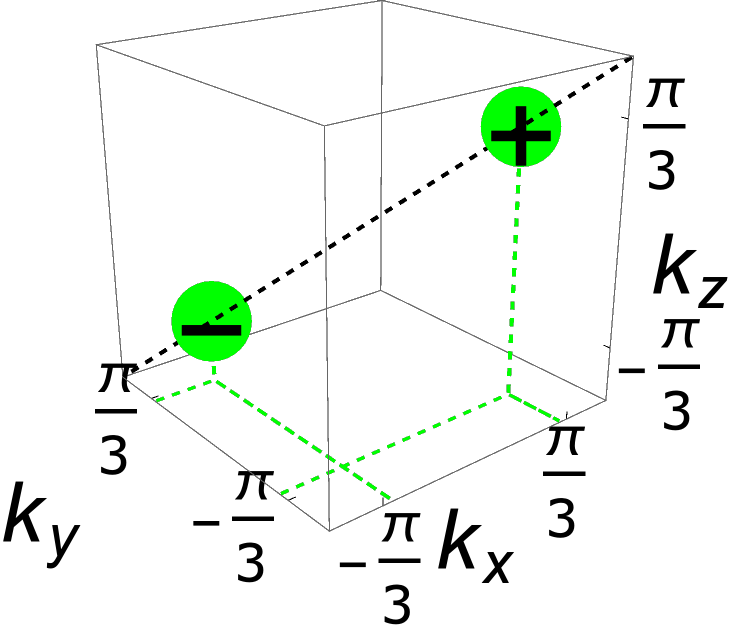}}
					\\ \hline
					{\Large $ \frac{\pi}{6} $} 
					&
					\parbox{3.2cm}{\includegraphics[width=3.2cm]{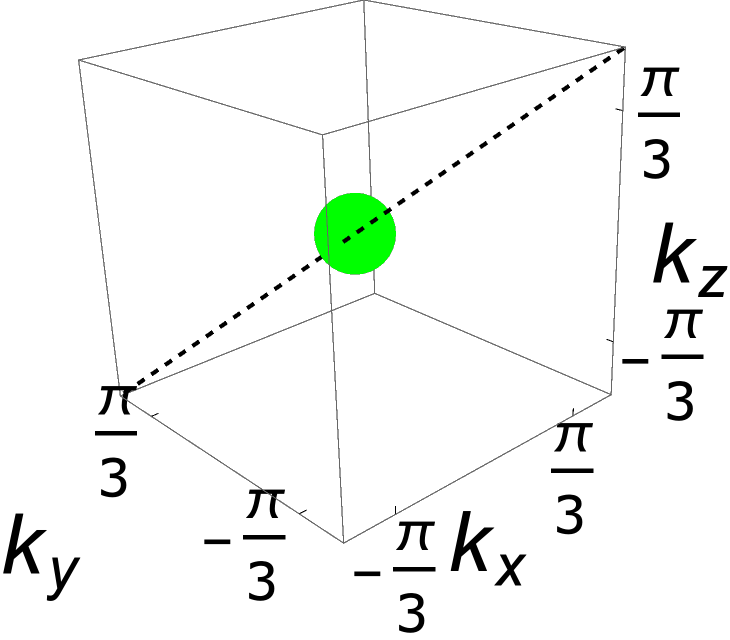}}
					\\ \hline
					{\Large $ \frac{\pi}{8} $} 
					&
					\parbox{3.2cm}{\includegraphics[width=3.2cm]{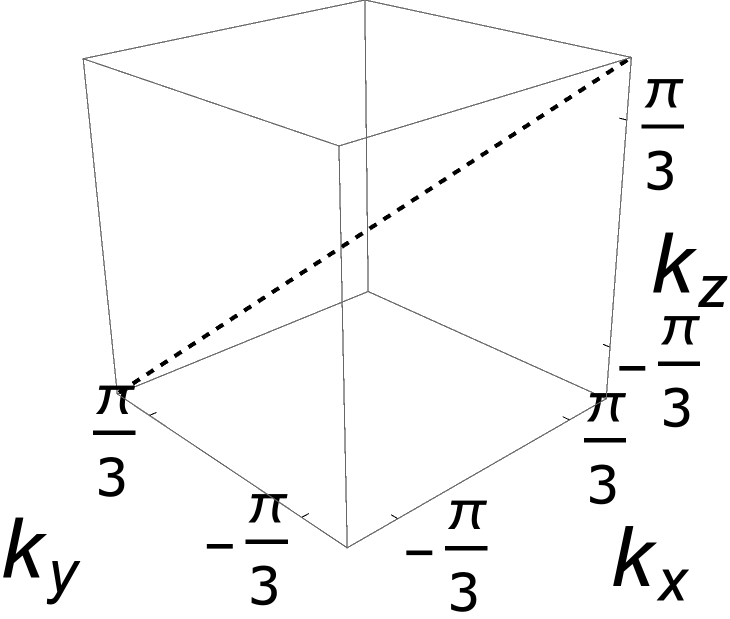}} \\ \hline
				\end{tabular}
			}\\ 
			(a)   Positions of Weyl points.  
		}
		\vspace{1cm}
		\parbox{3.3cm}{ \quad \\ \quad \\
				\includegraphics[width=3.2cm]{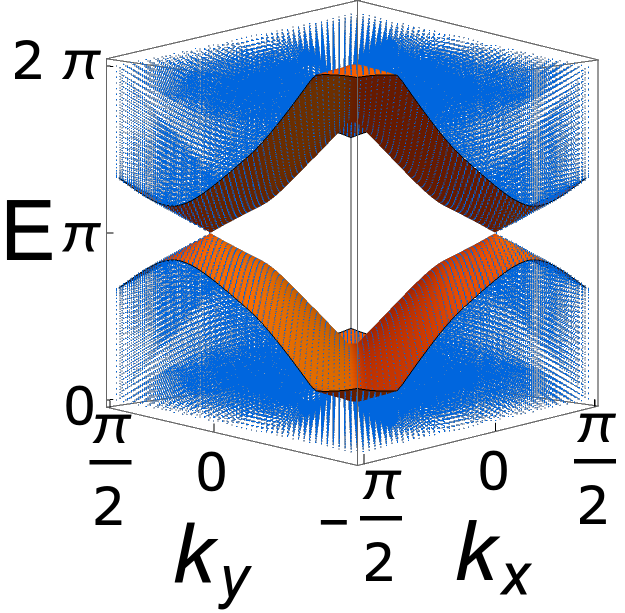}
			\\ (b) $ \phi=\pi/3 $, PBC $ x,y, z $
			\\ \quad
			\\
				\includegraphics[width=3.2cm]{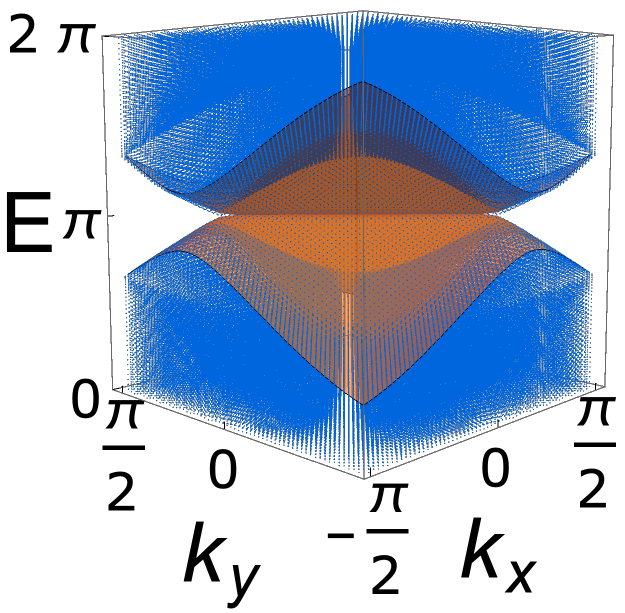}
			\\ (c) $ \phi=\pi/3 $, PBC $ x,y $  
			\\ \quad \\
				\includegraphics[width=3.2cm]{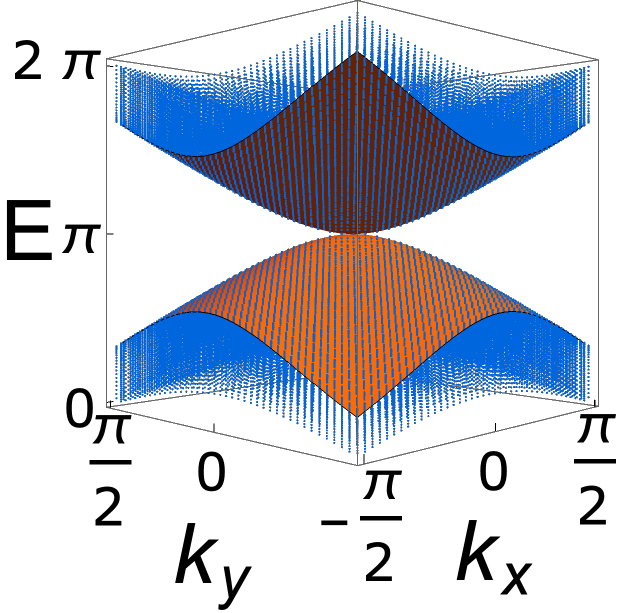}
			\\ (d) $ \phi=\pi/6 $, PBC $ x,y,z $
		}
		\caption{\label{fig:phase}  (a) The different phases can be distinguished by the presence or absence of Weyl points at quasienergy $\pi$. For $\phi=\pi/8$, in the metallic phase ($\phi<\pi/6$), no Weyl points are present. At the transition, 
$\phi=\pi/6$, the band touching point  appears at $\mathbf{k}= \mathbf{0}$ (green circle).  For  $\pi/6<\phi<\pi/2$ the band touching splits into two Weyl points of opposite charge (green circles with $\pm$ sign) that separate along the diagonal $k_x=- k_y=k_z$, as shown for $\phi=\pi/3$. At fine tuning $\phi=\pi/2$ the dispersion becomes flat in two directions and the Weyl points disappear to reappear for $\phi>\pi/2$ with reversed charges (signs in green dots correspond to the limit $\phi=\pi/2-0)$. 
(b-d) Quasienergy spectra versus $k_x, k_y$ for periodic boundary conditions in $x$ and $y$ and either periodic (b,d) or open (c) boundary conditions in $z$ direction. A surface Fermi arc can be observed for $\phi=\pi/3$ by comparing the spectra with periodic (b) and open (c) boundary conditions  the $ z $ direction. The orange surface  denotes the contour formed by quasi-energies closest to $ E=\pi $ at each $ (k_x, k_y) $. For $ \phi = \pi /6 $ (d) a pair of Weyl points  reduces to a single  band touching point at the quasi-energy $E=\pi$.
}
	\end{figure}

	\begin{figure}
		\parbox{4cm}{
		\includegraphics[width=4cm]{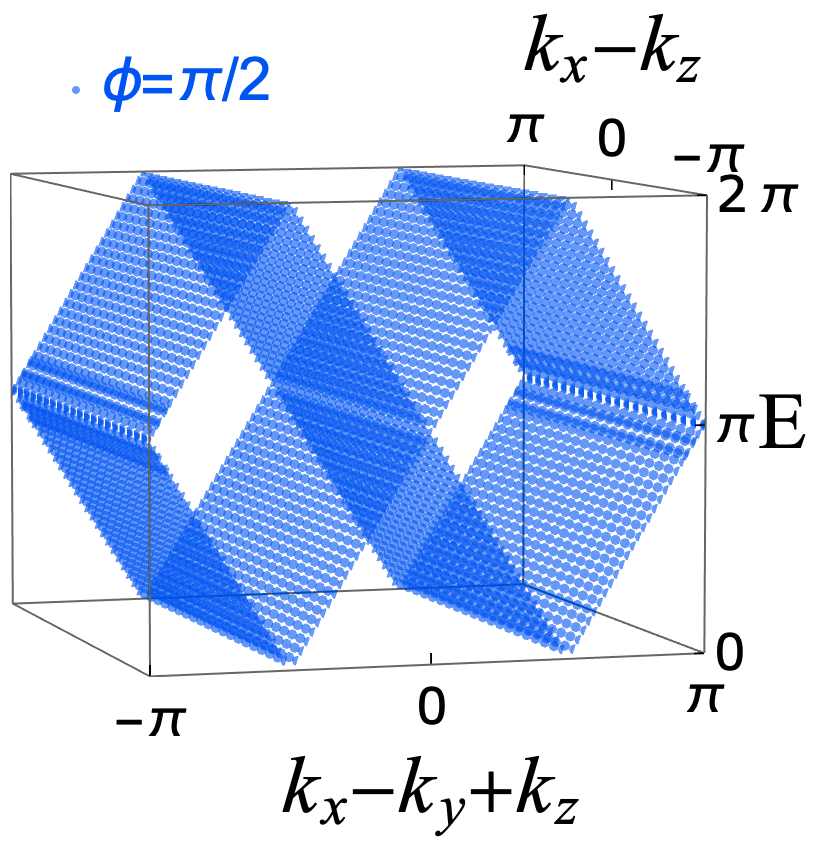}  \\ (a) Fine tuned point }
		\parbox{4cm}{
		\includegraphics[width=4cm]{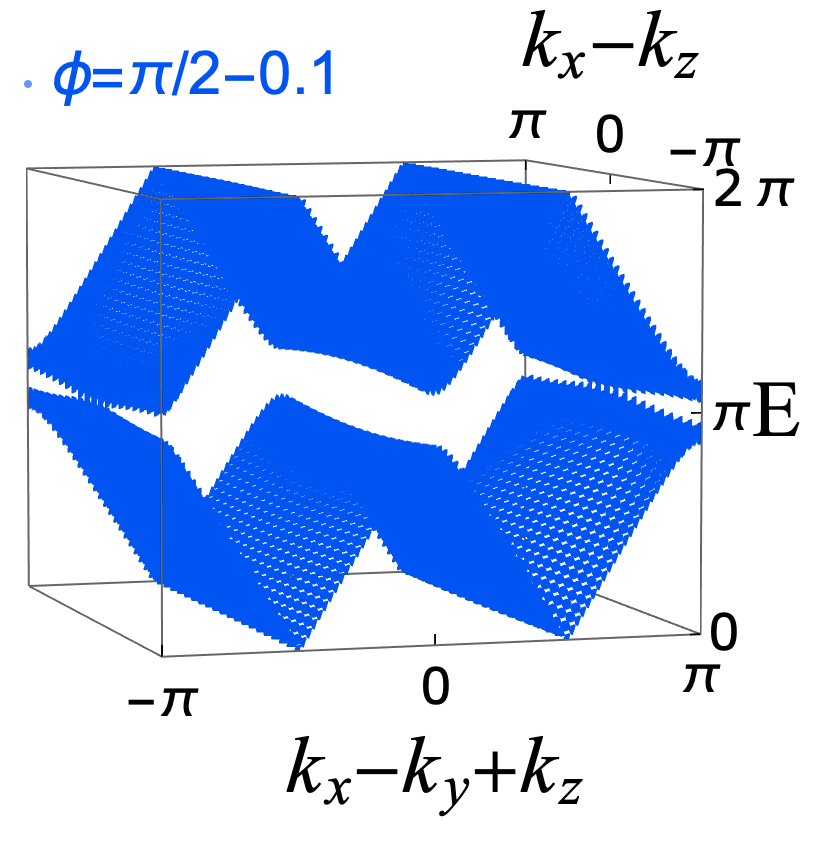} \\ (b) Slight deviation}
	
		\caption{\label{fig:aniso}  (a) Anisotropic 1D-like dispersion along the coordinate   $k_x-k_y+k_z$ at the fine-tuned point $ \phi=\pi/2 $. (b) For  a small detuning, $ \phi=\pi/2 - 0.1 $, the dispersion is no longer  completely flat in other two directions,  and a pair of  non-equivalent Weyl points is formed along the diagonal at $\mathbf{k} = \pm k_0(1, -1, 1)$. 
		Yet the dispersion remains highly anisotropic.
}
	\end{figure}

 	Before investigating the system with open boundary conditions and the emergence of chiral hinge modes, let us first consider the case of a translation invariant system with periodic  boundary conditions. Here, we would focus on the   better understood Weyl physics as an anchor point to draw the phase diagram, in order to use it as a reference frame to discuss the new hinge states in the next  Section. We clarify that the phase diagram for Weyl physics overlaps, but does {\em not} coincide completely with that for the hinge states.

	Let us begin with the topologically trivial high-frequency (weak driving) limit corresponding to $ \phi  \ll 1 $.  
In that case one can retain only the lowest order terms in $ \phi $ when expanding  the stroboscopic evolution operator $ U_F $  of Eq.~(\ref{eq:uf}), resulting in 
\begin{equation}
 U_F|_{\phi\rightarrow0} 
 \simeq e^{-i\phi 2(\cos k_x + \cos k_y + \cos k_z)\tau_1}\,.
 \end{equation}
 The spectrum $ \pm 2\phi \sum_{\mu=x,y,z} \cos k_\mu $ corresponds to that of a static simple cubic lattice artificially described by 2 sublattices, where the bands are folded as the Brillouin zone size is halved. 
 While the system remains gapless at quasienergy $0$ for arbitrary $\phi$, a  characteristic feature of the high-frequency  (weak driving) regime is a finite energy gap at quasienergy $\pi$, resulting from the fact that the band width is proportional to $\phi$ and thus  is small compared to the dimensionless driving energy $\hbar\omega=2\pi$. 
This behavior  can be observed in the spectrum for $\phi=\pi/8$  shown in Fig.~\ref{fig:spectrum} at the bottom of column (1). 
	
	 Increasing  the  driving strength $ \phi $, the band width grows relative to  $2\pi$.  When $\phi=\pi/6$  the Floquet band,  which is gapless at quasienergy $ 0 $, starts to touch  itself at quasienergy $\pi$ and momentum $\mathbf{k}=0$,   as one can see in Fig.~\ref{fig:phase} (d).
	 Going to the regime $\phi\in (\pi/6, \pi/2)$,  the band touching point splits  into a pair of Weyl points forming at quasienergy $ \pi $ with topological charges $\pm1$,  as shown in Fig.~\ref{fig:phase} (a). They are located at the quasimomenta $\mathbf{k} =   k_0 \mathbf{d}$ along the diagonal vector 
\begin{equation}
\mathbf{d} = (1, -1, 1), 
\end{equation}
with 
\begin{equation}
k_{0}=\pm \left(1/2\right)\arccos\left[\left(1/2-\sin^{2}\phi\right)/\sin^{2}\phi\right]\,\, \mathrm{modulo}\,\,\pi  
\end{equation}
(see Appendix \ref{Appendix:Evolution-operator-along-diagonal}), so that  $k_0\to \pm \pi/3 $ as $ \phi\to \pi/2$. 	We  observe the emergence of surface Fermi arc states connecting the Weyl points, when comparing the spectrum with full  periodic boundary conditions  to that with open boundary conditions along z-direction, as illustrated in  Fig.~\ref{fig:phase} (b) and (c) respectively. Note that the reversed process of increasing driving frequency (or equivalently, decreasing $ \phi $ towards $ \pi/6 $) would correspond to the usual scenario that two Weyl points merge together at $ \boldsymbol{k}=\boldsymbol{0} $ and   the spectrum becomes gapped out.

	As the driving strength approaches the fine tuned point,  $ \phi=\pi/2 - \varepsilon $ with $\varepsilon \ll1$, the Weyl dispersion acquires a highly anisotropic form  shown in  Fig.~\ref{fig:aniso} (b).  The dispersion remains steep along the diagonal coordinate  $\boldsymbol{k} \cdot \mathbf{d} = k_x-k_y+k_z$, but becomes increasingly flat in other two directions.
 Exactly at fine tuning, $\phi = \pi/2$,  the constituent evolution operators \eqref{eq:u_mu} reduce to $ U_{\mu\pm} \left(\mathbf{k}\right) = -i( \cos k_\mu \tau_1 \pm \sin k_\mu \tau_2) $,   and the Floquet stroboscopic operator takes the form $ U_F = -e^{-i2 \tau_3 \boldsymbol{k} \cdot \mathbf{d} }  $. 
	This provides the quasi-energies  
\begin{equation}
E_{\boldsymbol{k},\pm} = \pm 2 \boldsymbol{k} \cdot \mathbf{d} + (2m+1)\pi \,,
\end{equation}
where $ m\in\mathbb{Z} $ labels the Floquet bands,  and where the upper and lower branches labeled by $\pm$ now directly correspond to sublattices A and B, i.e. $\pm\rightarrow \tau_3$.
	 At this fine-tuned point the Weyl points  disappear and the dispersion $ E_{\boldsymbol{k},\pm} $ is completely flat for the momentum plane normal to $\mathbf{d} $,  as one can see in Fig.~\ref{fig:aniso} (a). Hence  a particle can only propagate  along the diagonal $\mathbf{d} $ with a dimensionless velocity $\bm{v_\pm}=\pm 2\bm{d}$, depending on whether  the particle occupies a site on the sublattice $A$ or $B$ at the beginning of a driving cycle. 
	  Note that for fine tuned driving  ($\phi=\pi/2$), the  Floquet Hamiltonian  
\begin{equation}	
 H_F \equiv -i\ln U_F = 2 \tau_3\mathbf{d}\cdot\mathbf{k}  +\tau_0 \pi   
\end{equation}
is periodic in  momentum space only  thanks to the periodicity in quasi-energies.  Such an effective Hamiltonian  is characteristic  for periodically driven systems, and does not have a static counterpart with finite range hopping.

The fine-tuned dispersion can be understood by considering the dynamics in real space. For $ \phi = \pi/2$ in each step $\mu\pm$ the particle  is fully transferred from a sublattice A site positioned at $\mathbf{r}_{A}$ to a neighboring site B situated at $\mathbf{r}_{B}=\mathbf{r}_{A}\pm \mathbf{e}_{\mu}$ or vice versa.  
During the six steps  composing the driving period, the particle follows the quasi-one-dimensional trajectory shown in Fig.~\ref{fig:model} (b). Thus, after completing each period
the particle located on a site of sublattice $A$ ($B$) is transferred by $+ 2\mathbf{d}$ ($- 2\mathbf{d}$) to an equivalent site of the same sublattice, giving rise to stroboscopic  motion along the diagonal directions  $\pm \mathbf{d}$
at the velocity $\bm{v}_\pm$.

Before leaving this  Section, we remark that the disappearing of Weyl points at $ \phi=\pi/2 $ is not because they are gapped out (unlike at $ \phi=\pi/6 $), but due to the accidental band-gap closure in a whole plane with Weyl points residing on it, as can be seen in Fig.~\ref{fig:aniso} at $ \phi=\pi/2 $.  Consequently there is no need for the two Weyl points to merge at the same location in Brillouin zone when they disappear in this way. More specifically, starting from $ \pi/6 < \phi < \pi/2 $ where Weyl points are well defined, one can enclose a Weyl point with a gapped surface (eigenstates from a certain band) surrounding it, where the Weyl charge is the total Berry flux penetrating out of the surface. At exactly $ \phi=\pi/2 $, any such surrounding surface would encounter a divergence of local Berry curvature as the two bands become gapless in a whole plane containing the Weyl point and   thus are intersecting   at any enclosing surfaces. Weyl points and their charges therefore lose definitions at such a fine-tuned point.  When the phase $ \phi $  exceeds $ \pi/2 $, the two Weyl points reappear with opposite  charges compared  to the $ \phi<\pi/2 $ situation. 

	\section{ Chiral hinge modes for open boundary conditions}
	\label{sec:Hinge}
	
	An intriguing effect shows up when the system is subjected to open boundary conditions with {\em at least two}  properly chosen boundary planes, where a special type of chiral modes form at certain hinges.  In that case eigenstates of the system  reorganize into a set states of the quasi-one-dimensional nature orienting parallel to each other and located at various distances from the hinge  when open boundaries are introduced.  To gain  an intuition, it will be useful to start from the fine-tuned point $ \phi=\pi/2 $ where  the stroboscopic motion of the particle  at the hinge  follows a classical trajectory in  real space.  Subsequently we will demonstrate numerically the robustness of such chiral hinge modes against parameter changes and  inclusion of local defects.

	Let us now fix  first the  notation for later use.  We will consider open boundary conditions with  a face of a  boundary plane oriented  perpendicular to a Cartesian axis  $\mu=x,y,z$. For full open boundaries  the atomic motion is restricted by planes  in all three Cartesian directions.
		 We will also investigate a semi-infinite system  restricted by two planes  orthogonal to   $x$ and $y$,  while keeping periodic boundary conditions  along the $z$ direction. Note  that the three-fold symmetry $ x\rightarrow y, y\rightarrow z, z\rightarrow -x $ would correspond to shifting the time origin  by $ T/6 $ for  the Floquet operator \eqref{eq:uf}, which is merely a gauge transformation and does not change  physical observables. Therefore, choosing  another combination of open/periodic boundary conditions along  different Cartesian directions  results in identical phenomena. 
	
	 Finally, since the boundaries are cut along the orthogonal directions for a cubic lattice, which is not terminating along the directions of Bravais  vectors, it is more convenient to label the lattice sites in both sublattices directly by their Cartesian coordinates
 $
 {\bm r}=(x,y,z)
 $,
 with  $x,y,z$ taking integer values  between $1$ and $L$ carrying units of  the lattice constant.  The sites belong to the sublattice  $A$  (or $B$) for even (or odd) values of $x+y+z$.

\subsection{ Fine-tuned driving}
\label{Subsec:Fine-tunned-driving}
\subsubsection{ Real-space picture}

  	     We will commence our discussion of the hinge dynamics at the fine tuned point using the real space picture.     Subsequently in Sec.~\ref{Subsubsec:Floquet-hinge} we will present a rigorous analysis of the hinge states by going to the momentum representation along the hinge direction.  	 
 Let us consider   the atomic motion  for $x\ge 1$ or $y\ge 1$  confined by a single boundary plane  at $ x=1 $  or $ y=1 $, as illustrated in Fig.~\ref{fig:model} (c).
	Starting from a site deep inside the bulk of, say, sublattice $B$,   the particle will propagate stroboscopically on this sublattice     covering a distance $-2\bm{d}$ in diagonal direction   in each driving period,  until   approaching the boundary $x=1$  or $y=1$.  At the boundary tunneling between the lattice sites $B$ and $A$ cannot occur during one of the six steps of the driving cycle. As a result, the particle changes the sublattice and starts to propagate on the $A$ sublattice in opposite direction. The two colors on each panel of Fig.~\ref{fig:model} (c) denote two possible types 
	 of such a reflection that are distinguished by the lattice site at which the driving cycle   starts  leading to the reflection.  
	 Specifically a tunneling event is impeded in the first (dark grey) or the  fifth (yellow) driving step  for a particle situated right at the surface plane  $ x=1$ ($y=1$) or next to it $ x=2$ ($y=2$),  as marked by small planes  in Fig.~\ref{fig:model} (c). 
  	
  	 Let us next consider  a similar motion of the particle  in the region $x\ge1$ and $y\ge1$  confined by two intersecting surface planes $ x=1 $ and $ y=1 $. After the  particle changes the sublattice  $B\rightarrow A$ at  the $ x=1 $ boundary, it travels in reversed direction in steps of $2\bm{d}$ until  it eventually reaches the $ y = 1 $ surface and  is again backreflected,  this time with the sublattice change $A\rightarrow B$.  In this way, the particle will move back and forth between the $x=1$ and the $y=1$ surfaces  which share a hinge.
  	
  	 It is now important to note that each reflection is accompanied   by a lateral shift along the surface planes, as can be seen in Fig.~\ref{fig:model} (c). Such lateral shifts during reflections  resemble the Goos-H\"{a}nchen effect occurring when light is reflected at the boundary between two media.
	 The lateral shift would have stronger influence when   the particle start from   a location closer to the   hinge $ x=y=1 $ or $ x=y=L $ where reflections occur more frequently, as illustrated in Fig.~\ref{fig:model} (d). 
	 In particular, for particles starting from sublattice $ B $ at $ x=y=1 $, the dynamics is {\em completely} given by the lateral shift along $ -z $ direction,   as one can see in the lower part of Fig.~\ref{fig:model} (d).   The corresponding eigenstate for Floquet operator would be a {\em single} chiral mode localized along the hinge propagating along $ -z $. Similarly, at $ x=y=L $, one can verify that another chiral mode associated with sublattice $ A $ propagates along $ +z $ direction. Note that these two modes are separated by the whole system size, unlike in the 1D case~\cite{Budich2017} with two modes of opposite chirality residing at the same location  and thus are subjected to backscattering on imperfections.

	 In the next  Sec.~\ref{Subsubsec:Floquet-hinge}  we will see through rigorous calculation that the whole $ xy $-plane can be divided into two parts by an off-diagonal line joining $ x=1, y=L $ and $ x=L, y=1 $, where eigenstates localized closer to $ x=y=1 $ or $ x=y=L $ hinges will carry  the chiral motion along $ -z $ or $ +z $ respectively. Therefore, under perturbations, it will be the eigenstates closest to the center ``chirality border line" to lose their chirality along $ z $ first,  and the hybridization 
	gradually  spreads towards the hinges as perturbation strengthens. Importantly, the eigenstates right at the hinges $ x=y=1 $ and $ x=y=L $  do not involve any  motion in the $ xy $-plane, as can be seen in Fig.~\ref{fig:model} (d). They  correspond to the genuine chiral hinge modes described by  the evolution operator $ U_F $ over a single  period. 
	That means the chirality of such a mode  does not depend on the ballistic trajectories finishing several Floquet periods in the $ xy $-plane,   and thus the closest hinge mode should be more robust against imperfections. This is verified numerically away from fine-tuned point in Fig.~\ref{fig:spectrum} (2nd column), and also in Figs.~\ref{fig:dyn} and \ref{fig:dyn_detuned} in the presence of local defects, as will be discussed more specifically later.  	
	
	  Prior to going to a rigorous analytical calculation, we make a side remark regarding the previous Weyl physics.   The general boundary reflections exchanging sublattices $ A $ and $ B $ break particle-hole symmetry $ \Gamma = \tau_3 K $. But that will not destroy the Weyl physics because it reduces the symmetry class from $ D $ to $ A $, which still hosts a $ \mathbb{Z} $ classification for point defect (i.e. Weyl points) in a 3D Brillouin zone~\cite{Chiu2016}.  Furthermore, the orthogonal boundary we choose still preserves the inversion symmetry, and therefore Weyl points and their associated Fermi arcs in the open boundary systems would still exhibit an inversion symmetric fashion as found previously. The only thing that might be slightly affected is the symmetry $ E_{1,\boldsymbol{k}} = - E_{2,\boldsymbol{k}} $ for Fermi arc on the surface, while the Weyl physics phase diagram concerns bulk spectrum and as $ L\rightarrow\infty $ surface effects would minimize. The hinge states discussed in this  Section do not rely on the particle-hole symmetry either.
	
	\subsubsection{Floquet   states} \label{Subsubsec:Floquet-hinge}

\begin{figure}[h]
\begin{centering}
\includegraphics[scale=0.3]{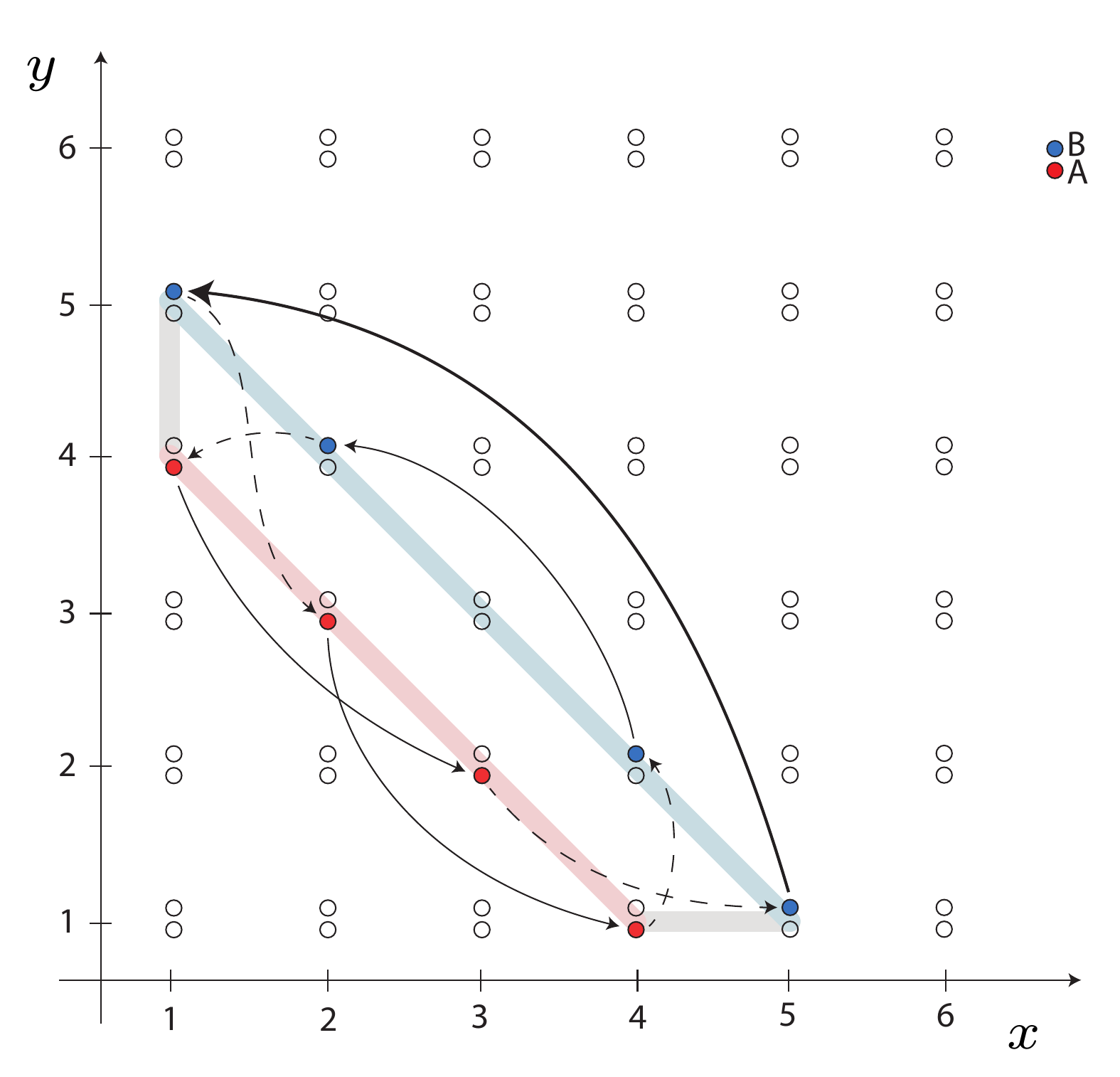} 
\par\end{centering}
\caption{\label{fig:hinge-stroboscop-1}An example of a fine-tuned stroboscopic
motion of a particle at the lower-left hinge. The picture shows
the projection of the particle's trajectory in the $xy$ plane.
The sites of the $B$ and $A$ sublattices are marked in blue and
red, respectively. The particle is initially at the site of the sublattice
$B$ characterized by  the coordinates $x=M+1$ and $y=1$, with even $M=4$. This corresponds
to the lower row ($y=1$) and the fifth column ($x=5$). Subsequently
the particle undergoes the stroboscopic evolution described by Eqs.
\eqref{eq:B--A1}-\eqref{eq:A--B2}  in Appendix \ref{Appendix:Hinge states}. Dashed lines with arrows show
stroboscopic reflections from the planes $x=1$ or $y=1$.
 Bulk ballistic trajectories over  one/two driving periods are indicated by thin/thick solid arrows. 
The particle returns to its initial site after
$2M+1=9$  periods. 
 }
\end{figure}

\begin{figure}[h]
\begin{centering}
\includegraphics[scale=0.3]{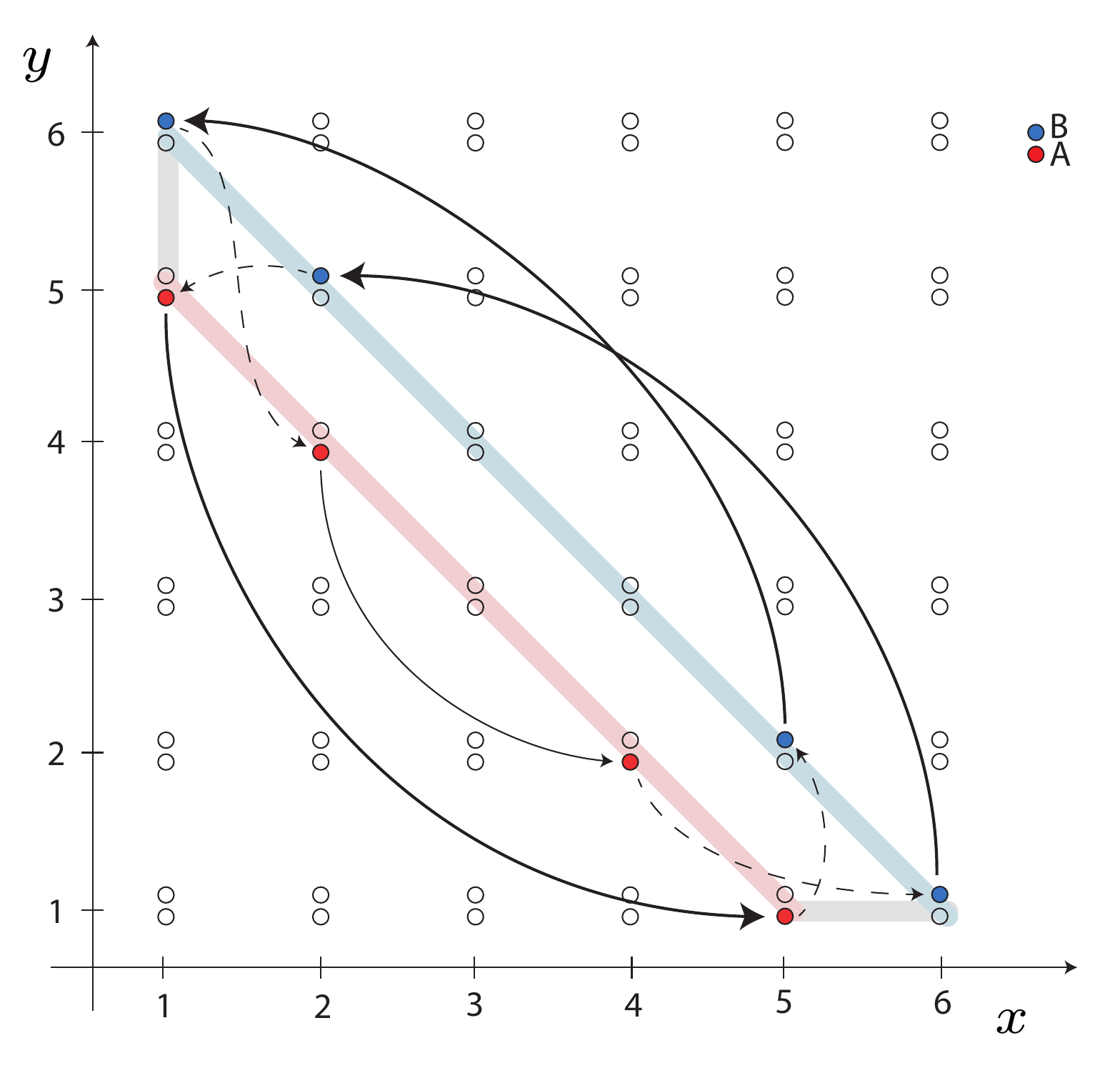} 
\par\end{centering}
\caption{\label{fig:hinge-stroboscop-2} Like in Fig.~\ref{fig:hinge-stroboscop-1}
the particle is initially at the site of the sublattice $B$ with $x=M+1$ and
$y=1$, but now with odd $M=5$.  The particle returns
to the initial site after $2M+1=11$ periods.
}
\end{figure}

	  Before carrying out  analytical calculations of the Floquet states at the fine-tuned $ \phi=\pi/2 $,   in Figs.~\ref{fig:hinge-stroboscop-1} and \ref{fig:hinge-stroboscop-2} we illustrate  the full path of a particle projected onto the $xy$-plane for reflections occurring at $ x=1 $ and $ y=1 $ surfaces.   One can see in these figures that the particle returns to the same  transverse position $(x,y)$ after 
	  making two loops.   Such a closed trajectory contains
   two reflections at each boundary $ x=1 $ and $ y=1 $, corresponding to the two different types of reflections depicted in Fig.~\ref{fig:model}(b).
	The back reflected particle propagates in a trajectory situated closer to the hinge or further away from it for the $B\rightarrow A$  or $A\rightarrow B$ reflections, respectively. This is similar to changing a track for  a train before sending it backwards. Because of such chiral Goos-H\"{a}nchen shifts, the particle visits a larger number of $B$ sites than $A$  sites  
	when traveling between the two surface planes.   Recalling that the ballistic motion   over the $B$   or $A$ sites    takes place along the diagonal   $-\bm{d}$ or $\bm{d}$, the   dynamics projected onto the $xy$ place is accompanied by  motion in   $-z$ or $+z$ direction   depending on the sublattice.   Thus  one arrives at an overall steady advance in $-z$ direction    after the particle   completes a closed two loop  trajectory in  the $ xy $ plane   containing more sites of the $B$ sublattice.   Therefore all eigenstates would occupy localized quasi-1D regions in $ xy $-plane parallel to each other, each carrying certain group velocity along $ z $. 
	
	More precisely, as demonstrated in Appendix \ref{Appendix:Hinge states}, after four reflections by the surfaces, a particle comes back to the original point in the $ xy $-plane but shifts by $-2$  units in $z$-direction to an equivalent lattice site. 
	One can similarly verify that a particle  comes back to the same $ (x,y) $  transverse point after four reflections by  the opposite $ x=L $ and $ y=L $ planes  accompanied by a shift by  $ +2 $  lattice unites along $ z $ direction.  
	
	In this way, the particle's trajectory starting from any lattice site would roughly covers a quasi-1D region in the $ xy $-plane along $ \boldsymbol{e}_x - \boldsymbol{e}_y $ directions, whose length is proportional to the distance away from the hinge $ x=y=1 $. All complete   double loop trajectories in the $ xy $-plane would be associated with a shift by $ -2 $ or $ +2 $ along $ z $, depending on whether they are closer to the $ x=y=1 $ or $ x=y=L $ hinge.   Although such a picture applies to a particle situated further away from the hinges, the advance in the $-z$ or $+z$ direction takes place  also for trajectories situated very close to the hinge $ x=y=1 $ or $ x=y=L $ where the particle is reflected simultaneously from both hinge planes, as illustrated in Fig.~\ref{fig:model}(d) for the  $ x=y=1 $ hinge.

	The fact that trajectories   represent closed loops in the $ xy $-plane allows us to construct analytically the eigenstates of  the stroboscopic evolution operator $ U_F $ with open boundaries in $ x,y $ and periodic boundary along $ z $. The details are elaborated in Appendix~\ref{Appendix:Hinge states} while we outline the results and derivations below. We would chiefly describe the hinge related to $ x=y=1 $, while mention certain results for $ x=y=L $ directly.
	
	Suppose the particle initially occupies a site of the sublattice $B$  with transverse coordinates $y=1$ and $x=M+1$,  so the particle is situated at the  hinge surface $ x=1 $ and is  $M\ge0$ sites away from  the other  $ y=1 $ hinge surface, as illustrated in  Figs.~\ref{fig:hinge-stroboscop-1} and \ref{fig:hinge-stroboscop-2} for $M=4$ and $M=5$. In that case, it takes $2M+1$ driving periods for the particle to come back to the initial position $\left(M+1,1\right)$ in the $xy$ plane,  while having shifted by $-2$ lattice units in $z$ direction, i.e.
	\begin{equation}
		\left(U_{F}\right)^{2M+1}\left|B,M+1,1,z\right\rangle =-\left|B,M+1,1,z-2\right\rangle\,.
		\label{Full_cycle_hinge-state_coordinate_representation}
	\end{equation} 
 	After averaging over such a full cycle   involving $2M+1$ driving periods, the particle travels with  an $M$-dependent mean velocity of $v_{M-} = -2/\left(2M+1\right)$  along the   hinge axis  $z$
	(see Appendix \ref{Appendix:Hinge states} for more details). 
 	Here   $|s,x,y,z\rangle$  describes a particle  located at site $(x,y,z)$  belonging to sublattice   $s=B=-1$ or   $s=A=1$   with $s=(-1)^{x+y+z}$.
 
	Let us now consider periodic boundary conditions in $z$ direction  with $ z = 1,2,\dots 2N_z $, i.e.  $\left|s,x,y,z+2N_z\right\rangle = \left|s,x,y,z\right\rangle$, while keeping open boundary conditions in  $x$ and $y$. It is  then convenient to introduce hybrid position-momentum basis states 
	\begin{equation}
		\left|{s,x,y,k_{z}}\right\rangle' =\frac{1}{\sqrt{N_{z}}}\sum_z e^{ik_{z}z}\left|s,x,y,z\right\rangle \,.
		\label{eq:Momentum-basis-z}
	\end{equation} 
 	  A basis vector $\left|{s,x,y,k_{z}}\right\rangle'$ is characterized by a quasimomenta $k_z=m\pi/N_z$ with $m=0,1,\dots,N_z-1$,    where $k_z$  is defined modulo $\pi$,   as the lattice periodicity equals to 2 length units in $z$ direction.  Equation~\eqref{Full_cycle_hinge-state_coordinate_representation} yields
	\begin{equation}
		\left(U_{F}\right)^{2M+1} \left|{B,M+1,1,k_{z}}\right\rangle'=-\left|{B,M+1,1,k_{z}}\right\rangle' e^{2ik_{z}}\,.
	\end{equation} 
	
	The above results are for a particle starting from the hinge surface $ y=1, x=M+1 $, which traces out a closed loop for its classical trajectory visiting $ 2M+1 $ sites at different stroboscopic moments. In a similar manner,  if the particle  starts at any intermediate site of the trajectory at stroboscopic moment $ pT $,
	\begin{equation}
		\label{eq:M,k_z,s}
 		\left|M,k_{z},p\right\rangle \equiv \left(U_{F}\right)^{p}\left|{B,M+1,1,k_{z}}\right\rangle'\,,
 	\end{equation}
	it will  also finish the loop after $2M+1$ driving periods:  
	\begin{equation}
		\left(U_{F}\right)^{2M+1} \left|M,k_{z},p\right\rangle   = - \left|M,k_{z},p\right\rangle e^{2ik_{z}}\,,
		\label{eq:U_F^2M+1--B,M+1,1,k_z}
	\end{equation}
	with $p=0,1,\ldots2M$.  
 	By superimposing the states $ \left|M,k_z,p\right\rangle $  corresponding to all sites involved in  the stroboscopic trajectory,  one can construct the  Floquet hinge states  representing eigenstates of the Floquet evolution operator $U_F$: 
	\begin{equation}
		\overline{\left|M,k_z,q\right\rangle} =\frac{1}{\sqrt{2M+1}} \sum_{p=0}^{2M} \left|M,k_{z},p\right\rangle \exp\left(\mathrm{i}\frac{2\pi q-2k_{z}+\pi}{2M+1}s\right)\,,\label{eq:M,q,k-Floquet-states}
	\end{equation}
 where the index  $q=0,1,\ldots,2M$ labels these modes.  The corresponding quasienergies for the eigenstates \eqref{eq:M,q,k-Floquet-states}   situated closer to the $ x=y=1 $   reads	\begin{equation}
		E_{M,k_{z},q}=\left(\frac{2\pi q-2k_{z}+\pi}{2M+1}\right)\textrm{ mod }2\pi \,.
		\label{eq:E_Mqk--finge1}
	\end{equation}

 	An analogous dispersion but with an opposite slope  (opposite sign for  the group velocity along $ z $)  can be obtained for the states formed closer to the opposite hinge  $x=y=L$.  The dispersion  branches for  both types of hinge modes  reproduce the spectrum  shown in the row 1 and column (2) of Fig.~\ref{fig:spectrum}.  The corresponding eigenstates are quasi-1D in $ xy $-plane along $ \boldsymbol{e}_y-\boldsymbol{e}_x $ and are  plane-waves along $ z $. 
 	
	Such a spectrum of the system with open boundary conditions in $x$ and $y$ directions looks completely different from the one for full periodic boundary conditions  shown in column (1) of Fig.~\ref{fig:spectrum}  or Fig.~\ref{fig:aniso} (a),  where  in the latter case all the bulk modes have the same positive or negative dispersion slope (group velocity)  $v_{z}= \pm2$.
	In contrast, for the beam geometry  [column (1) of Fig.~\ref{fig:spectrum}] the spectrum is now organized into Eq.~(\ref{eq:E_Mqk--finge1}), representing different quasi-1D eigenstates $ M $ sites away from the hinge $ x=y=1 $. Each eigenstate is characterized by a different group velocity 
	\begin{equation}
		\label{V-M-pm}
		v_{M\pm} = \pm\frac{2}{2M+1}
	\end{equation}
	decreasing with the distance $M$ from the hinge,  where ``$ - $" and ``$ + $" correspond  to the states located relatively closer to  $x=y=1$ and  $x=y=L$ respectively. The  red / blue colors  in Fig.~\ref{fig:spectrum} indicate the mean distance of each mode from the two relevant hinges. The dark  red (blue)  mode associated with $M=0$  is localized directly at  the hinge  $x=y=1$ ($x=y=L$) and  propagates at the largest velocity in negative (positive) $z$ direction. Modes with a smaller slope have larger $M$ and thus are located further away from the particular hinge, as indicated by the color. The real-space density of four different hinge modes at $\phi=\pi/2$ is illustrated in the real-space plot shown in  row 1 and column (2) of Fig.~\ref{fig:spectrum}. 
	 In addition, a measure for the degree of localization of a mode $|\psi\rangle$ is the inverse participation ratio $\text{IPR} = \sum_j |\langle j|\psi\rangle|^4$, with real-space site  states $|j\rangle$ and the Floquet eigenstates $ |\psi\rangle $. It is shown in the spectra of Fig.~\ref{fig:spectrum} via the dot size roughly indicating the inverse of the number of sites a mode is spread over. Thus modes localized near the hinges have larger IPR than those that are more spread over in $ xy $-plane locating further away from the hinges, indicating that the  latter modes  with smaller chiral group velocities are less narrowly localized in $ xy $-plane than the  former faster modes located closer to the hinge. 
	
	 In this way,  the group velocity $ v_{M\pm} $  along the $z$ direction with open boundaries along $ x,y $  is very different from the  projected group velocity 
	$ v_{z} = \mathbf{v} \cdot  \mathbf{e}_z$ 
	 of a full periodic system.  This is due to  the unidirectional bulk group velocity $\mathbf{v}=\pm 2 \mathbf{d}$ and can be intuitively understood as follows. Each eigenstate of the system  with open boundary conditions  involves 4 boundary reflections, as can be inferred from Figs.~\ref{fig:hinge-stroboscop-1} and \ref{fig:hinge-stroboscop-2}. 
Such an eigenstate is composed of  two pairs of  counterpropagating plane waves representing the eigenstates of the original periodic system  propagating  along the cubic diagonal $ \pm \boldsymbol{d} $  with  the $z$ projection of the group velocity $ v_z=\pm2$.  Consequently the original group velocity in the bulk gets neutralized, and  the overall group velocity  $ v_{M\pm} $ along $ z $ is due to the combined Goos-H\"{a}nchen shift over 4 times of boundary reflections.  As the Goos-H\"{a}nchen process occurs more frequently near the hinges $ x=y=1 $ and $ x=y=L $,   the magnitude of group velocity  $ v_{M,\pm} = \pm2/(2M+1) $  is maximum  at the hinges for $M=0$, while close to the central part  where $ M\propto L $,  the group velocity $ v_{M\pm} $ becomes rather small, and eventually vanishes in the thermodynamic limit $ v_{M\pm} \propto 1/L\rightarrow 0 $. This is quite different from a usual crystal where modes closer to the  center of the bulk should be more insensitive to any boundary effects, while in our case one arrives at another situation where   $ v_z=\pm2 $  for the periodic boundary conditions while  for open boundary conditions one has $ v_{M\pm}\rightarrow 0 $ for the modes near the bulk center.
		
	 The observation that the whole   (Floquet) spectrum and    the corresponding eigenstates of   our system are completely reorganized, when switching from periodic to open boundary conditions,  
	 resembles the extensive accumulation of boundary modes featured in the non-Hermitian skin effect~\cite{Yao18PRL,budich2020RMP,Ueda2020AdvPhys,Kawabata2020PRB}. 
	 Therefore, the formation of an extensive number of reconstructed chiral   hinge modes in our Floquet system might be called  \emph{chiral second-order Floquet skin effect},  in analogy to the terminology used for non-Hermitian systems  \cite{Kawabata2020PRB}.  A subtle difference is that the eigenstates of a  unitary Floquet evolution operator are orthogonal to each other, unlike eigenstates of non-Hermitian Hamiltonians that  that are generally non-orthogonal. It implies the counting rule that on average, there can be at most one Floquet eigenstate per lattice site, so an accumulation of states right at the boundary is not possible in a Floquet system.
  Note also that
the non-Hermitian skin effect is associated with the exceptional points of the non-Hermitian Hamiltonian when the boundaries are introduced 
 \cite{budich2020RMP,Ueda2020AdvPhys},  whereas no exceptional points are formed for periodically driven systems described by the unitary Floquet evolution operators.
 More details on these issues are available in Appendix~\ref{Appendix:Non-Hermit}.

	\subsection{Beyond fine-tuned driving}

	\begin{figure}[t]
		\includegraphics[width=2cm]{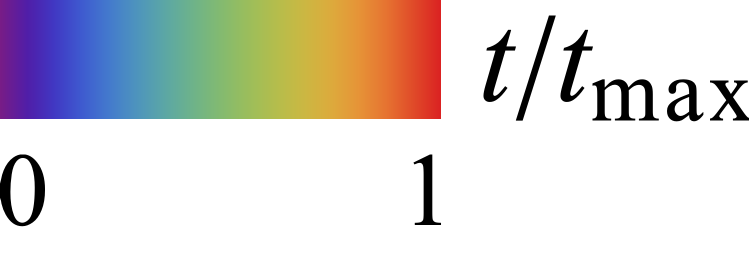}
		\qquad\qquad
		\includegraphics[width=2cm]{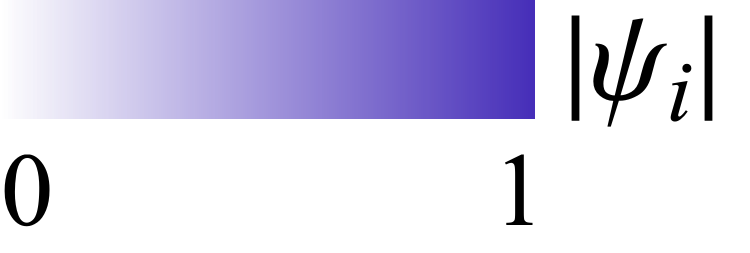}		
		\\
			\parbox{2.8cm}{
				\includegraphics[width=2.7cm]{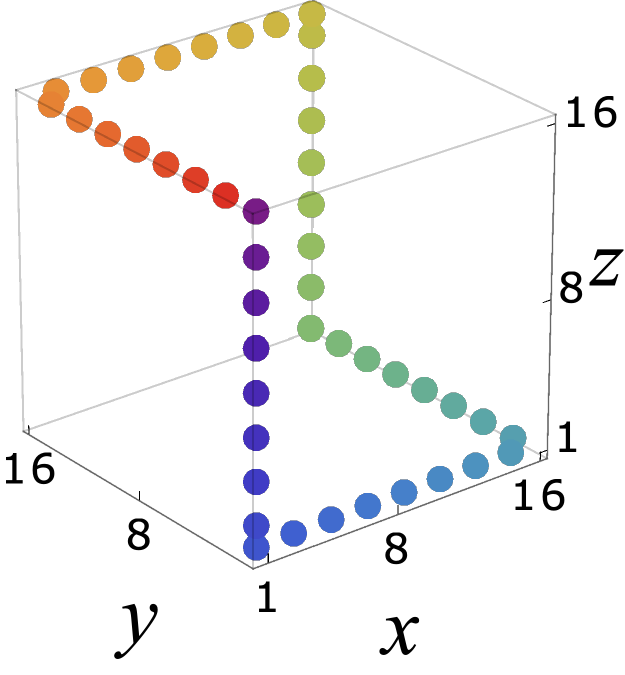} \\ (a) without defect }  
		\boxed{
			\parbox{5.2cm}{
				\parbox{2.6cm}{
					\includegraphics[width=2.8cm]{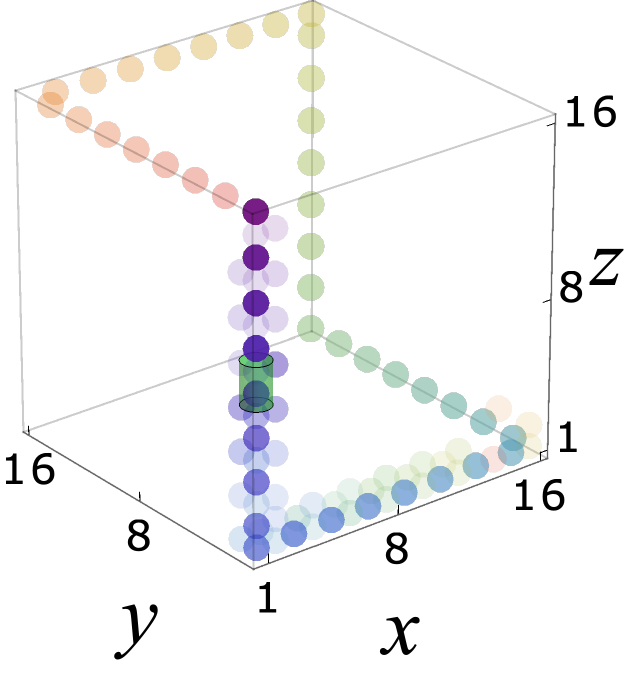}   }
				\parbox{2.4cm}{
					\includegraphics[width=2.5cm]{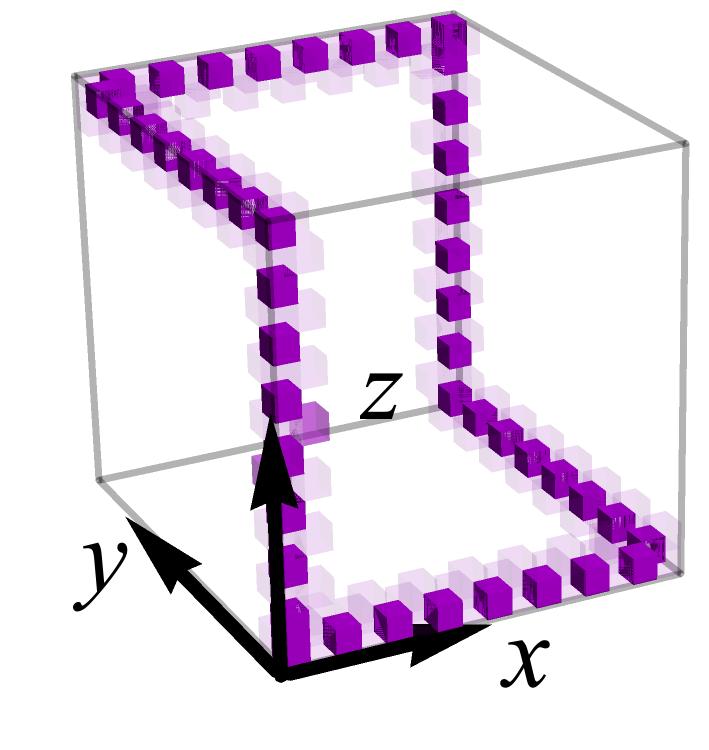} \\ \quad    }
				\\ (b) with defect }  
		}
		\\
		\parbox{8.6cm}{
			\parbox{2.8cm}{
			\includegraphics[width=2.8cm]{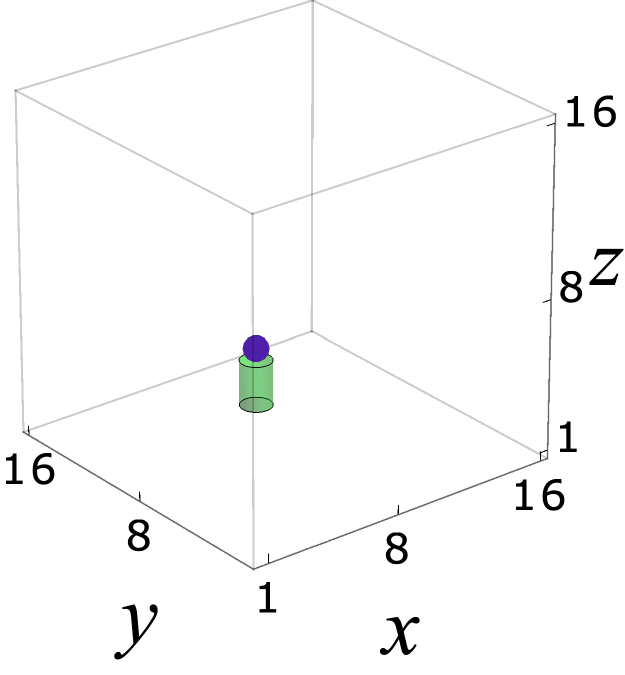}
			\\ (c1) $ t=3 $ }
			\parbox{2.8cm}{
			\includegraphics[width=2.8cm]{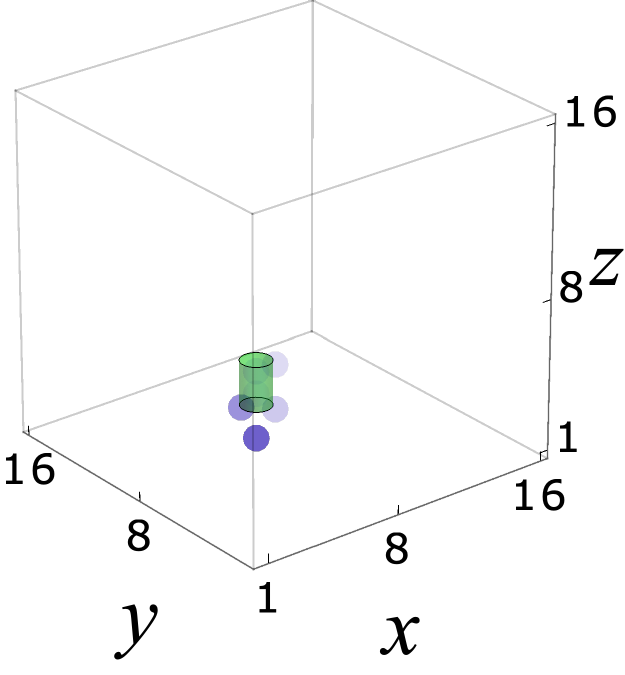}
			\\ (c2) $ t=5 $ } 
			\parbox{2.8cm}{
			\includegraphics[width=2.7cm]{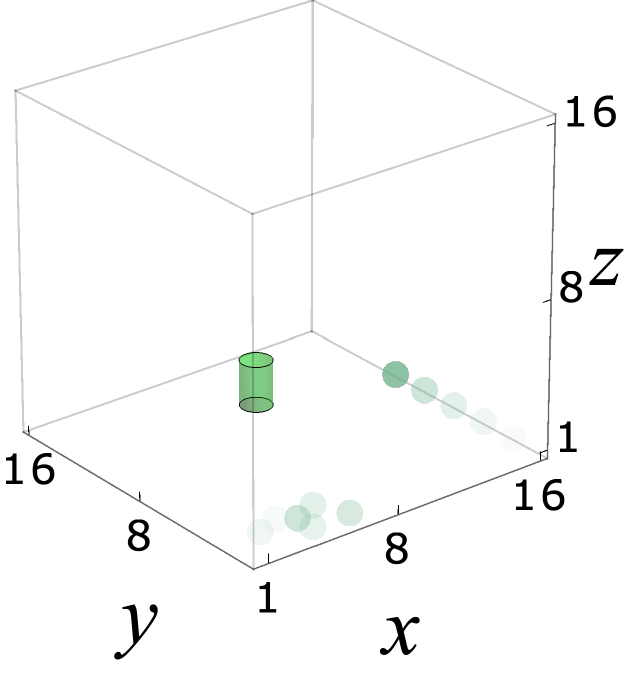}
			\\ (c3) $ t=20 $ }
		}    
		\caption{\label{fig:dyn}  The dynamics of  a particle initially localized at the site $ (x,y,z)=(1,1,16) $ for a system of $16\times16\times16$ sites with full open boundary conditions and fine tuned $\phi=\pi/2$. The squared wave function at different times is reflected in the opacity of the plotted dots. Different colors indicate time. (a) Without defect. (b) In the presence of a potential defect of energy $\Delta = 3\pi $ at the two sites marked by the green tube. Additionally, we plot the squared wave function of one hinge mode. (c) Snapshots of the time evolution in the presence of the defect at different times. 
}
	\end{figure}

	\begin{figure}
		[t]
		\includegraphics[width=2cm]{fig3_legend1}
		\qquad\qquad
		\includegraphics[width=2cm]{fig3_legend2}		
		\\
			\parbox{2.8cm}{
				\includegraphics[width=2.7cm]{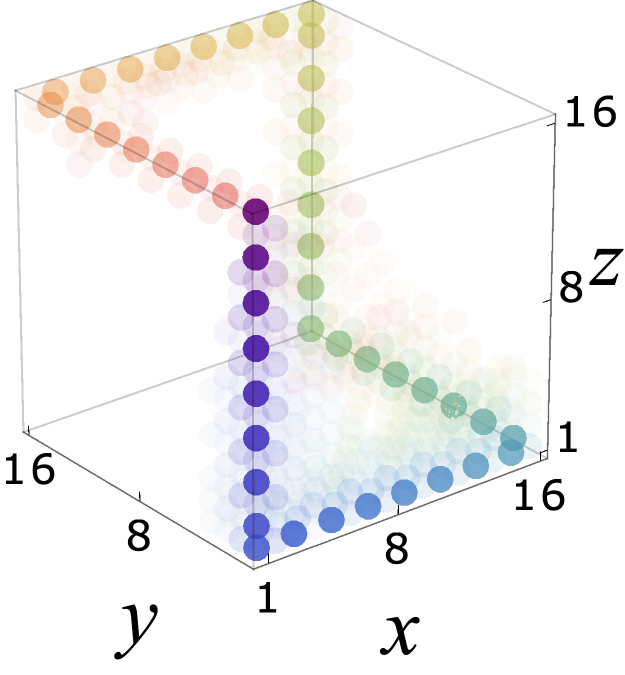} \\ (a) without defect }   
		\boxed{
			\parbox{5.2cm}{
				\parbox{2.6cm}{
					\includegraphics[width=2.8cm]{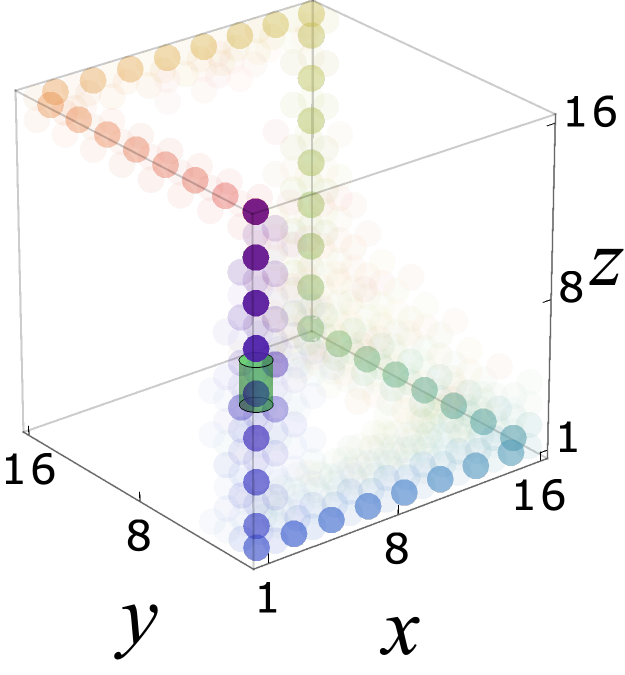}   }
				\parbox{2.4cm}{
					\includegraphics[width=2.4cm]{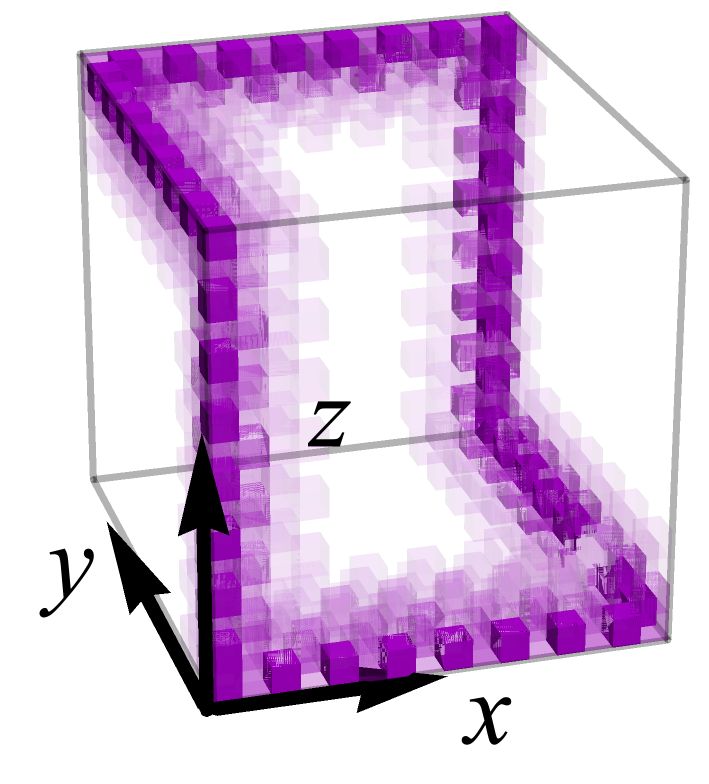} \\ \quad    }
				\\ (b) with defect}    
		}
		
		\parbox{8.6cm}{
			\parbox{2.8cm}{
				\includegraphics[width=2.8cm]{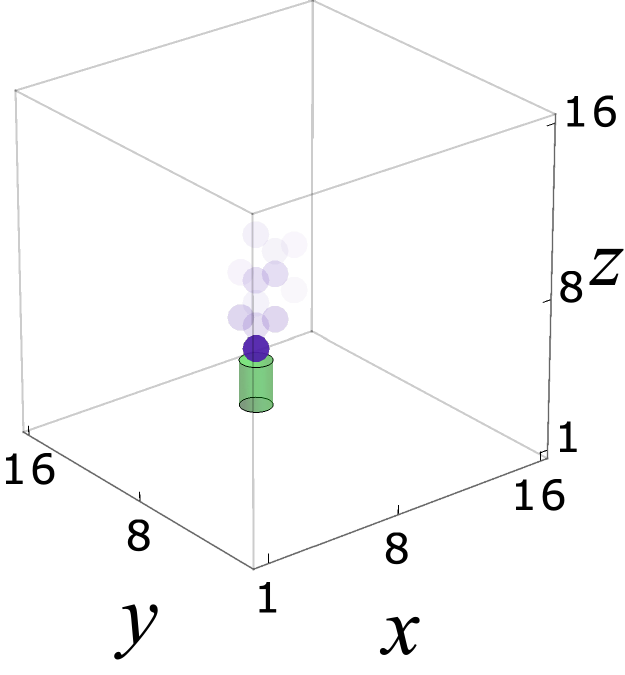}
				\\ (c1) $ t=3 $ }
			\parbox{2.8cm}{
				\includegraphics[width=2.8cm]{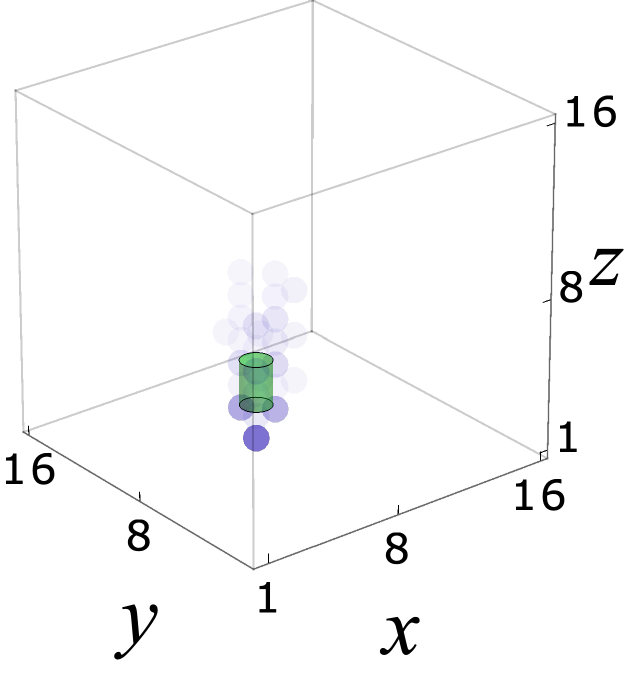}
				\\ (c2) $  t=5 $ }
			\parbox{2.8cm}{
				\includegraphics[width=2.7cm]{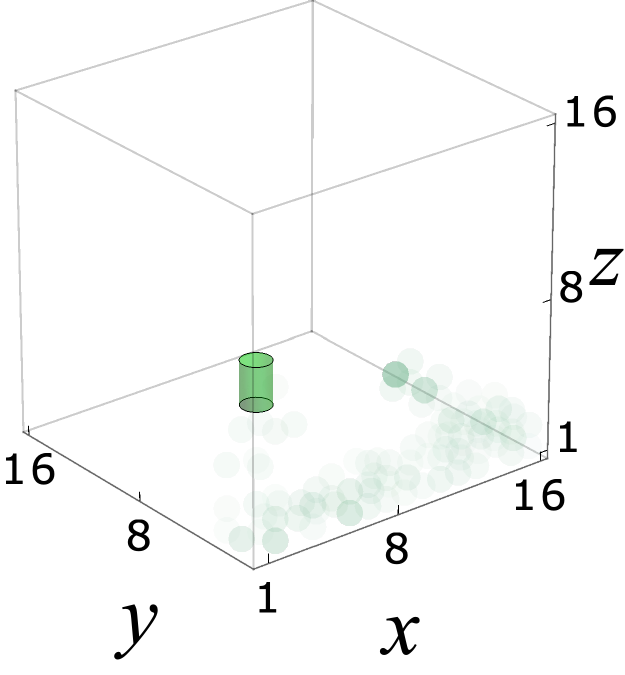}
				\\ (c3) $  t=20 $ }
		}    
		\caption{\label{fig:dyn_detuned} As Fig.~\ref{fig:dyn}, but for $ \phi = 0.9\times\pi/2 $, away from fine tuning.}
	\end{figure}
	
	Although the above  analysis is based on ballistic trajectories at the fine-tuned driving parameter $ \phi=\pi/2 $, we expect the chirality of the hinge modes to be robust also against perturbations and tuning away from $\phi=\pi/2$.   This applies especially  to the hinge states with larger chiral velocities, which are situated closer to the hinge    and thus are spatially well separated from counter-propagating modes at the opposite hinge. 
	The chiral hinge modes  persist for a rather wide range of $ \phi $ beyond $ \phi=\pi/2 $.
	This can be observed from the example of $ \phi=\pi/3 $ ($ 33.3\% $ detuning, half-way across the Weyl phase transition) displayed in Fig.~\ref{fig:spectrum} column (2) around $ k_z=\pi/2$ and $E=0 $. 
 	We can see that the hinge modes at smaller distances $M$ from the hinge still preserve their chirality 
	 along the hinge, as expected previously. In turn, hinge modes with larger $M$ that are closer to the sample center, are gradually mixed with modes of opposite chirality   and lose chiral nature together with localization in the $ xy $-plane when $ \phi $ deviates away from $ \phi=\pi/2$.

 	We  have also considered the exemplary eigenstates  for a cube geometry with full open boundary conditions along all Cartesian axes $x$, $y$ and $z$ presented in column (3) of Fig.~\ref{fig:spectrum} showing that
	for  $ \phi =\pi/2$ and $ \phi=\pi/3$ the chiral  modes at  certain hinges are joined to form a closed loop respecting inversion symmetry of the system [see also Fig.~\ref{fig:dyn} (a)]. The six hinges not participating in this closed loop do not carry hinge modes, since their two boundary planes are not connected along the diagonal direction  $\bm{d}$. Meanwhile, non-hinge modes  representing the bulk dynamics all center along the cubic diagonal [column (3) of Fig.~\ref{fig:spectrum}].	
	
\begin{figure}
	[h]
	\includegraphics[width=2cm]{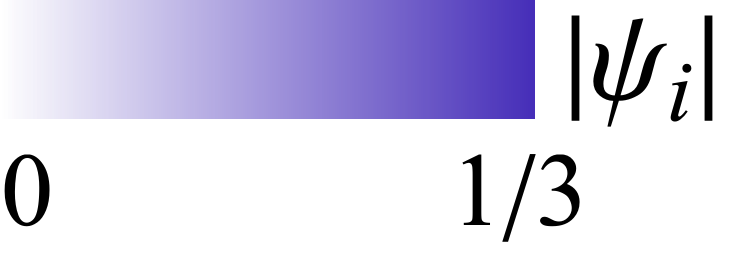}
	\\
	\parbox{4cm}{\includegraphics[width=4cm]{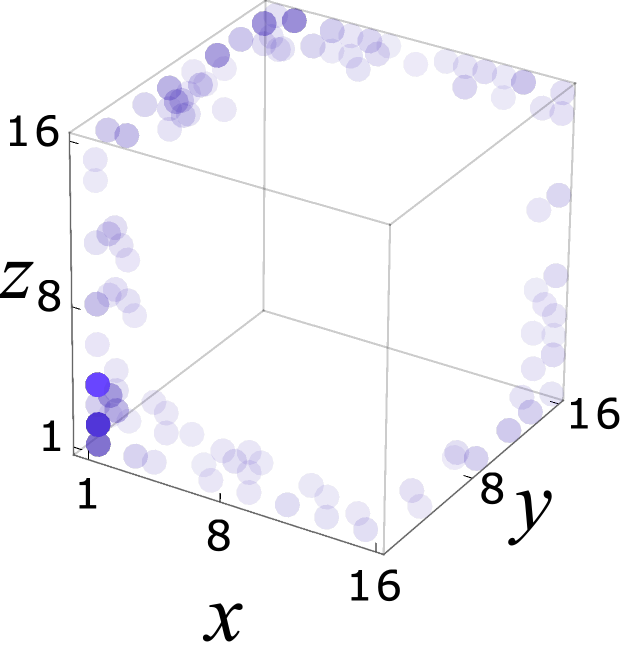} \\ (a) $ \Delta=0 $} \quad 
	\parbox{4cm}{\includegraphics[width=4cm]{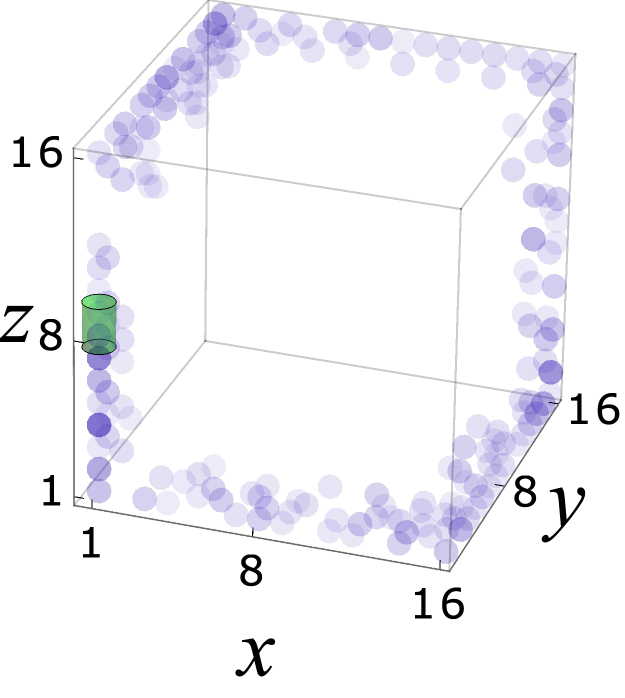} \\ (b) $ \Delta = 3\pi $}
	\caption{\label{fig:long_time}  Density distribution of a particle initially localized at the corner site
	$ (x,y,z) = (1,1,16)  $ after an evolution over $100$ driving cycles for the non-fine-tuned parameter 
	$\phi=0.9\times \pi/2$, without defect (a) and with a defect (b) [corresponding to the parameters of 
	Fig.~\ref{fig:dyn_detuned} (b)]. The densities are representative for late-time states in the limit $t\to\infty$ as
	similar   distribution is found also after 1000 driving cycles.}
\end{figure}

	 In the previous Sections  we have made arguments that  at the fine tuning, $ \phi=\pi/2 $,  the chiral   modes locating right at   the hinge $ x=y=1 $ (or $ x=y=L $)  do not involve  the ballistic motion in the $ xy $-plane over many driving periods, and therefore their chiral transporting feature should be preserved even if they become hybridized with nearby modes carrying different   or no chirality.  We now further test such an expectation by checking the robustness of chiral hinge transports in the presence of local defects in  Figs.~\ref{fig:dyn} and~\ref{fig:dyn_detuned}. Here, we simulate the dynamics of a particle in the presence of  a defect for a system with open boundary conditions in all three directions  representing a generalization  of the  defect-free situation shown in column (3) of Fig.~\ref{fig:spectrum}. Figure~\ref{fig:dyn} illustrates the dynamics of a particle initially located at a corner $ (x,y,z) = (1,1,16)  $ of the system, where  two orthogonally oriented transporting hinges meet, (a) for the fine tuned situation without  a defect and (b)    in the presence of a strong potential offset of $\Delta = 3\pi$ on two neighboring hinge sites (indicated by a green tube) at $ (x,y,z) = (1,1,8), (1,1,9) $, respectively.   The corresponding plots for non-fine-tuned driving with $\phi=0.9(\pi/2)$ are presented in Fig.~\ref{fig:dyn_detuned}.
  	We find that despite  this strong defect  the chiral nature of the hinge modes   ensures that no backscattering occurs at the defect and the majority of the   wave-packet continues to follow  chiral trajectories along the hinges. In Figs.~\ref{fig:dyn} (b) and~\ref{fig:dyn_detuned} (b) we also plot representative eigenstates of the system with defects.   These eigenstates remain localized  at the hinge,  with only a small distortion compared  to the situation without defect shown in column (3) of Fig.~\ref{fig:spectrum}. That means the  localization properties  of  the particle carrying  the chiral motion   along the hinge  are not just an ephemeral phenomena. Once the original particle distribution overlaps with such an eigenstate, certain portion of those particles will always be localized along the hinge undergoing constant chiral  motion. This is further verified in Fig.~\ref{fig:long_time} for   evolution over a long time. It shows the time-evolved state after 100 driving cycles for the non-fine-tuned system ($\phi=0.9\times \pi/2$) both without defect (a) and with defect (b). Very similar   distribution is also found after even longer evolution, e.g.\ over 1000 driving cycles; the densities are, thus, representative for late-time states in the limit $t\to\infty$. Note that such a process  implies the particle have encountered the defects for many (or infinite) times when they loop over the connected hinges repeatedly, without losing their localization  the hinge and  being scattered away.

	\begin{figure}
		[h]
		\parbox{2.3cm}{
			\begin{tabular}{ccccccc}\hline\hline
				steps & 1 & 2 & 3 & 4 & 5 & 6
				\\
				\hline
				$ \alpha_{00} $ & 0 & 0 & 0 & 1 & 1 & 1
				\\
				\hline
				$ \alpha_{10} $ & 1 & 0 & 1 & 0 & 1 & 0
				\\
				\hline
				$ \alpha_{01} $ & 1 & 1 & 0 & 0 & 0 & 1
				\\
				\hline
				$ \alpha_{11} $ & 0 & 1 & 1 & 1 & 0 & 0
			\end{tabular}
		}\\
		\parbox{8.5cm}{
			\boxed{
			\includegraphics[width=3.8cm]{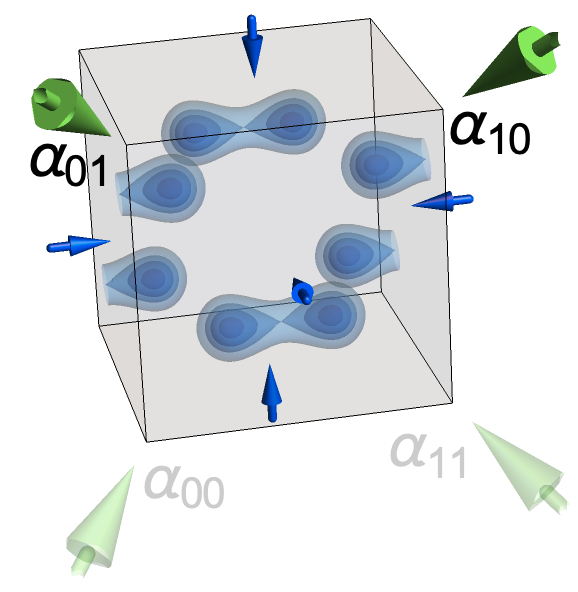} 1
			}
			\boxed{
			\includegraphics[width=3.8cm]{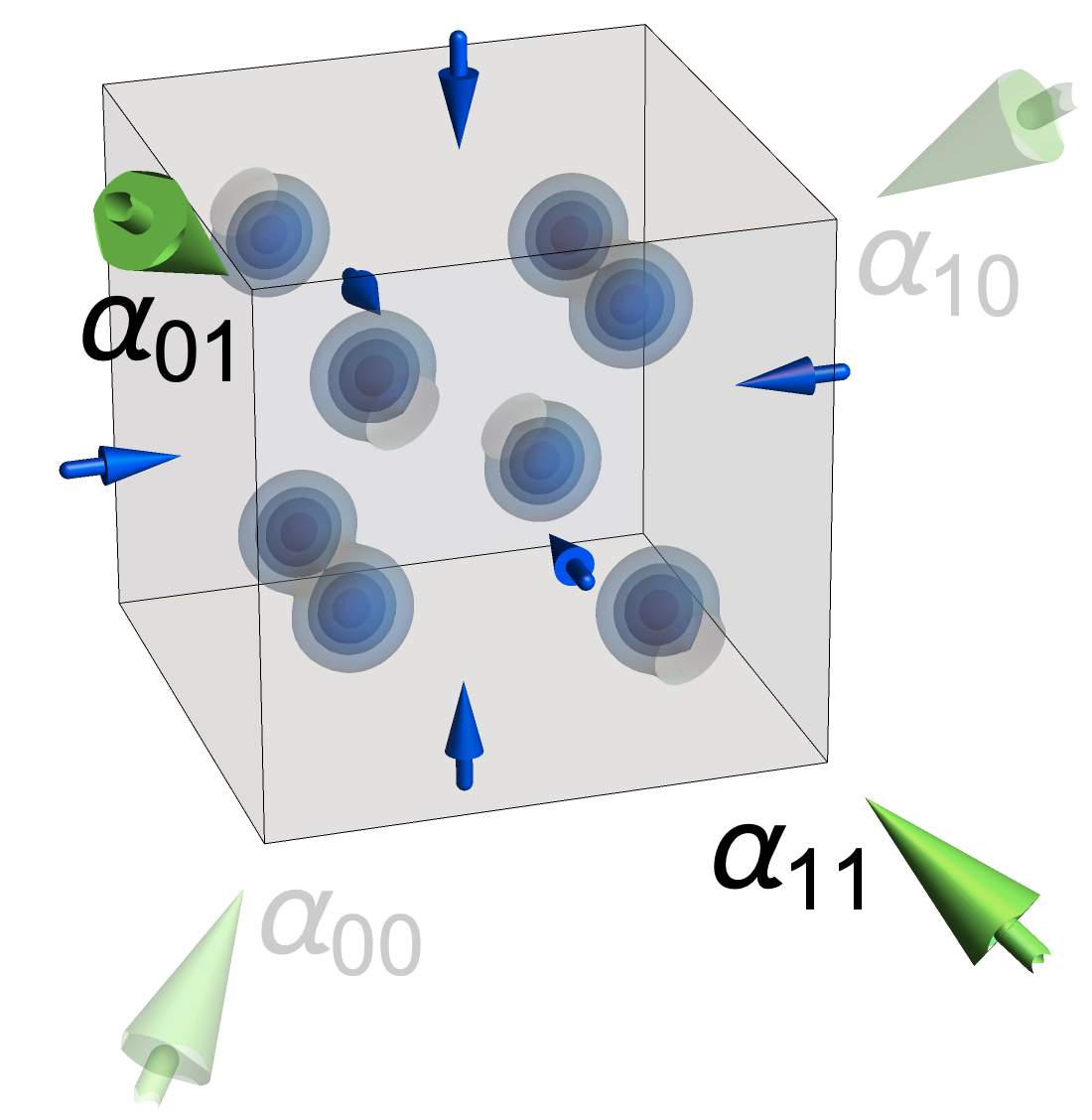} 2 }\\
			\boxed{
			\includegraphics[width=3.8cm]{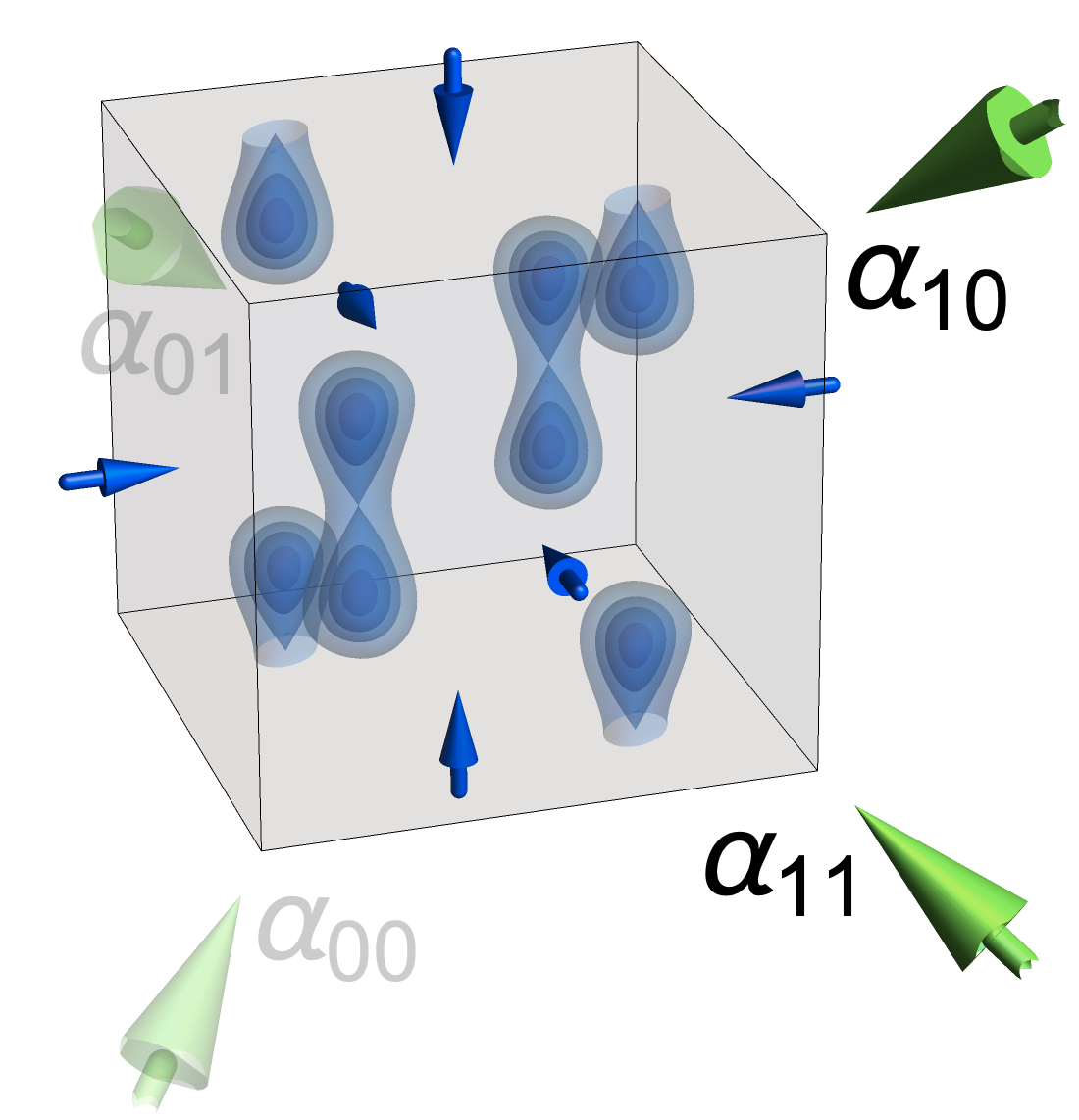} 3} 
			\boxed{
			\includegraphics[width=3.8cm]{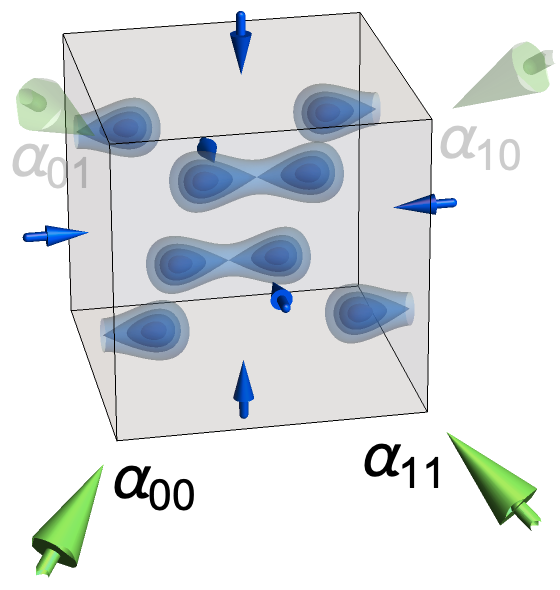} 4} \\
			\boxed{
			\includegraphics[width=3.8cm]{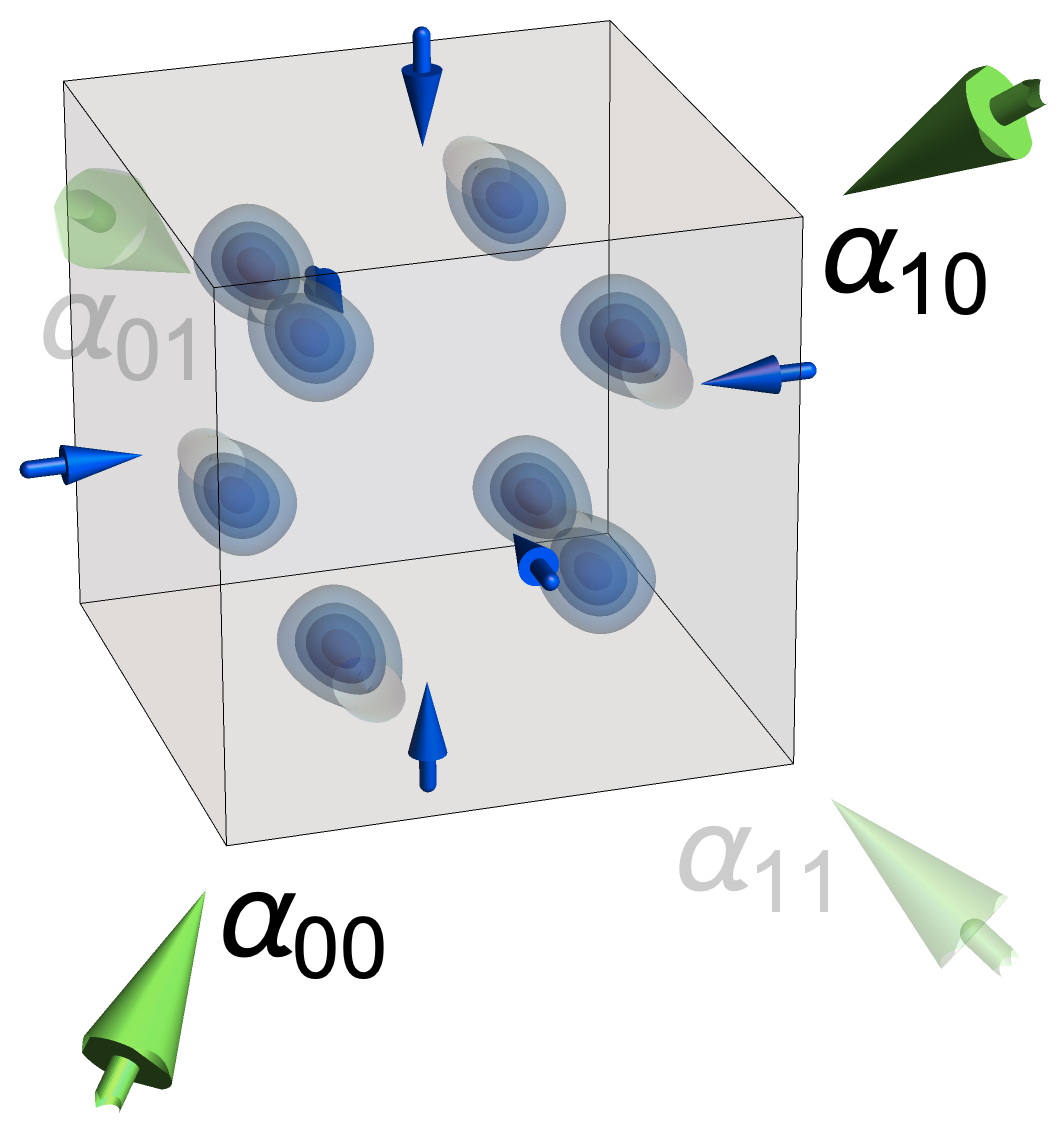} 5} 
			\boxed{
			\includegraphics[width=3.8cm]{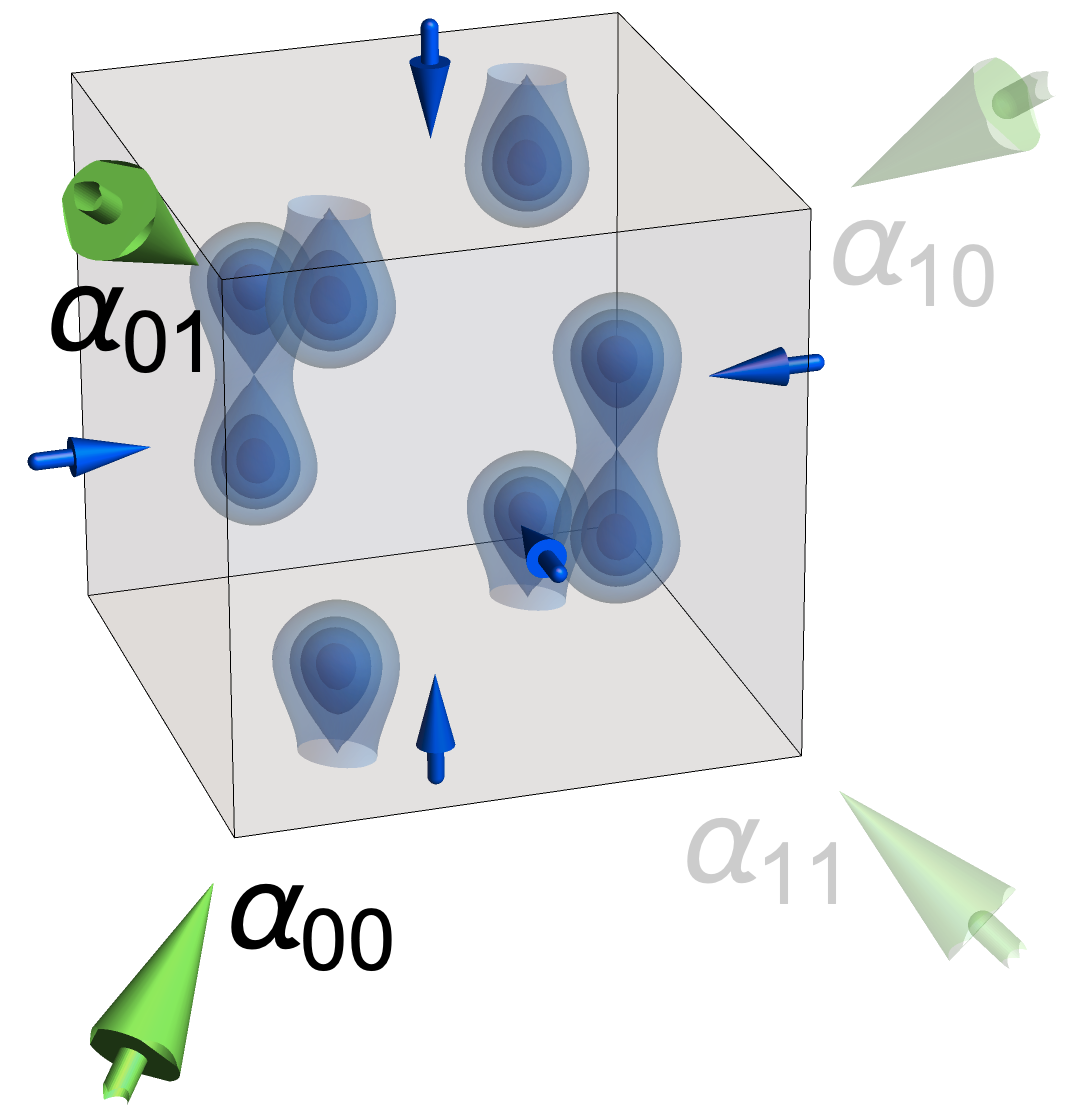} 6}
			
		}
		\caption{\label{fig:expt}  The stepwise modulation of the dimensionless superlattice amplitudes $\alpha_{ab}$ according to the protocol given in the table, gives rise to different dimerizations of the cubic lattice in each driving step, that enables tunneling along the desired bonds. }
	\end{figure}

	\section{Experimental Realization with ultracold atoms in optical lattices
	\label{sec:Experimental}}
	\subsection{Engineering of the  driven lattice}
 Above, we have shown that the proposed modulation of tunneling gives rise to a  variety of phenomena, including the robust creation of a pair of Weyl points, unidirectional bulk transport, chiral Goos-H\"{a}nchen-like shifts, and the macroscopic accumulation of chiral hinge modes for open boundary conditions corresponding to a chiral second-order Floquet skin effect. The model itself is, nevertheless, rather simple and its implementation with ultracold atoms in optical lattices can be accomplished using standard experimental techniques. All what is needed is a static cubic host lattice  potential of equal depth $V_0$ in each  Cartesian direction and a superlattice potential, whose amplitudes along various diagonal lattice directions are modulated  in a stepwise fashion in time  in order to suppress/allow tunneling along the six different bonds specified by our protocol.  
   This can be achieved using the  following optical lattice potential:
	\begin{align}\nonumber
	&   V  (\boldsymbol{r}) = V_0 \sum_{\mu=x,y,z} \cos^2 (2k_L r_\mu) 
	\\
	&
	+ V_1\sum_{a,b=0,1} 
	\alpha_{ab}(t) \cos^2 k_L (x+(-1)^{a}y + (-1)^{b}z )\,,
	\end{align}
where only two of the four modulating lasers $ \alpha_{ab}$ with $a,b = 0,1$ are turned on   in each driving step, as shown in Fig.~\ref{fig:expt}. Such a modulation provides the required six-stage driving of the cubic lattice. Note that a similar modulation has recently been implemented in two dimensions \cite{Wintersperger2020}.

	\subsection{Detection of hinge dynamics}
	
	To observe the dynamics associated with the hinge modes, one can apply the boxed potential achieved in recent experiments~\cite{Gaunt2013,Navon2016,Mukherjee2017a,Lopes2017,Lopes2017a,Eigen2017, Ville2018}. There,  thin sheets of laser beams penetrate through the quantum gases creating a steep potential barrier.  Three pairs of such beams are imposed in a three-dimensional system, creating the sharp ``walls" for the box potential while leaving the central part of the gases homogeneous. 
		
\begin{figure}
		[h]
		\parbox{2.7cm}{\includegraphics[width=2.7cm]{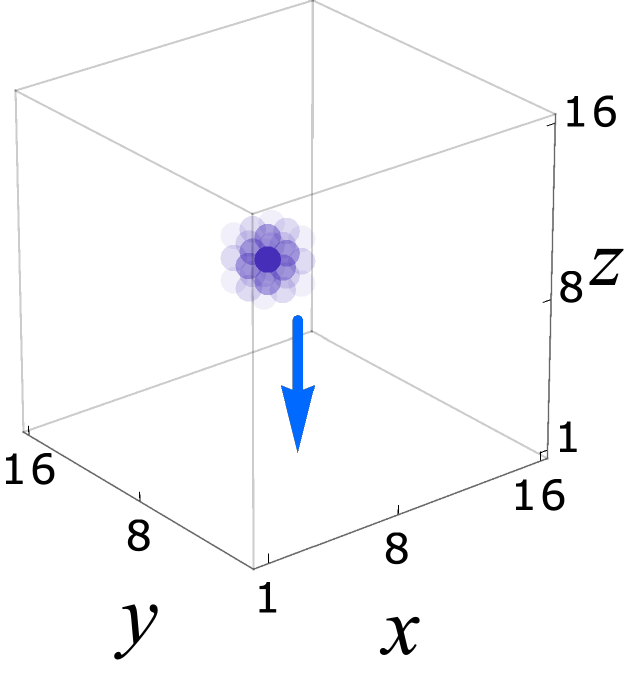} \\ (a) $ |\psi_i(t=0)| $}
		\parbox{5.7cm}{  \includegraphics[width=2.8cm]{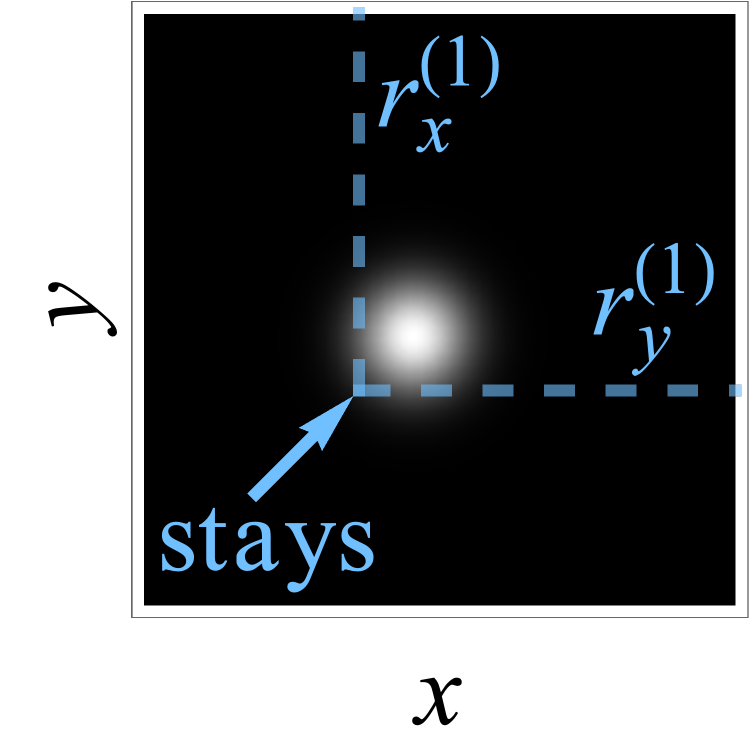}  \includegraphics[width=2.8cm]{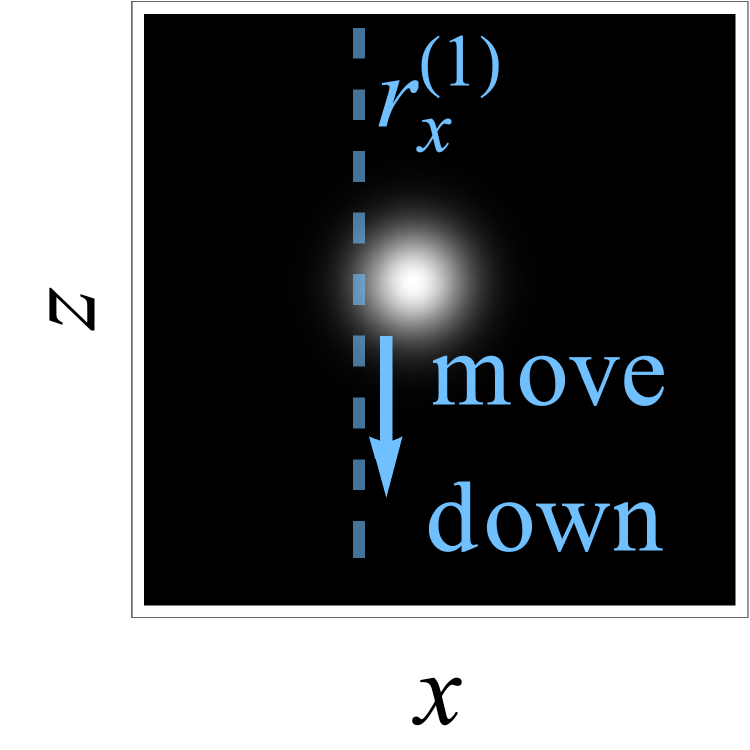} \\ (b) Projection to $ x $-$ y $ (left) and $ x $-$ z $ (right)  }
		\\ \quad \\ \, 
		\boxed{
			\parbox{3.7cm}{\includegraphics[width=3.2cm]{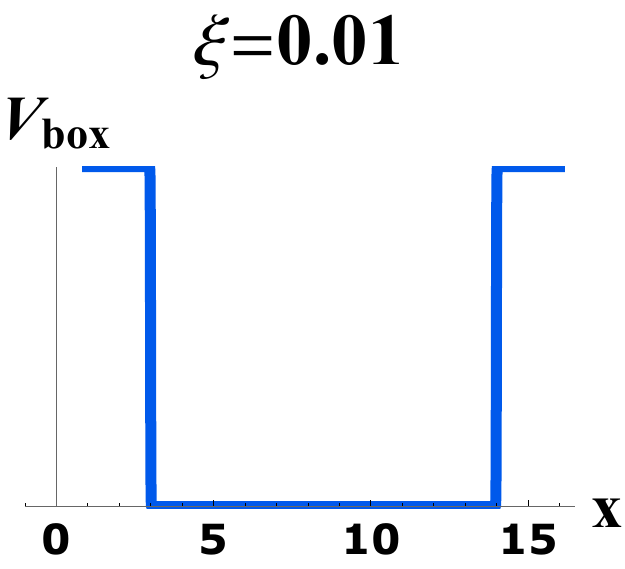} \\ (c1) Ideal boundary
				\includegraphics[width=3.7cm]{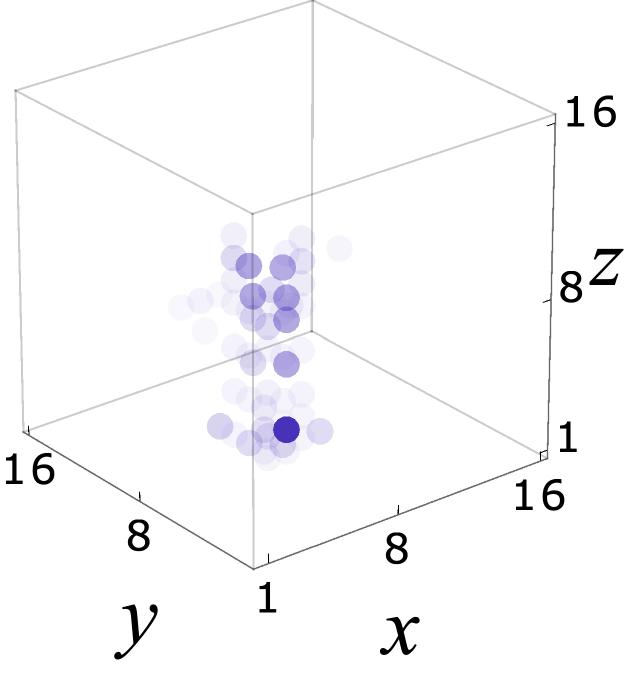} \\ (c2) $ |\psi_{i}(t=5)| $
				\includegraphics[width=3.7cm]{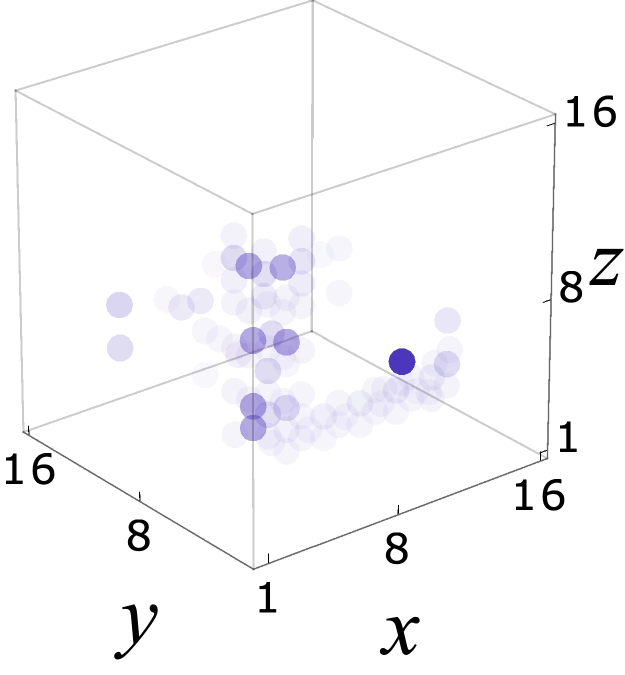}\\ (c3) $ |\psi_i(t=10)| $
		} } \quad
		\boxed{
			\parbox{3.7cm}{\includegraphics[width=3.2cm]{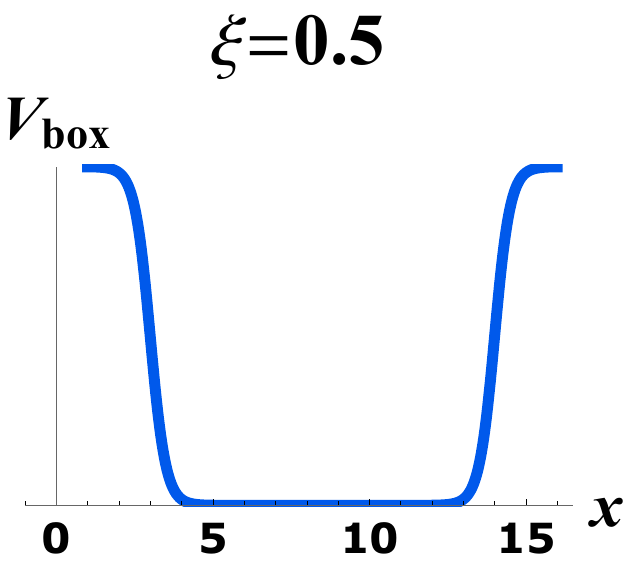}\\ (d1) Softer boundary
				\includegraphics[width=3.7cm]{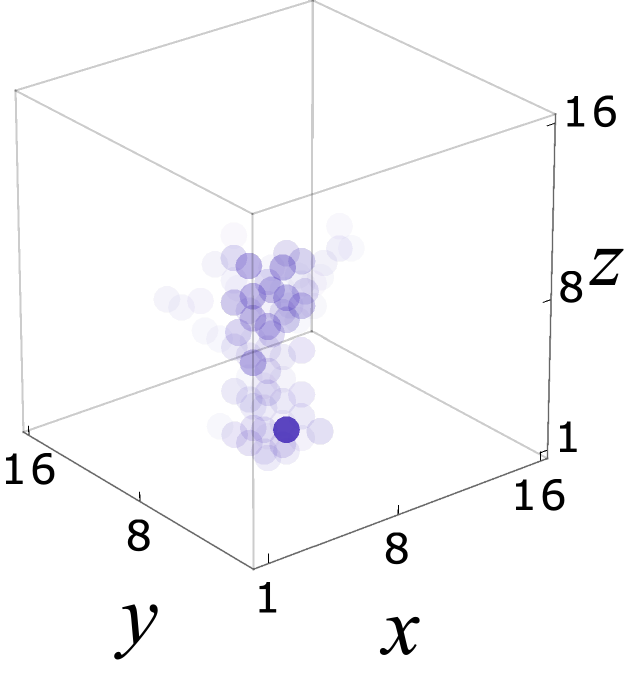} \\ (d2) $ |\psi_i(t=5)| $
				\includegraphics[width=3.7cm]{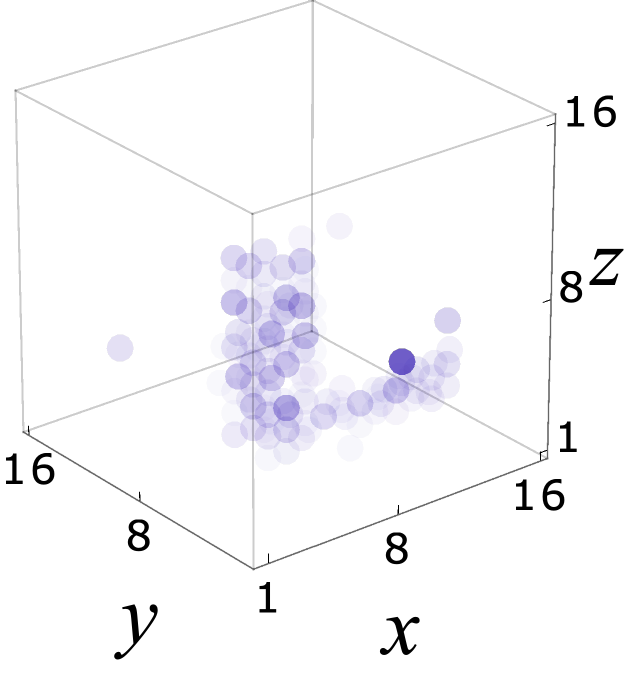}\\ (d3) $ |\psi(t=10)| $
		} }
		\caption{\label{fig:expt_dyn} The dynamics of particles in a box potential with different softness of the boundary. The initial density distribution takes a Gaussian profile spreading over several lattice sites. The parameters are $ \phi=\pi/2, V_bT/6\hbar = 7.5\pi $. The initial Gaussian profile has the center $ (x_0,y_0,z_0) = (4,4,12) $ and width $ s_0=0.75 $.}
	\end{figure}

	Essentially, such a potential combined with our lattice driving scheme immediately leads to the particle dynamics described in Fig.~\ref{fig:dyn} and Fig.~\ref{fig:dyn_detuned}. To take into account realistic experimental situations, two modifications are adopted in our following simulations. First, we consider the effect of a relatively ``softer" wall for the box potential with
	\begin{align}
		V_{\text{box}}(\boldsymbol{r}) = \frac{V_b}{2} \sum_{\mu=x,y,z} \left( 2 + \tanh\frac{r_\mu^{(1)} - r_\mu  }{\xi} + \tanh\frac{r_\mu - r_\mu^{(2)} }{\xi} \right),
	\end{align}
	where the potential ramps up over a finite distance of roughly $ 4\xi $ near the boundaries $ r_\mu^{(1,2)} $, see Fig.~\ref{fig:expt_dyn} for instance. 
The second modification we adopt is that the initial state  is not taken to be localized on a single lattice site but described by a gaussian wave packet of finite width, 
	\begin{align}
		\psi_{i=(x,y,z)}(t=0) = e^{-[(x-x_0)^2+(y-y_0)^2+(z-z_0)^2]/2s_0^2} .
	\end{align}

	The results of the dynamics  are presented in Fig.~\ref{fig:expt_dyn} (c1)--(c3) and (d1)--(d3), for the ideal sharp boundary (as in Fig.~\ref{fig:dyn}) and the realistic softer boundaries in experimental setting respectively. We see that the chiral motion snapshots for the ideal/softer boundaries exhibit qualitatively the same characters, signaling that a softer boundary does not cause significant changes. This is in consistent with the previous simulation showing the robustness of hinge states and their resulting chiral dynamics against local defects in Fig.~\ref{fig:dyn} and Fig.~\ref{fig:dyn_detuned}. The major difference from previous cases, then, derives from the initial state that overlaps with more than one set of eigenstates near the hinge, each with different group velocities as given in Eq.~\eqref{V-M-pm}. Finally, we mention that some portion of initial particle distribution would reside within the region with significant changes in $ V_{\text{box}} $. That portion of the particles could be permanently confined to the initial  hinge due to a mechanism similar to  Wannier-Stark localization. However, the majority of the particles are still traveling into the connecting hinges, as shown in Fig.~\ref{fig:expt_dyn} (c3) and (d3). 
	
	In cold atom experiments, the density profiles are usually detected by taking a certain projection plane, where the integrated (column-averaged) densities are observed. To this end, we point out that the hinge dynamics can be confirmed by observing the density profiles in two perpendicular planes. A schematic plot is given in Fig.~\ref{fig:expt_dyn} (b), corresponding to the dynamics along the hinge $ x=y\sim 1 $. The density profile taken at $ x-y $ plane (i.e. the ``top" view) would show a localized distribution at the corner, verifying the particles only locate at $ x=y\sim1 $. Meanwhile, the profile at $ x-z $ plane (i.e. ``side" view) indicates the movement/spreading along $ z $. In a more general situation, i.e. at long time limit with all 6 hinges populated as in Fig.~\ref{fig:long_time}, additional image projection planes could be exploited. We also mention that a simultaneous implementation of multiple imaging planes have been applied in experiments~\cite{Lu2020,Wang2020}.

	\subsection{Detection of Floquet Weyl points}  \label{Detection-Weyl}
	Weyl physics has been explored in recent cold atom experiments and theoretical proposals~\cite{Lu2020,Wang2020,Zhang2015,Dubcek2015,He2016}, and also extensively in solid state systems  \cite{Armitage2018}. Here, we discuss a scheme closely related to a recent experiment~\cite{Uenal2019,Wintersperger2020} detecting the spacetime singularities in anomalous Floquet insulators.

	\begin{figure}
		[h]
		\parbox{4.3cm}{\includegraphics[width=3.6cm]{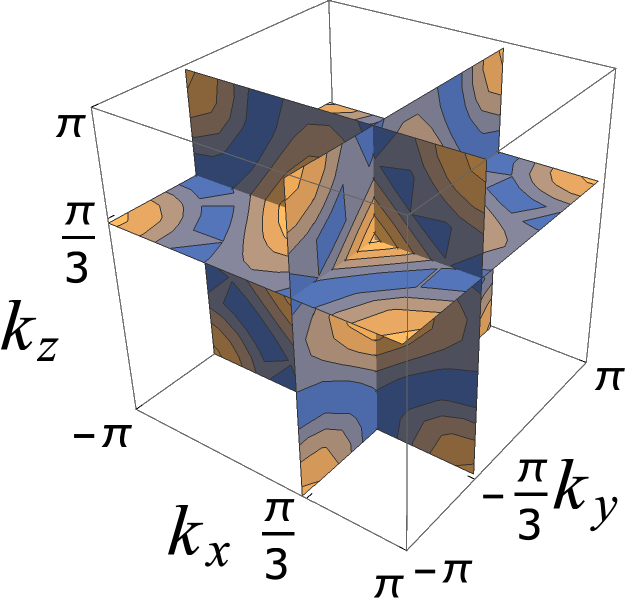}
			\includegraphics[width=0.5cm]{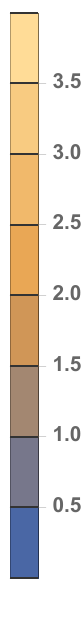} \\
			(a1) $ \Delta^{(0)} $ at $ \phi=\pi/8 $
		}
		\parbox{4.2cm}{\includegraphics[width=3.7cm]{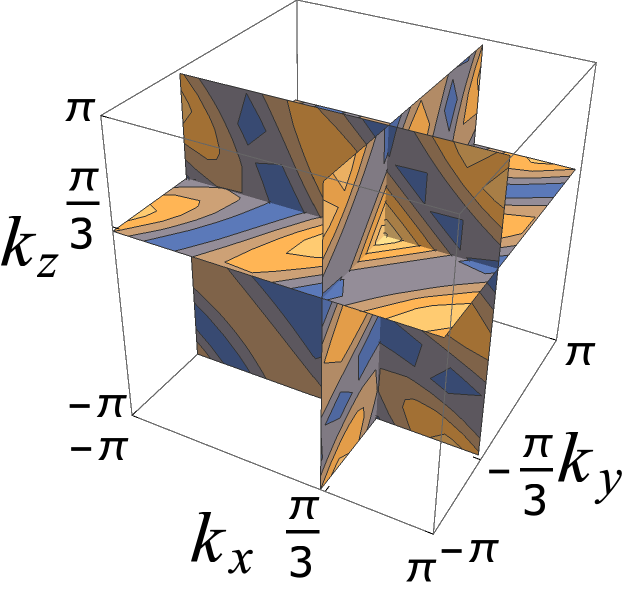}
			\includegraphics[width=0.4cm]{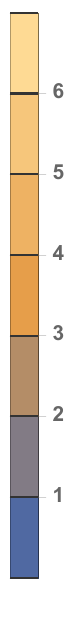} \\
			(a1) $ \Delta^{(0)} $ at $ \phi=\pi/3 $
		}\\ \quad \\ \quad \\
		\parbox{8cm}{\includegraphics[width=8cm]{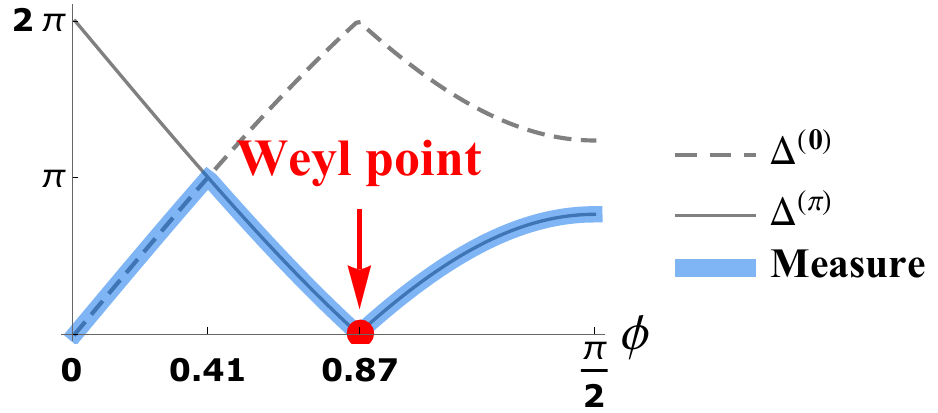} (b) Exemplary gap at $ \boldsymbol{k}_0=k_0(1,-1,1) $ with $ k_0=\pi/3-0.2 $.}
		\caption{\label{fig:gap} Simulation of gap measurements using St\"{u}ckelberg interferometry. (a1) and (a2) Contours for quasi-energy gap at $ E=0 $, for $ \phi=\pi/8 $ and $ \phi=\pi/3 $ respectively. (b) The gaps at $ E\sim0 $ and $ E\sim\pi $, and the measured gap which takes the smaller one of the two.  }
	\end{figure}
	First, the band touching at Weyl points can be verified using the St\"{u}ckelberg interferometry  \cite{Arimondo10PRA,Weitz10PRL,Nori10PR,Wintersperger2020}. Such a method measures the smaller gap for the two bands at $ E\sim0 $ and $ \pi $, see Ref.~\cite{Wintersperger2020} for details. Compared with the experiments for insulators, a difference here is that the two bands are, overall, always gapless at $ E=0 $. That means if a global gap is measured, it will prevent us from gaining information about the gaps or band touching at quasienergy $ E=\pi $. But fortunately, there exists a finite region neighboring to $ \boldsymbol{k}_0=\pi/3(1,-1,1) $ where the bands are always gapped at $ E=0 $ for all $ \phi $, see Fig.~\ref{fig:gap} (a1), (a2) for example. Then a {\em local} gap closure can be measured near $ \boldsymbol{k}_0 $. 
	
	The specific measurement for our case can be performed in the following way. An example for $ \boldsymbol{k}_0=(\pi/3-0.2)(1,-1,1) $ is presented in Fig.~\ref{fig:gap} (b). Let us denote the quasi-energy of the two Floquet bands at $ \boldsymbol{k}_0 $ with $ E_\pm(\boldsymbol{k}_0) = \pm E_0 $. Here, we use the branch cut along $ \pi $ in taking the logarithm of Floquet eigenvalues $ e^{-iE_\pm (\boldsymbol{k}_0)} $.  They have the same magnitudes and opposite signs due to particle-hole and inversion symmetry as explained in Sec. \ref{sec:Model}. Then, the local gap around $ E\sim 0  $ is $ \Delta^{(0)} \equiv 2E_0 $, while the other gap around the Floquet Brillouin zone boundary $ E\sim \pi $ is $ \Delta^{(\pi)} \equiv 2\pi - 2E_0 $. Therefore, $ \Delta^{(0)} = \Delta^{(\pi)} $ can only occur at $ 0, \pi  $ mod $ 2\pi $.
	In experiments, one can start from the high frequency limit ($ \phi\rightarrow0 $) where the band width is small   compared to $2\pi$ and therefore the measured gap always corresponds to $ \Delta^{(0)} $. Slowing down the driving, the gap $ \Delta^{(\pi)} $ shrinks while the other gap $ \Delta^{(0)} $ expands. At some point, the two gaps coincide with their magnitudes, as shown in Fig.~\ref{fig:gap} (b). Since it is always the smaller one of $ \Delta^{(0)} $ and $ \Delta^{(\pi)} $ that will show up in experimental measurement, one will observe a cusp shape of the measured gap, i.e. near $ \phi \approx 0.41 $ in Fig.~\ref{fig:gap} (b). One could then imply from the occurrence of the cusp that for $ \phi>0.41 $, the experimental data starts to reveal $ \Delta^{(\pi)} $, whose vanishing at $ \phi\approx 0.87 $ shows the existence of the Weyl point around $ E\sim \pi $. Similar measurements can be performed for $ \boldsymbol{k} $ slightly deviating from $ \boldsymbol{k}_0 $, which will show that at $ \phi=0.87 $, $ \Delta^{(\pi)} $ remains finite, proving that the band closure around $ E\sim \pi $ is a point contact.
	When  the designated $ \phi $ is slowly approached, one can perform a measurement of the gap at a certain $ \boldsymbol{k} $. A shortcut for our system is that focusing on  quasimomenta along the diagonal $ \boldsymbol{k}_0 = k_0(1,-1,1) $ is sufficient to determine the Weyl point,  as discussed in Sec.\ref{sec:Periodic}.

\begin{figure}
		[h]
		\parbox{6cm}{\includegraphics[width=5cm]{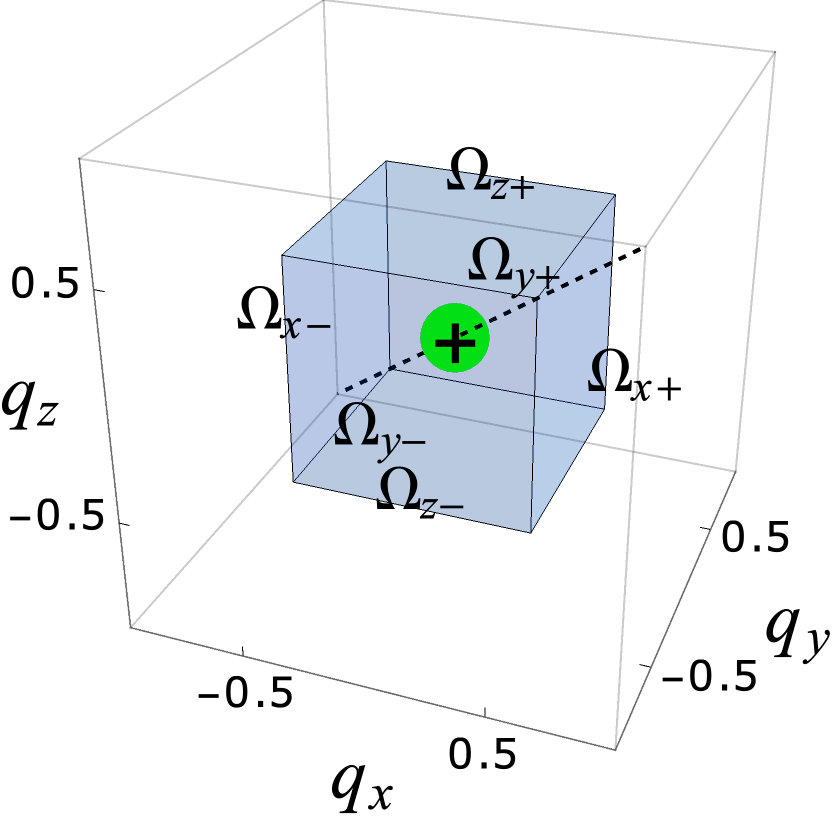} \\ (a) The surfaces wrapping a Weyl point}
		\parbox{1cm}{\includegraphics[width=1.2cm]{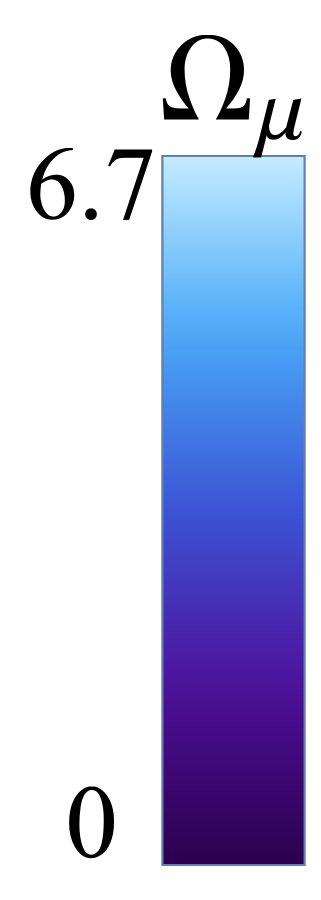}}
		\\
		\parbox{2.7cm}{\includegraphics[width=2.7cm]{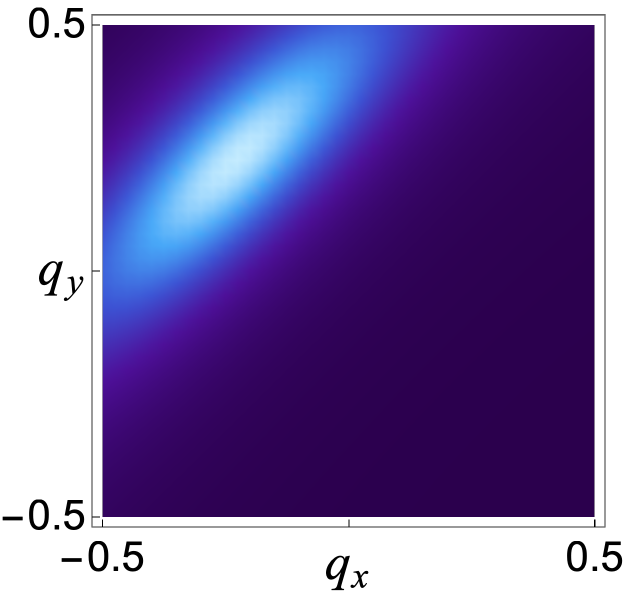} \\ (b1) $ \Omega_{z+} $, net 1.284}
		\parbox{2.7cm}{\includegraphics[width=2.7cm]{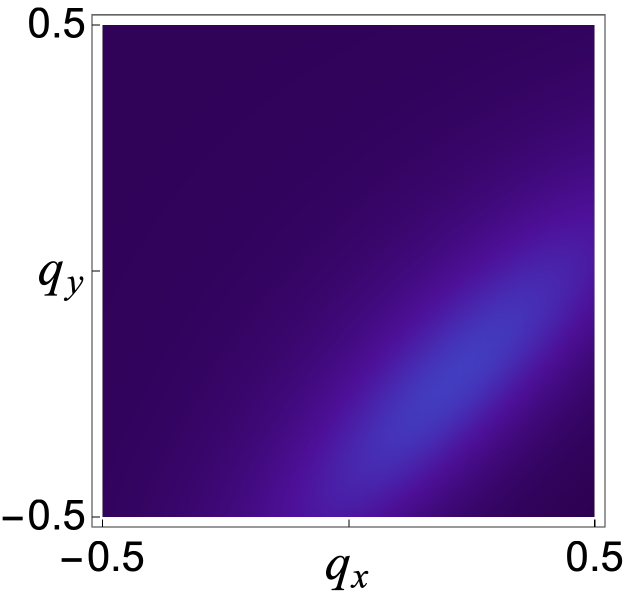} \\ (b2) $ \Omega_{z-} $, net 0.866}
		\parbox{2.7cm}{\includegraphics[width=2.7cm]{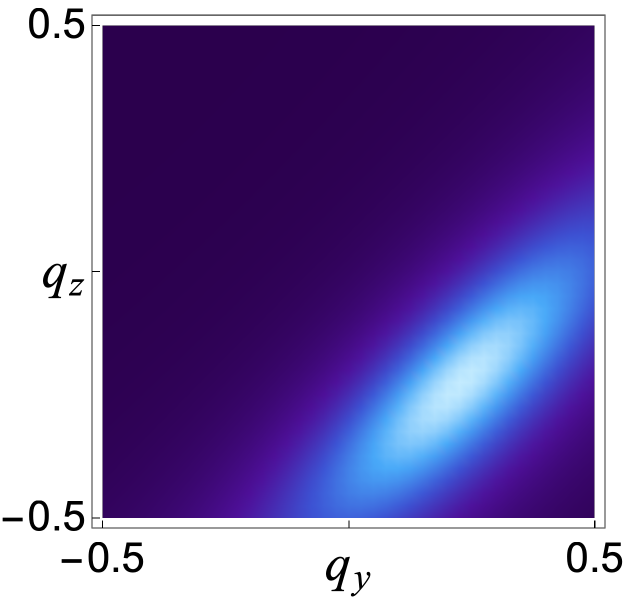} \\ (b3) $ \Omega_{x+} $, net 1.284}
		\parbox{2.7cm}{\includegraphics[width=2.7cm]{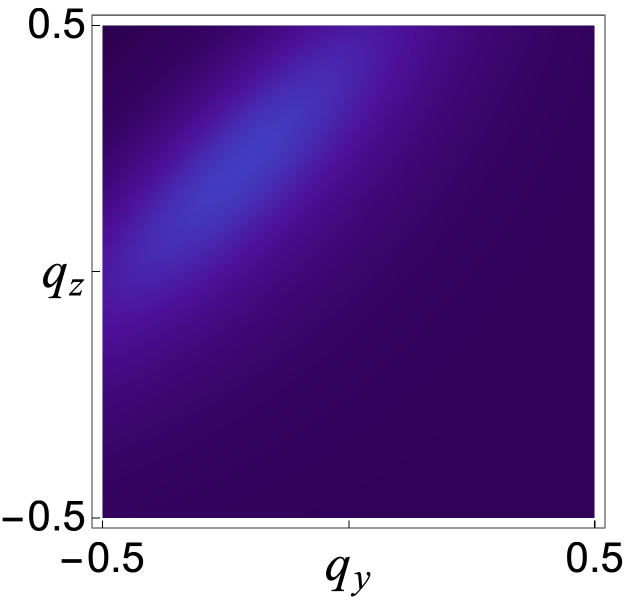} \\ (b4) $ \Omega_{x-} $, net 0.866}
		\parbox{2.7cm}{\includegraphics[width=2.7cm]{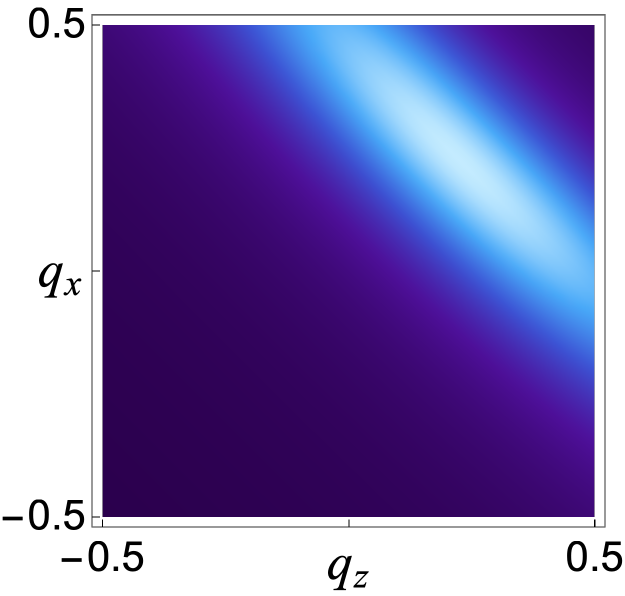} \\ (b5) $ \Omega_{y+} $, net 1.469}
		\parbox{2.7cm}{\includegraphics[width=2.7cm]{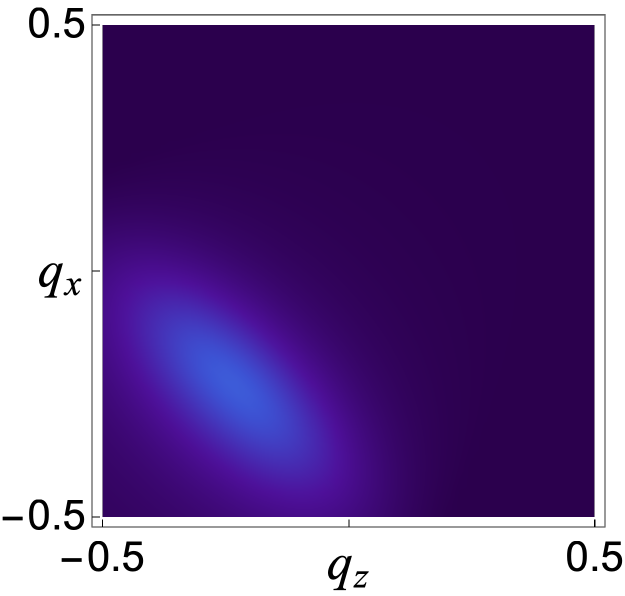} \\ (b6) $ \Omega_{y-} $, net 0.513}
		\caption{\label{fig:berry} The  Berry curvatures for the surfaces wrapping a Weyl point. Adding the net Berry curvatures up we have $ 2\pi $.}
	\end{figure}

	With the Weyl points determined, one could further apply band tomography~\cite{Hauke2014,Flaeschner2016} method for momentum states surrounding a certain Weyl point in order to determine its charge. Note that one does not need the eigenstate information throughout the whole Brillouin zone as the two bands are gapless in certain regions, except for just an arbitrarily small surface wrapping a Weyl point $ \boldsymbol{k}^{\text{(Weyl)}} $ determined previously. 
	As shown before, near the Weyl points in our model, there exists a finite region where the two bands are fully gapped in both $ E\sim0 $ and $ \pi $, which allows for populating eigenstates with bosons at a certain band~\cite{Flaeschner2016,Wintersperger2020}. As an example,  in Fig.~\ref{fig:berry} (a) we illustrate the surfaces formed by 6 faces $ q_{x,y,z} = \pm 0.5 $ of a cube, where $ \boldsymbol{q} = \boldsymbol{k} - \boldsymbol{k}^{(\text{Weyl})} $, with $ \boldsymbol{k}^{(\text{Weyl})} \approx 0.955\times(1,-1,1) $ for $ \phi=\pi/3 $, as in Fig.~\ref{fig:spectrum}. From the full information of the  Floquet eigenstates  $ |u_{n,\boldsymbol{k}} \rangle  $  given by Eq.~\eqref{eq:uf-eigen-equation}, the Berry curvature penetrating out of a plane normal to the unit vector $ \boldsymbol{e}_\mu $ can be computed as $ \Omega_\mu (\boldsymbol{k}) = \pm i \sum_{\nu\rho}\varepsilon_{\mu\nu\rho}  \left(\langle \partial_{k_\nu}u_{n,\boldsymbol{k}} | \partial_{k_\rho} u_{n,\boldsymbol{k}} \rangle \right) $, where $ \varepsilon_{\mu\nu\rho} $ is the Levi-Civita symbol, and $ \pm $ sign denotes that the unit vector penetrating out of the cube is along $ \pm \boldsymbol{e}_\mu $ directions.  Figure~\ref{fig:berry} shows momentum resolved Berry curvatures in each wrapping surface and their net fluxes $ \int_{\text{surf}} d\boldsymbol{k} \Omega_\mu(\boldsymbol{k}) $ in that plane.

	\section{Conclusion \label{sec:Conclusion}}
	
	In this paper, we have shown that three-dimensional periodically driven lattice systems can  experience a complete reconstruction of its eigenstates with drastically different features, when subjected to open boundary conditions. This corresponds to a chiral second-order Floquet skin effect. An intuitive understanding of this effect was given by considering the system at a fine-tuned point of the periodic driving, where the bulk motion can only take place forwards or backwards along a single diagonal direction. As a consequence, for open boundary conditions,  the particle is reflected back and forth between hinge-sharing surface planes, indicating a combination of  counter-propagating original  bulk eigenstates carrying opposite group velocity  into new chiral Floquet states associated with the hinge.  Thus the original bulk  motion  becomes neutralized, while the new group velocities are associated with a Goos-H\"{a}nchen shift during boundary reflections, which  are most significant close to hinges where  the particles is reflected more frequently. 
	
	This effect offers a alternative mechanism to generate hinge modes in addition to the more well-understood cases such as (Floquet) higher-order topological phases. The robustness of the chiral hinge modes here is more relevant to the special dispersion characteristic of Floquet systems, which allows for all eigenstates to be localized in a quasi-one-dimensional region close to fine-tuned point, while not necessarily require a bulk gap. While we briefly discuss in Appendix~\ref{Appendix:topo} a topological quantity specific to the fine-tuned point $ \phi=\pi/2 $, it will be an interesting future direction to explore whether such chiral hinge modes would have a stable and rigorous topological nature.   The previous analysis~\cite{Titum2016} have shown that the requirement of bulk quasi-energy gap becomes less important to characterize topologically protected chiral modes (i.e. Fig.~4 (a) (ii) in \cite{Titum2016}).  
	While weak disorder could mix different momentum states and destroy the hinge modes, it will be interesting to see whether a set of more robust hinge modes would emerge in the strongly disordered regime where all modes are localized, and to check the possible crossover or transitions when one increases the disorder strengths. 
	Our current approach and discussions would also be useful for considering observable effects in experiments, similar to the case of Weyl semimetals where surface  defects/disorder would partly destroy the Fermi arc but the chiral transport will be partially preserved therein~\cite{Wilson2018}.
	
	 It is noteworthy that even modes close to bulk center suffer from a complete reconstruction by open boundaries, hosting very different group velocity in the thermodynamic limit. Such an effect   resembles the accumulation of boundary modes in systems described by a non-hermitian Hamiltonian.
	 Yet different from this non-Hermitian skin effect, in our system we can have at most one state per lattice site on average, as Floquet states are orthogonal to each other.  Thus in our system the accumulation corresponds to the emergence of hinge-bound modes at increasing distance from the hinge. 
	Another interesting aspect is the competition or interplay between the hinge modes and the emergence of robust Wely points in our system,  so the hinge states can co-exist with the Fermi arc surface states.  The implementation of the model featuring both the second-order Floquet skin effect and the Weyl physics is straightforward with ultracold atoms in optical superlattices.

	\section{Acknowledgment}
	The authors thank E. Anisimovas and F. Nur \"{U}nal for helpful discussions. We acknowledge funding by the  European Social Fund under grant No. 09.3.3-LMT-K-712-01-0051 and the Deutsche Forschungsgemeinschaft (DFG) via the Research Unit FOR 2414 under Project No. 277974659. 
	
	\appendix

\section{Evolution operator and quasienergies along the diagonal 
\label{Appendix:Evolution-operator-along-diagonal}}

\subsection{Evolution operator}

In the bulk the stroboscopic evolution operator $U_{F}$ is generally given by Eqs.~\eqref{eq:uf}-\eqref{eq:u_mu} in the main text. Let us consider the operator $U_{F}$ for wave-vectors $\mathbf{k}$ along the cubic diagonal direction
$\boldsymbol{d}=(1,-1,1)$,
 for which

\begin{equation}
	\boldsymbol{k}= k_{0}\boldsymbol{d} \quad\text{and, thus,} \quad |\boldsymbol{k}|=\sqrt{3}|k_{0}|\,.  \label{eq:diagonal-Appendix}
\end{equation}
 In that case Eqs.~\eqref{eq:uf}-\eqref{eq:u_mu} simplify to 
\begin{equation}
 U_{F}=\left[U\left( k_{0}\right)U\left(- k_{0}\right)\right]^{3}\,,  \label{eq:uf-1}
\end{equation}
where 
\begin{equation}
U\left(k_{0}\right)
=\tau_0 \cos\phi-\mathrm{i}\tau_{k_{0}}\sin\phi\,, \label{eq:U_-k_0U_k_0}
\end{equation}
with $ \tau_{k_0} = \tau_1 \cos{k_0} + \tau_2 \sin {k_0}$,  where $\tau_{1,2,3}$ are Pauli matrices for the sublattice freedom,  $\tau_{0}$ being a unit $2\times2$ matrix. Explicitly one, thus, has 
\begin{equation}
   	U\left(k_{0}\right)U\left(-k_{0}\right)
	=\left[\cos^{2}\phi-\sin^{2}\phi \cos\left(2k_{0}\right)\right]\tau_0+\mathrm{i}b\,, 
\label{eq:U_-k_0U_k_0-1}
\end{equation}
with
\begin{equation}
 b=\sin^{2}\phi \sin\left(2k_{0}\right)\tau_{3}-\sin(2\phi) \cos k_{0}\tau_{1}. 
\label{eq:d}
\end{equation}

\subsection{Quasi-energies}

The evolution operator $U_{F}=e^{-\mathrm{i}H_{F}}$ defines the quasienergies
representing the eigenvalues of the of the Floquet Hamiltonian
$H_{F}$, which describes the stroboscopic time evolution  in multiples of the driving period $T=1$.
Using Eqs. \eqref{eq:uf-1} and \eqref{eq:U_-k_0U_k_0-1} for the
evolution operator, one arrives at the following equation for the
quasi-energies $E_{\mathbf{k}}$ 
\begin{equation}
\cos\left(E_{\mathbf{k}}/3\right)=\cos^{2}\phi-\sin^{2}\phi\cos\left(2k_{0}\right)\,.\label{eq:quasi-energy_equation}
\end{equation}
This provides the dispersion (modulo $2\pi$) along the diagonal
$k_{x}=-k_{y}=k_{z}=k_{0}$ 
\begin{equation}
E_{k_0\boldsymbol{d},\gamma}=3\gamma\arccos\left[\cos^{2}\phi-\sin^{2}\phi\cos\left(2k_{0}\right)\right],\,\,\mathrm{with}\,\,\gamma=\pm1\,.\label{eq:quasi-energy_equation-1}
\end{equation}

In particular, quasienergies  $E_{k_0\boldsymbol{d},\gamma}=\pi$ (modulo $2\pi$) correspond to \begin{equation}
\cos^{2}\phi-\sin^{2}\phi\cos\left(2k_{0}\right)=1/2\,,\label{eq:pi-point-equation}
\end{equation}
and thus 
\begin{equation}
\cos\left(2k_{0}\right)=\frac{1/2-\sin^{2}\phi}{\sin^{2}\phi}\,,\label{eq:pi-point-cos(2k_0)}
\end{equation}
giving (modulo $\pi$)
\begin{equation} k_{0}=\pm\left(1/2\right)\arccos\left[\left(1/2-\sin^{2}\phi\right)/\sin^{2}\phi\right]\,.  \label{eq:k_0}
\end{equation}
At the fine tuned point ($\phi=\pi/2$) the condition Eq.\eqref{eq:pi-point-equation}
reduces to 
\begin{equation}
 \cos\left(2k_{0}\right)=-1/2\,,\quad\mathrm{giving}\quad k_{0}=\pm\pi/3\,. \label{eq:pi-point-cos(2k_0)-phi--pi/2}
\end{equation}
On the other hand, at $\phi=\pi/6$ one has $\sin^{2}\phi=1/4$, so
that 
\begin{equation}
\cos\left(2k_{0}\right)=1,\,\quad\mathrm{giving}\quad k_{0}=0\,.\label{eq:pi-point-cos(2k_0)-phi--pi/6}
\end{equation}

In this way, two band touching points are formed at quasienergy $\pi$ 
for $\pi/6<\phi<\pi/2$, as well as for $\pi/2<\phi<5\pi/6$ (beyond
the fine tuning point at $\phi=\pi/2$). By taking $\phi<\pi/6$ or
$\phi>5\pi/6$, Eq.\eqref{eq:pi-point-equation} can no longer be
fulfilled, so a band gap is formed at quasienergy $\pi$.

\section{ Stroboscopic hinge motion at fine tuning\label{Appendix:Hinge states}}

 In this  Appendix we give a detailed description of the stroboscopic real-space dynamics of the system at fine tuning, 
$\phi=\pi/2$, giving rise to chiral hinge-bound Floquet modes.  We will consider the hinge that is shared by the two 
surface planes oriented in the $-x$ and $-y$ direction, which is parallel to the $z$-axis. 
The projection of the particle's trajectory in the $xy$ plane is illustrated in Figs. \ref{fig:hinge-stroboscop-1} and
 \ref{fig:hinge-stroboscop-2}. 
 A particle of sublattice $s=+1,-1\equiv A, B$ is translated by $2s\bm{d}$ during each driving cycle, provided $x+2s\ge1$ and $y-2s\ge1$  to ensure it does not hit  any the boundary plane.
In that case the state-vector $\left|s,x,y,z\right\rangle $
transforms according to the following rule after a single driving
period:
\begin{equation}
U_{F}\left|s,x,y,z\right\rangle =-\left|s,x+2s,y-2s,z+2s\right\rangle \,.\label{eq:free_z}
\end{equation}
The particle thus propagates with a stroboscopic velocity $\mathbf{v}=2s\left(1,-1,1\right)$
in opposite directions $s=\pm 1$ for different sublattices $A$ and $B$.

Suppose initially the particle occupies a site of the sublattice $B$
at the boundary $y=1$ situated $M$ sites away from the
hinge ($x=M+1$) with odd $M+z$, so that  $s=B=-1$. The corresponding
initial state vector is $\left|s,M+1,1,z\right\rangle \equiv\left|B,M+1,1,z\right\rangle $.
 The subsequent stroboscopic trajectory projected to the $xy$ plane
is shown in Fig. \ref{fig:hinge-stroboscop-1} for $M=4$ and
in Fig. \ref{fig:hinge-stroboscop-2} for $M=5$. Generally it takes
$\left(2M+1\right)$ driving periods for the system to return to its
initial state $\left|M+1,1,z\right\rangle $. To see this, consider
the stroboscopic evolution of the particle with an even $M>2$ and
odd $z$. The stroboscopic motion of the particle then splits
into four bulk and four boundary segments illustrated in Fig. \ref{fig:hinge-stroboscop-1}
for $M=4$. 

During the first $M/2$ driving periods the particle undergoes the
bulk ballistic motion along the sites of the $B$ sublattice, and the
state vector transforms as $\left|B,M+1,1,z\right\rangle \rightarrow\left|B,1,M+1,z-M\right\rangle $.
Subsequently the particle is reflected from the plane $x=1$ to
a site of the sublattice $A$ situated closer to the hinge, $\left|B,1,M+1,z-M\right\rangle \rightarrow\left|A,2,M-1,z+2-M\right\rangle $,
as shown in Fig.~\ref{fig:model}(c) of the main text. During the
next $M/2-1$ driving periods the particle propagates ballistically
along the sites of the $A$ sublattices, giving $\left|A,2,M-1,z+2-M\right\rangle \rightarrow\left|A,M,1,z\right\rangle $.
The subsequent reflection from the plane $y=1$ brings the particle
to a site of the $B$ sublattice situated further away to from the
hinge, $\left|A,M,1,z\right\rangle \rightarrow\left|B,M,2,z-2\right\rangle $.
The evolution takes place in the similar way during final
four segments. Explicitly the full stroboscopic dynamics is given
by: 
\begin{align}\label{eq:free-B1}
&\left(U_{F}\right)^{M/2}\left|B,M+1,1,z\right\rangle =\left(-1\right)^{M/2}\left|B,1,M+1,z-M\right\rangle \,,
\\ \label{eq:B--A1}
&U_{F}\left|B,1,M+1,z-M\right\rangle =-\mathrm{i}\left|A,2,M-1,z+2-M\right\rangle \,,
\\ \nonumber
&\left(U_{F}\right)^{M/2-1}(-\mathrm{i})\left|A,2,M-1,z+2-M\right\rangle \\ \label{eq:free-A1}
& =\mathrm{i}\left(-1\right)^{M/2}\left|A,M,1,z\right\rangle \,,
\\ \label{eq:A--B1}
&U_{F}\mathrm{i}\left|A,M,1,z\right\rangle =\left|B,M,2,z-2\right\rangle \,,
\\
\label{eq:free-B2}
&\left(U_{F}\right)^{M/2-1}\left|B,M,2,z-2\right\rangle =-\left(-1\right)^{M/2}\left|B,2,M,z-M\right\rangle \,,
\\
\label{eq:B--A2}
&U_{F}(-1)\left|B,2,M,z-M\right\rangle =\mathrm{i}\left|A,1,M,z-M\right\rangle \,,
\\ \nonumber
&\left(U_{F}\right)^{M/2-1}\mathrm{i}\left|A,1,M,z-M\right\rangle 
\\ \label{eq:free-A2}
& =(-\mathrm{i})\left(-1\right)^{M/2}\left|A,M-1,2,z-2\right\rangle \,,
\\  \label{eq:A--B2}
&U_{F}(-\mathrm{i})\left|A,M-1,2,z-2\right\rangle =-\left|B,M+1,1,z-2\right\rangle \,,
\end{align}
In this way, after $\left(2M+1\right)$ driving periods the particle
returns back to the initial position $\left(M+1,1\right)$ in the $xy$
plane and is shifted by $2$ lattice units to an equivalent point  of the sublattice $B$
in the direction opposite to the $z$ axis. The same holds for the
initial state vector $\left|B,M+1,1,z\right\rangle $ characterized
by an odd $M$ and even $z$ (see Fig. \ref{fig:hinge-stroboscop-2}
for $M=5$), as well as for a particle situated closer to the hinge
($0\le M\le3$) where the reflections can take place simultaneously from
both planes $x=1$ and $y=1$, as illustrated in Fig.~\ref{fig:model}(d)
in the main text. Thus one can write for any distance $M\ge 0$ from the
hinge: 
\begin{equation}
\left(U_{F}\right)^{2M+1}\left|B,M+1,1,z\right\rangle =-\left|B,M+1,1,z-2\right\rangle \,, \label{eq:U_F^2M+1-coordinate}
\end{equation}
This means the particle propagates along the hinge in the $-z$ direction
with the stroboscopic velocity equal to  $-2/\left(2M+1\right)$. The relations
analogous to Eq.\eqref{eq:U_F^2M+1-coordinate} hold for all $2M+1$
states of the stroboscopic sequence featured in Eqs. \eqref{eq:B--A1}-\eqref{eq:A--B2}

The origin of such chiral hinge states can be explained as follows.
 The particle in the sublattice $B$ is reflected to a site of the
$A$ sublattice situated closer to the hinge, whereas the particle
in the sublattice A is reflected to a site of the sublattice $B$
situated further away from the hinge, as one can see in Figs. \ref{fig:hinge-stroboscop-1}
and \ref{fig:hinge-stroboscop-2}, as well as in Eqs.~\eqref{eq:B--A1},
\eqref{eq:A--B1}, \eqref{eq:B--A2}, \eqref{eq:A--B2}. Consequently
the number of $B$ sites visited over all $2M+1$ driving periods
($M+1$) exceeds the corresponding number of $A$ sites ($M$). The
four reflections do not yield any total shift of the particle in the
$z$ direction. On the other hand, the ballistic motion between sites
the same sublattice $B$ ($A$) is accompanied by a shift by $2$
lattice sites in the $z$ ($-z$) direction for each driving period.
This leads to the overall shift of the particle to an equivalent site
in the $-z$ direction is due to the difference in the number of the
visited $B$ and A sites after $2M+1$ driving periods.

\section{Non-Hermitian Hamiltonian corresponding to stroboscopic operator  \label{Appendix:Non-Hermit}}

Recently it was suggested \cite{Bessho2020Duality} to associate a
non-Hermitian Hamiltonian $H_{NH}\left(\mathbf{k}\right)$ to the
momentum space stroboscopic evolution operator $\mathrm{i}U_{F}\left(\mathbf{k}\right)$.
Let us consider such a non-Hermitian Hamiltonian for our 3D periodically
driven lattice
\begin{equation}
H_{NH}=\mathrm{i}U_{F}\label{eq:H_NF}\,.
\end{equation}
 For the fine tuned driving ($\phi=\pi/2$) the bulk stroboscopic
evolution operator corresponds to a non-Hermitian Hamiltonian describing
a unidirectional transfer between the lattice sites along the diagonal
$\mathbf{d}=(1,-1,1)$
and in the opposite direction $-\mathbf{d}$ for the sublattices $A$
and $B$, respectively: 
\begin{equation}
H_{NH}^{bulk}=-\mathrm{i}\sum_{\mathbf{r}_{A}}\left|A,\mathbf{r}_{A}+2\mathbf{d}\right\rangle \left\langle A,\mathbf{r}_{A}\right|-\mathrm{i}\sum_{\mathbf{r}_{B}}\left|B,\mathbf{r}_{B}-2\mathbf{d},\right\rangle \left\langle B,\mathbf{r}_{B}\right|\,.\label{eq:H_NH-bulk-r}
\end{equation}

The open boundary conditions for the hinge corresponding to $x\ge1$
and $y\ge1$ are
obtained by imposing a constraint on the state-vectors entering the
real space non-Hermitian Hamiltonian \eqref{eq:H_NH-bulk-r} 
\begin{equation}
\left|s,\mathbf{r}_{s}\right\rangle =0\quad\mathrm{for}\quad\mathbf{r}_{s}\cdot\mathrm{e}_{x,y}\le0,\,\,\mathrm{with}\,\,s=A,B\,.\label{eq:OBC}
\end{equation}
The bulk non-Hermitian Hamiltonian \eqref{eq:H_NH-bulk-r} supplied with
the open boundary conditions \eqref{eq:OBC} describes a unidirectional
coupling between unconnected linear chain of the $A$ or $B$ sites
terminating at the hinge planes. The eigenstates of each linear chain
represent non-Hermitian skin modes which are localized at one end
of the chain depending on the direction of asymmetric hopping \cite{budich2020RMP,Ueda2020AdvPhys}.
In the present situation such skin modes would be trivially localised on
different planes of the hinge for the chains comprising different
sublattice sites $A$ or $B$, and no chiral motion is obtained along
the hinge.

Yet the open boundary conditions \eqref{eq:OBC} are not sufficient
to properly represent the boundary behavior of a particle in our periodically driven
lattice.
In fact, bulk non-Hermitian Hamiltonian~\eqref{eq:H_NH-bulk-r}  supplied with
the boundary conditions \eqref{eq:OBC}  is no longer a unitary operator. Thus one can not associate such an non-Hermitian
operator with the evolution operator, in 
contradiction with Eq.~\eqref{eq:H_NF}.
The unitarity is restored by adding to 
$H_{NH}^{bulk}$ extra terms 
[in addition to the conditions \eqref{eq:OBC}]
to include effects of the chiral backward reflection at the hinge
planes to a neighboring linear chain composed of the sites of another
sublattice. These terms are described by Eqs.\eqref{eq:B--A1}, \eqref{eq:A--B1},
\eqref{eq:B--A2} and \eqref{eq:A--B2}, and correspond to the dashed lines in  
Figs.~\ref{fig:hinge-stroboscop-1}-\ref{fig:hinge-stroboscop-2}. The chiral reflection
appears due to the micromotion during the driving period involving six steps (see Fig.~\ref{fig:model}(b)), and 
is a characteristic feature of the periodically driven 3D lattice.
 As discussed in the previous Appendix \ref{Appendix:Hinge states},
the backreflection leads to  the chiral motion along the hinge.

		\section{Tentative discussions on topological origins}\label{Appendix:topo}
		
		 An interesting question is whether the robust chiral hinge states are a consequence of topological properties of the driven system.  However, as we see from column (2) of Fig.~\ref{fig:spectrum}, the quasi-energies of hinge states are fully mixed with the bulk spectrum, and therefore no traditional topological band theories for gapped or semimetallic systems apply. Here, we  apply an approach similar to  that used in refs.~\cite{Kitagawa2010} and \cite{Budich2017} for computing topological invariant at a fine-tuned parameter point, where eigenstates are analytically obtainable, and a topological invariant based on eigenstate projections can be calculated.
		
		In   section \ref{Subsec:Fine-tunned-driving} we have obtained equation \eqref{eq:U_F^2M+1--B,M+1,1,k_z}  describing the evolution  over $2M+1$ driving periods of the $M$th hinge state 
		$ \left|M,k_{z},p\right\rangle $  at the fine tuned point.	Using this equation one can define the  the quasienergy winding number \cite{Kitagawa2010} for the $M$th hinge state  via the Floquet evolution operator over the $2M+1$ driving periods:  
		\begin{equation} \label{eq:W_M}
			W_M=\frac{1}{2\pi i}\int_{0}^{\pi} dk_z U_{k_{z},2M+1}^{-1}  \partial_{k_z}  U_{k_{z},2M+1}  = 1 \,,
		\end{equation} 
		where 
		\begin{equation} \label{eq:U_k_z,M}
			U_{k_{z},2M+1} = \left\langle M,k_{z},p\right| \left( U_{F} \right)^{2M+1}  \left|M,k_{z},p\right\rangle =-e^{2ik_{z}}  \,,
		\end{equation} 
		 with $p=0,1,\ldots2M $.  In Eq.~\eqref{eq:W_M}  the integration over $k_z$ extends  over one Brillouin zone of width $\pi$, as the distance between two non-equivalent lattice sites equals to $2$ in $z$ direction.  A similar procedure can be applied to the opposite hinge at $ (x,y)=(L,L) $, where the hinge modes shown in blue in column  (2) of Fig.~\ref{fig:spectrum} are characterized by the opposite  group velocity and thus the winding number  is  opposite $ W_M=-1 $.

		The rigorous quantization of the topological invariant $W_M$ is associated with  the  fine tuning, $ \phi=\pi/2 $.	
		However, the spatial separation between hinge modes of opposite chirality  allows to preserve their chiral character  also away from  the fine tuning point $\phi=\pi/2$,  as one can see in  column (2) of Fig.~\ref{fig:spectrum}.    Thus, the formation of bulk states via the mixing of hinge modes of opposite chirality happens mostly in the center of the system, where hinge modes corresponding to large $M$ and small chiral velocities lie close by to their counter propagating partners associated with the opposite hinge. 
		In turn, the states with the largest chiral velocity, which are situated close to the hinge and far away from counter-propagating modes  of the opposite hinge, are  much less affected by a small detuning.  
		
		In this way, tuning away from $\phi=\pi/2$ we find a crossover (rather than a topological transition) in which the chiral hinge modes are gradually destroyed, as can be observed in the real-space plots in columns (2) and (3) of Fig.~\ref{fig:spectrum}. 
		Note that already a small deviation from the fine tuned point $ \phi=\pi/2 $ destroys the chiral hinge states  in a narrow region near $ k_z=0$ and $ E= \pi $, as can be seen from the spectrum shown for $\phi=\pi/3$ in column (2) of Fig.~\ref{fig:spectrum}.  In this spectral surface Fermi arc states are formed, which equally provide definite chiral transport,  yet around the surfaces rather than the hinges. 
		Thus a fraction of the chiral hinge states is transformed into Fermi arc surface states in the vicinity of the Weyl points. The latter states extend to a larger and larger spectral area as the detuning increases.


	%

\end{document}